\documentclass[extra,mreferee]{gji}

\usepackage{fancyhdr}
\usepackage{amsmath,amssymb,mathrsfs}
\usepackage{setspace}
\usepackage{graphicx}
\usepackage{color,xcolor}
\usepackage{lineno,hyperref}
\hypersetup{colorlinks=true,
            linkcolor=black,
            citecolor=black}
\DeclareMathAlphabet\mathbfcal{OMS}{cmsy}{b}{n}

\modulolinenumbers[5]
\allowdisplaybreaks[2]

\makeatletter
\newdimen\rh@wd
\newdimen\rh@hta
\newdimen\rh@htb
\newbox\rh@box
\def\rh@measure#1{\setbox\rh@box=\hbox{$#1$}
    \rh@wd=\wd\rh@box \rh@hta=\ht\rh@box}

\def\widecheck#1{\rh@measure{#1}%
  \setbox\rh@box=\hbox{$\widehat{\vrule height \rh@hta width\z@ \kern\rh@wd}$}%
  \rh@htb=\ht\rh@box \advance\rh@htb\rh@hta \advance\rh@htb\p@
  \ooalign{$\vrule height \ht\rh@box width\z@ #1$\cr
           \raise\rh@htb\hbox{\scalebox{1}[-1]{\box\rh@box}}\cr}}
\makeatother

\newcommand{\Ndof}{N_\mathrm{dof}^{D^e}}
\newcommand{\dd}{\,\mathrm{d}}
\def\mathbi#1{\textbf{\em #1}}
\newcommand{\superscript}[1]{\ensuremath{^{\textrm{#1}}}}
\newcommand{\etal}{et al.}

\newcommand{\innprod}[1]{(#1)}

\newcommand{\innprodDeS}[1]{\innprod{#1}_{\DeS}}
\newcommand{\innprodDeF}[1]{\innprod{#1}_{\DeF}}

\newcommand{\Os}{{\Omega_\mathrm{S}}}
\newcommand{\Of}{{\Omega_\mathrm{F}}}
\newcommand{\rSS}{{\scriptscriptstyle\mathrm{SS}}}
\newcommand{\rFF}{{\scriptscriptstyle\mathrm{FF}}}
\newcommand{\rFS}{{\scriptscriptstyle\mathrm{FS}}}
\newcommand{\rSF}{{\scriptscriptstyle\mathrm{SF}}}
\newcommand{\Ssf}{{\Sigma_{\rSF}}}
\newcommand{\Sfs}{{\Sigma_{\rFS}}}
\newcommand{\Sff}{{\Sigma_{\rFF}}}
\newcommand{\Sss}{{\Sigma_{\rSS}}}

\newcommand{\tsC}{\mathbi{C}}

\newcommand{\tsE}{\mathbi{E}}
\newcommand{\tsS}{{(\tsC \tsE\,)}}
\newcommand{\tsH}{\mathbi{H}}
\newcommand{\tsA}{\mathcal{A}}

\newcommand{\tsQ}{{\mathbfcal{Q}}}
\newcommand{\tsLa}{{\Lambda}}

\newcommand{\ttsA}{\widetilde{\mathcal{A}}}

\newcommand{\ttsQ}{\widetilde{\tsQ}}
\newcommand{\ttsLa}{\widetilde{\tsLa}}

\newcommand{\vev}{\mathbi{v}}
\newcommand{\vew}{\mathbi{w}}
\newcommand{\vef}{\mathbi{f}}
\newcommand{\veg}{\mathbi{g}}
\newcommand{\ven}{\mathbi{n}}
\newcommand{\veq}{\mathbi{q}}
\newcommand{\vep}{\mathbi{p}}

\newcommand{\verr}{{\mathbi{e}}}
\newcommand{\tverr}{{\widetilde\verr}}
\newcommand{\vxi}{{\boldsymbol\eta}}
\newcommand{\veta}{{\boldsymbol\epsilon}}
\newcommand{\tvxi}{\widetilde{\boldsymbol\eta}}
\newcommand{\tveta}{\widetilde{\boldsymbol\epsilon}}
\newcommand{\tvev}{\widetilde{\mathbi{v}}}
\newcommand{\tvew}{\widetilde{\mathbi{w}}}

\newcommand{\tveg}{\widetilde{\mathbi{g}}}
\newcommand{\tveq}{\widetilde{\mathbi{q}}}
\newcommand{\tvep}{\widetilde{\mathbi{p}}}
\newcommand{\trho}{\widetilde\rho}
\newcommand{\tlambda}{\widetilde\lambda}

\newcommand{\tE}{\widetilde E}
\newcommand{\tP}{\tlambda\tE}
\newcommand{\tH}{\widetilde H}

\newcommand{\tQ}{\tlambda\tH}

\newcommand{\tf}{\widetilde f}
\newcommand{\Dt}[1]{\frac{\partial #1}{\partial t}}
\newcommand{\De}{{D^{\mathrm{e}}}}

\newcommand{\DeS}{{D^{\mathrm{e}}_{\scriptscriptstyle\mathrm{S}}}}
\newcommand{\DeF}{{D^{\mathrm{e}}_{\scriptscriptstyle\mathrm{F}}}}
\newcommand{\Tss}{{\Sigma_{\rSS}^{\mathrm{e}}}}
\newcommand{\Tfs}{{\Sigma_{\rFS}^{\mathrm{e}}}}
\newcommand{\Tsf}{{\Sigma_{\rSF}^{\mathrm{e}}}}
\newcommand{\Tff}{{\Sigma_{\rFF}^{\mathrm{e}}}}
\newcommand{\avg}[1]{\{\!\{#1\,\}\!\}}
\newcommand{\jmp}[1]{{\,[\![#1\,]\!]}}

\newcommand{\Flux}{\mathcal{F}}
\newcommand{\tFlux}{\widetilde{\Flux}}

\newcommand{\Fluxc}{\mathcal{F}^\mathrm{C}}
\newcommand{\tFluxc}{\widetilde{\Flux}^\mathrm{C}}
\newcommand{\Fluxp}{\mathcal{F}^\mathrm{P}}
\newcommand{\tFluxp}{\widetilde{\Flux}^\mathrm{P}}

\newcommand{\Energy}{{\mathcal{E}}}

\newcommand{\norm}[1]{{\left\Vert #1 \,\right\Vert}}
\newcommand{\Norm}[1]{{\Vert #1 \,\Vert}}

\newcommand{\ddt}[1]{\frac{\partial #1}{\partial t}}
\newcommand{\DDt}[1]{\frac{\dd #1}{\dd t}}

\newcommand{\mm}{{-}}
\newcommand{\pp}{{+}}
\newcommand{\resv}{\mathrm{res}_{\scriptscriptstyle\mathrm{S}}}
\newcommand{\ressS}{\mathrm{res}_\rSS}
\newcommand{\ressF}{\mathrm{res}_\rSF}
\newcommand{\tresv}{\widetilde{\mathrm{res}}_{\scriptscriptstyle\mathrm{F}}}
\newcommand{\tressS}{\widetilde{\mathrm{res}}_\rFS}
\newcommand{\tressF}{\widetilde{\mathrm{res}}_\rFF}

\newcommand{\normVS}[1]{\Norm{#1}\hspace{-1mm}\phantom{|}_{L^2(\DeS;\,\tsQ,\tsLa)}}
\newcommand{\normVF}[1]{\Norm{#1}\hspace{-1mm}\phantom{|}_{L^2(\DeF;\,\ttsQ,\ttsLa)}}
\newcommand{\NormVS}[1]{\norm{#1}_{L^2(\DeS;\,\tsQ,\tsLa)}}
\newcommand{\NormVF}[1]{\norm{#1}_{L^2(\DeF;\,\ttsQ,\ttsLa)}}
\newcommand{\normSS}[1]{\Norm{#1}_{L^2(\Tss)}}
\newcommand{\normFF}[1]{\Norm{#1}_{L^2(\Tff)}}
\newcommand{\normSF}[1]{\Norm{#1}_{L^2(\Tsf)}}
\newcommand{\normFS}[1]{\Norm{#1}_{L^2(\Tfs)}}

\newcommand{\Sumss}{{\sum_{\mathrm{e}}}}
\newcommand{\Sumff}{{\sum_{\mathrm{e}}}}
\newcommand{\Sumfs}{{\sum_{\mathrm{e}}}}
\newcommand{\Sumsf}{{\sum_{\mathrm{e}}}}
\newcommand{\SumS}{{ \sum_{\mathrm{e}}}}
\newcommand{\SumF}{{ \sum_{\mathrm{e}}}}
\newcommand{\JmpRd}{\hspace{-3mm}}
\newcommand{\Half}{\tfrac12}

\newcommand{\ColorRed}{\color{black}}

\title[\texttt{DG method for acousto-elastic waves} ]
 {A Discontinuous Galerkin method with a modified 
  penalty flux for the propagation and
  scattering of acousto-elastic waves}

\author[R. Ye \etal]{ 
\small Ruichao Ye$^1$, Maarten V. de Hoop$^{2}$, Christopher L. Petrovitch$^3$, 
Laura J. Pyrak-Nolte$^4$, Lucas C. Wilcox$^5$ \\
$^1$ Department of Earth Science, Rice University, Houston TX, USA\\
$^2$ Simons Chair in Computational and Applied Mathematics and Earth Science, Rice University, Houston TX, USA \\
$^3$ Applied Research Associates, Inc., Raleigh-Durham NC, USA\\
$^4$ Department of Physics, Purdue University, West Lafayette IN, USA \\
$^5$ Department of Applied Mathematics,
        Naval Postgraduate School, Monterey CA, USA
}

\begin{document}

\maketitle 

\begin{summary}
We develop an approach for simulating acousto-elastic wave phenomena,
including scattering from fluid-solid boundaries, where the solid is
allowed to be anisotropic, with the Discontinuous Galerkin method. We
use a coupled first-order elastic strain-velocity, acoustic
velocity-pressure formulation,
and append penalty terms based on
interior boundary continuity conditions to the numerical (central)
flux so that the consistency condition holds for the discretized
Discontinuous Galerkin weak formulation. We incorporate the
fluid-solid boundaries through these penalty terms and obtain a stable
algorithm.
Our approach avoids the diagonalization into
polarized wave constituents such as in the approach based on solving
elementwise Riemann problems.
\end{summary}



\section{Introduction}

The accurate computation of waves in realistic three-dimensional Earth
models represents an ongoing challenge in local, regional, and global
seismology. Here, we focus on simulating coupled acousto-elastic wave
phenomena including scattering from fluid-solid boundaries, where the
solid is allowed to be anisotropic, with the Discontinuous Galerkin
method. Of particular interest are applications in geophysics, namely,
marine seismic exploration and global Earth inverse problems using
earthquake-generated seismic waves as the probing field. In the first
application, we are concerned with the presence of the ocean bottom
and in the second one with 
the core-mantle-boundary (CMB) 
and inner-core-boundary (ICB).
Our formulation closely follows the
analysis of existence of (weak) solutions of hyperbolic first-order
systems of equations by 
\cite{Blazek2013}. We use an
unstructured tetrahedral mesh with local refinement to accommodate
highly heterogeneous media and complex geometries, which is also an
underlying motivation for employing the Discontinuous Galerkin method
from a computational point of view.

In the past three decades, a wide variety of numerical techniques has
been employed in the development of computational methods for
simulating seismic waves. 
The most widely used one is based on the finite difference method
[e.g., 
\cite{Madariaga1976} and
\cite{Virieux1986}]. This method has been applied to
computing the wavefield in three-dimensional local and regional models
[e.g., 
\cite{Graves1996} and 
\cite{Ohminato1997}]. The use of optimal or compact
finite-difference operators has provided a certain improvement [e.g.,
\cite{Zingg1996} and 
\cite{Zingg2000}]. Methods
that resort to spectral and pseudospectral techniques based on global
gridding of the model have also been used both in regional [e.g.,
\cite{Carcione1994}] and global [e.g., 
\cite{Tessmer1992} and 
\cite{Igel1999}] seismic wave
propagation and scattering problems. However, because of the use of
global basis functions (polynomial: Chebyshev or Legendre, or
harmonic: Fourier), these techniques are limited to coefficients which
are (piecewise) sufficiently smooth. The finite difference method
suffers from a limited accuracy in the presence of 
a free surface or surface discontinuities with topography
within the model [e.g., 
\cite{Robertsson1996} and 
\cite{Symes2009}]. 
A procedure for
the stable imposition of free-surface boundary conditions for a
second-order formulation can be found in 
\cite{Appelo2009}. Another approach, belonging to a broader family of
interface methods, handles both free surfaces [e.g., 
\cite{Lombard2008}] and fluid-solid interfaces [e.g., 
\cite{Lombard2004}] in such a way, conjectured by the authors, that
enables higher-order accuracy to be obtained. 
\cite{Kozdon2013} use summation-by-parts finite difference
operators along with a weak enforcement of boundary conditions to
develop a multi-block finite difference scheme which achieves
higher-order accuracy for complex geometries.

A key development in the computation of seismic waves has been based
on the spectral element method (SEM) 
[\cite{Komatitsch2002}]. In its
original formulation, in terms of displacement 
[\cite{Komatitsch1998}], continuity of displacement and velocity is
enforced everywhere within the model. In the case of a boundary
between an inviscid fluid and a solid, however, the kinematic boundary
condition is perfect slip; therefore, only the normal component of
velocity is continuous across such a boundary, and thus this
formulation is not applicable. Some classical finite-element 
methods (FEMs) alternatively introduce coupling conditions on fluid-solid
interfaces between
displacement in the solid and pressure in the fluid 
[e.g. \cite{Zienkiewicz1978, Berm'udez1999}].

The FEM and SEM are commonly (but not exclusively) 
based on the second-order form of the system
of equations describing acousto-elastic waves. In this case, the
acousto-elastic interaction is effected by coupling the respective
wave equations through appropriate interface conditions. To resolve
the coupling, a predictor-multicorrector iteration at each time step
has been used 
[\cite{Komatitsch2000}, 
\cite{Chaljub2003}]. A computationally more efficient time
stepping method for global seismic wave propagation accommodating the
effects of fluid-solid boundaries, as well as transverse isotropy with
a radial symmetry axis and radial models of attenuation, was proposed
in 
\cite{Komatitsch2005}. 
It uses a velocity potential
formulation and a second-order accurate Newmark time integration, in
which a time step is first performed in the acoustic fluid and then in
the elastic solid using interface values based on the fluid
solution. Currently the SEM is used in a variety of
implementations in global and regional seismic simulation, with the
effects of variations in elastic parameters, density, ellipticity,
topography and bathymetry, fluid-solid interfaces, anisotropy, and
self-gravitation included [e.g. \cite{Carrington2008}].

In contrast to classical finite element discretizations, the
Discontinous Galerkin (DG) method imposes continuity of approximate
solutions between elements only weakly through a numerical flux. The
Discontinuous Galerkin method has been employed for solving
second-order wave equations in both the acoustic and elastodynamic
settings [e.g. 
\cite{Rivi`ere2000}, 
\cite{Grote2006}, 
\cite{Chung2006} and 
\cite{DeBasabe2008}]. 
\cite{Etienne2010} employ a central numerical flux in a
DG scheme combined with a leap-frog time integration for the
velocity-stress elastic-wave formulation. 
\cite{Dumbser2006, Kaser2008} developed a non-conservative
formulation with an upwind numerical flux using 
only the material properties from the side of the interface
that is opposite to the outer normal direction.
\cite{Wilcox2010} derived an upwind numerical flux by solving
the exact Riemann problem on interior boundaries of each element with
material discontinuities based on a velocity-strain formulation of the
coupled acousto-elastic equations. 

{\ColorRed
In this work, we essentially extend the upwind flux, given by 
\cite{Warburton2013} for hyperbolic systems, to a penalty flux
based on the boundary continuity condition for general 
fluid-solid interfaces. 
}
The novelties of our approach are the following: we
\begin{enumerate}
\item
use a coupled \textit{first-order} elastic strain-velocity, acoustic
velocity-pressure formulation,
\item
obtain a self-consistent Discontinuous Galerkin weak formulation
\textit{without diagonalization} into polarized wave constituents,
\item
\textit{append penalty terms}, derived from interior boundary
continuity conditions, with an appropriate weight \textit{to the
  numerical (central) flux} so that the consistency condition holds
for the discretized Discontinuous Galerkin weak formulation,
\item
\textit{incorporate fluid-solid boundaries} through the mentioned
penalty terms.
\end{enumerate}
We note that the DG method is naturally adapted to well-posedness, 
in the sense that it makes use of coercivity of the operator 
defining the part of the system containing the spatial derivatives 
\textit{separately} in the solid and fluid regions.

\section{The system of equations describing acousto-elastic waves}

We consider a bounded domain $\Omega \subset \mathbb{R}^3$ which is
divided into solid and fluid regions, $\Os$ and $\Of$,
respectively. 
The interior boundaries include solid-solid interface $\Sss$, 
fluid-fluid interface $\Sff$, and fluid-solid interface $\Sfs$, $\Ssf$
(where we distinguish whether the fluid or solid is on a particular side). 
We present the weak form of the coupled acousto-elastic system of equations.

Hooke's law in an elastodynamical system is expressed by relating
stress, $S_{ij}$, and strain, $E_{kl}$.  Assuming small deformations
gives a linear relationship, that is, $S_{ij} = c_{ijkl} E_{kl}$,
where $c_{ijkl}$ is the stiffness tensor. Through the relevant
symmetries, this tensor only contains 21 independent components. 
We use the Voigt notation which simplifies the writing of tensors 
while introducing
$\mathbi{S} = (S_{11}, S_{22}, S_{33}, S_{23} ,
S_{12}, S_{13})^T$ and $\mathbi{E} = (E_{11}, E_{22}, E_{33}, E_{23} ,
E_{12}, E_{13})^T$. 
In this notation the stiffness tensor takes the form of
a 6 by 6 matrix, $\mathbi{C}$, defined by,
\begin{equation}
\mathbi{S} = \mathbi{C} \mathbi{E} ,\quad
\tsC = \begin{bmatrix}
	C_{11} & C_{12} & C_{13} &  2 C_{14} &  2 C_{15} &  2 C_{16} \\[-1mm]
	C_{12} & C_{22} & C_{13} &  2 C_{24} &  2 C_{25} &  2 C_{26} \\[-1mm]
	C_{13} & C_{23} & C_{33} &  2 C_{34} &  2 C_{35} &  2 C_{36} \\[-1mm]
	C_{14} & C_{24} & C_{34} &  2 C_{44} &  2 C_{45} &  2 C_{46} \\[-1mm]
	C_{15} & C_{25} & C_{35} &  2 C_{45} &  2 C_{55} &  2 C_{56} \\[-1mm]
	C_{16} & C_{26} & C_{36} &  2 C_{46} &  2 C_{56} &  2 C_{66} 
\end{bmatrix} .
\label{eqn:stiffness}
\end{equation}
The isotropic case is obtained by setting all of the $C_{ij}$
components to zero except for $C_{11} = \lambda+2\mu$, $C_{12} =
C_{13} = C_{23} = \lambda$, $C_{44} = \mu$, $C_{55} = \mu$, and
$C_{66} = \mu$; $(\lambda,\mu)$ are the \emph{Lam\'e}
parameters. 
Furthermore, $\rho$ denotes the density. The anisotropic
elastodynamical equations are written in terms of the strain,
$\mathbi{E}$, and the particle velocity, $\mathbi{v}$,
\begin{equation}\label{eq:elasticPDE}
   \Dt{\tsE} 
   = \tfrac12 \left( \nabla \vev + \nabla \vev^T \right) ,
   \quad
   \rho \, \Dt{\vev} 
   = \nabla \cdot (\tsC \tsE) + \vef
\end{equation}
in $\Os$. In fluid regions, $\Of$, we use the pressure-velocity
formulation,
\begin{equation}\label{eq:acousticPDE}
   \Dt{\tE} 
   =   \nabla \cdot \tvev 
  -\frac{\tf}{\tlambda}  ,
  \quad
   \widetilde{\rho} \, \Dt{\tvev} 
   = \nabla (\tlambda \tE) .
\end{equation}
Here, $\widetilde P=-\tP$ is the pressure, 
while we use $\ \widetilde{ }\ $ to
distinguish acoustic field quantities and material parameters from the
elastic ones. 
In the above, $\tf$ denotes a volume source density of injection and
$\vef$ denotes a volume source density of force.

The solid-solid, fluid-solid and fluid-fluid
boundary conditions are given by
\begin{subequations}\label{eq:BC}
\begin{align}
   \vev^+ - \vev^- = 0 &
   \quad\quad \mbox{ and } &
   \ven \cdot \tsS^+ - \ven \cdot \tsS^- =0 &
   \quad\quad\text{ on } \Sss ,
\label{eq:BC1}\\
   \ven \cdot (\vev^\pm - \tvev^\mp) = 0 &
   \quad\quad \text{ and } & 
   \ven \cdot \tsS^\pm - (\tP)^\mp \ven = 0 &
   \quad\quad\text{ on } \Ssf \mbox{ and } \Sfs,
\label{eq:BC2}\\
   \ven \cdot (\tvev^+ - \tvev^-) = 0 &
   \quad\quad \text{ and } &
   (\tP)^+ -(\tP)^- =0 &
   \quad\quad\mbox{ on } \Sff .
\label{eq:BC3}
\end{align}
\end{subequations}
The $\pm$ convention is determined by the direction of the interface
normal, $\ven$. 
The outer normal vector points in the direction of the ``$+$'' side of the 
interface.

We introduce test functions (tensors)
$\tsH, \vew$ in the solid regions and
$\tvew, \tH$ in the fluid regions, which are assumed to be contained
in the same spaces and satisfy the same boundary conditions
as $\tsE, \vev, \tvev$ and $\tE$. Using
(\ref{eq:elasticPDE}) and (\ref{eq:acousticPDE}), we find that
\begin{subequations}\label{eq:Weakform}
\begin{align}
   \int_{\Os} \ddt{\tsE} : (\tsC \tsH\,) \dd\Omega 
&= \int_{\Os} \tfrac12 (\nabla \vev + \nabla \vev^T) :
               (\tsC \tsH\,) \dd\Omega ,
\label{eq:Weakform1}\\
   \int_{\Os} \rho\, \ddt{\vev} \cdot \vew \dd\Omega 
&= \int_{\Os} (\nabla \cdot (\tsC \tsE\,))
     \cdot \vew \dd\Omega + \int_{\Os} \vef \cdot \vew \dd\Omega ,
\label{eq:Weakform2}\\
   \int_{\Of} \ddt{\tE} \tlambda\, \tH \dd\Omega 
&= \int_{\Of} (\nabla \cdot \tvev) \tlambda\, \tH \dd\Omega
      - \int_{\Of} \tf \, \tH \dd\Omega ,
\label{eq:Weakform3}\\
   \int_{\Of} \trho\, \ddt{\tvev} \cdot \tvew \dd\Omega 
&= \int_{\Of} \nabla (\tP) \cdot \tvew \dd\Omega .
\label{eq:Weakform4}
\end{align}
\end{subequations}
Assuming an outer traction-free boundary condition in (\ref{eq:Weakform2})
and an outer pressure-free boundary condition in (\ref{eq:Weakform3}),
and applying an integration by parts, we obtain
\begin{subequations}
	\begin{align}
   \int_{\Os} \rho\, \ddt{\vev} \cdot \vew \dd\Omega 
 =& -\int_{\Os} \tsS : \nabla \vew  \dd\Omega
   + \int_{\Sfs} (\ven \cdot \tsS^\mm) \cdot \vew^\mm \dd\Sigma 
   + \int_{\Os} \vef \cdot \vew \dd\Omega ,
\label{eq:elasticWeak0}\\
  \int_{\Of} \ddt{\tE} \tlambda\, \tH \dd\Omega 
 =& -\int_{\Of} \tvev\cdot\nabla (\tlambda\tH) \dd\Omega
   + \int_{\Sfs} (\ven \cdot \tvev^\mm)(\tlambda\tH)^\mm  \dd\Sigma
   - \int_{\Of} \tf \, \tH \dd\Omega .
\label{eq:acousticWeak0}
	\end{align}
\end{subequations}
We use the fluid-solid boundary conditions (\ref{eq:BC2}),
replacing the fluid-solid surface integrals in
  (\ref{eq:elasticWeak0}) and (\ref{eq:acousticWeak0}) 
  by taking the average of both sides
  consistent with a central flux scheme, and obtain
\begin{subequations}\label{eq:WeakCouple}
\begin{align}
   \int_{\Os} \rho\, \ddt{\vev} \cdot \vew \dd\Omega 
&= -\int_{\Os} \tsS : \nabla\vew \dd\Omega
\nonumber\\ &
   + \int_{\Sfs} \tfrac12((\tP)^\pp \ven+\ven \cdot \tsS^\mm) 
       \cdot \vew^\mm \dd\Sigma
   + \int_{\Os} \vef \cdot \vew \dd\Omega ,
\label{eq:WeakCouple1}\\
  \int_{\Of} \ddt{\tE} \tlambda\tH \dd\Omega 
&=-\int_{\Of} \tvev \cdot\nabla(\tlambda\tH) \dd\Omega
\nonumber\\ &
   + \int_{\Sfs} {\tfrac12(\ven \cdot \vev^\mm+\ven \cdot \tvev^\pp)}\,
       (\tlambda\tH)^\mm \dd\Sigma
   - \int_{\Of} \tf \, \tH \dd\Omega .
\label{eq:WeakCouple2}
\end{align}
\end{subequations}
This form of the equations is analogous to the one used in the spectral
element method, see 
\cite{Chaljub2004}. Applying
an integration by parts, again, in (\ref{eq:WeakCouple}), we recover
the coupled strong formulation,
\begin{subequations}\label{eq:StrongCouple}
\begin{align}
   \int_{\Os} \rho\, \ddt{\vev} \cdot \vew \dd\Omega 
&= \int_{\Os} (\nabla \cdot \tsS) \cdot \vew \dd\Omega
\nonumber\\ &
   + \int_{\Sfs} {\tfrac12}((\tP)^\pp \ven - \ven \cdot \tsS^\mm)
         \cdot \vew^\mm \dd\Sigma
   + \int_{\Os} \vef \cdot \vew \dd\Omega ,
\label{eq:StrongCouple1}\\
   \int_{\Of} \ddt{\tE} \tlambda\tH \dd\Omega 
&= \int_{\Of} (\nabla \cdot \tvev) \tlambda\tH \dd\Omega
\nonumber\\ &
   + \int_{\Sfs} {\tfrac12}(\ven \cdot (\vev^\pp - \tvev^\mm))\,
         (\tlambda\tH)^\mm \dd\Sigma
   - \int_{\Of} \tf \, \tH \dd\Omega .
\label{eq:StrongCouple2}
\end{align}
\end{subequations}
We use this system of equations together with (\ref{eq:Weakform1}) and
(\ref{eq:Weakform4}) to develop our Discontinuous Galerkin method
based approach.

\section{Discontinuous Galerkin method with fluid-solid boundaries}
\label{sec:DG-acel}

The domain is partitioned into elements, $D^e$. We
distinguish elements, $\DeS$, in the solid regions from elements,
$\DeF$, in the fluid regions. 
Correspondingly, we distinguish
fluid-fluid ($\Tff$), solid-solid ($\Tss$) and 
fluid-solid ($\Tfs, \Tsf$) faces for each element; 
thus the interior boundaries are decomposed as
\[
	\Sigma_{\ast\,\bullet}=\cup\Sigma_{\ast\,\bullet}^\mathrm{e},
	\quad \ast,\,\bullet\in\{\mathrm{S},\mathrm{F}\},
\]
and so are the elements'
boundaries: $\partial\DeS = \Tss \cup \Tsf$ and
$\partial\DeF = \Tff \cup \Tfs$. 
The mesh size, $h$, is defined as the
maximum radius of each tetrahedral's inscribed sphere.

We introduce the  broken polynomial space $ V_h = \bigoplus_{\De}V_h^{\De}$
where the local space is defined elementwise as 
$V_h^{\De}=\mathrm{span}\{\phi_n(\De)\}_{n=1}^{\Ndof}$, with $\phi_n$ 
a set of polynomial basis further discussed in Section \ref{sec:nodal}.
The subscript ``$h$'' indicates the refinement of $V_h$ with decrease in
mesh size.
The semi-discrete time-domain, discontinuous Galerkin formulation
using a central flux yields: Find $\tsE_h, \vev_h, \tvev_h, \tE_h $,
with each component for each one of them in $ V_h $ 
such that 
\begin{equation}\label{eq:DG1}
\begin{split}
&  \int_{\DeS} \ddt{\tsE_h} : (\tsC \tsH_h) \dd\Omega 
   + \int_{\DeS} \rho\, \ddt{\vev_h} \cdot \vew_h \dd\Omega 
\\
&\hspace*{0.3cm}
   - \int_{\DeS} \tfrac12 (\nabla \vev_h + \nabla \vev_h^T) :
                 (\tsC \tsH_h) \dd\Omega
   - \int_{\DeS} (\nabla \cdot (\tsC \tsE_h)) \cdot \vew_h \dd\Omega
\\
&\hspace*{0.3cm}
   - \int_{\Tss} \tfrac12 \jmp{\vev_h}_\rSS \cdot 
            (\ven \cdot (\tsC \tsH_h)^\mm) \dd\Sigma
   - \int_{\Tsf} {\color{black}\tfrac12} \jmp{\vev_h}_\rSF
            \cdot (\ven \cdot (\tsC \tsH_h)^\mm) \dd\Sigma
\\
&\hspace*{0.3cm}
   - \int_{\Tss} \tfrac12 \ven\cdot(\jmp{\tsC\tsE_h}_\rSS)
             \cdot \vew_h^\mm \dd\Sigma
   - \int_{\Tsf} {\tfrac12} \ven\cdot(\jmp{\tsC\tsE_h}_\rSF)
   			 \cdot \vew_h^\mm \dd\Sigma 
   = \int_{\DeS} \vef_h \cdot \vew_h \dd\Omega ,
\end{split}
\end{equation}
and
\begin{equation}\label{eq:DG2}
\begin{split}
&  \int_{\DeF} \ddt{\tE_h} \tlambda\tH_h \dd\Omega
   + \int_{\DeF} \trho\, \ddt{\tvev_h} \cdot \tvew_h \dd\Omega 
\\
&\hspace*{0.3cm}
   - \int_{\DeF} (\nabla \cdot \tvev_h) \, \tQ_h \dd\Omega
   - \int_{\DeF} \nabla (\tP_h) \cdot \tvew_h \dd\Omega
\\
&\hspace*{0.3cm}
   - \int_{\Tff} \tfrac12 (\ven\cdot\jmp{\tvev_h}_\rFF) \,
             (\tQ_h)^\mm \dd\Sigma
   - \int_{\Tfs} {\tfrac12} (\ven \cdot \jmp{\tvev_h}_\rFS) \,
   			 (\tQ_h)^\mm \dd\Sigma
\\
&\hspace*{0.3cm}
   - \int_{\Tff} \tfrac12 \jmp{\tP_h}_\rFF
   				 (\ven \cdot \vew_h^\mm) \dd\Sigma
   - \int_{\Tfs} {\tfrac12} \jmp{\tP_h}_\rFS
                 (\ven \cdot \vew_h^\mm) \dd\Sigma
   = -\int_{\DeF} \tf_h \, \tH_h \dd\Omega ,
\end{split}
\end{equation}
hold for each element $\DeS$ or $\DeF$,
for all test functions $\tsH_h, \vew_h, \tvew_h, \tH_h 
{\in V_h}$. The notations $\vef_h$ and $\tf_h$ indicate 
polynomial approximation of $\vef$ and $\tf$.
Here,
\begin{subequations}\label{eq:jump_solid}
\begin{align}
&\begin{array}{rl}
   \jmp{\vev}_\rSS =& \vev^+ - \vev^- \\
   \jmp{\tsC\tsE}_\rSS =& \ven\,(\ven \cdot (\tsC \tsE)^+ 
          - \ven \cdot (\tsC \tsE)^-)
\end{array}
&\quad \mbox{ on } \Tss , \\
&\begin{array}{rl}
\jmp{\vev}_\rSF=&(\ven\cdot(\tvev^+ - \vev^-))\ven \\
\jmp{\tsC\tsE}_\rSF=&\ven\,((\tP)^+ \ven 
- \ven \cdot (\tsC \tsE)^-)
\end{array}
&\quad \mbox{ on } \Tsf
\end{align}
\end{subequations}
in the solid regions, while
\begin{subequations}\label{eq:jump_fluid}
\begin{align}
&\begin{array}{rl}
\jmp{\tvev}_\rFF=& (\ven \cdot (\tvev^+ - \tvev^-))\ven \\
\jmp{\tP}_\rFF=& (\tP)^+ - (\tP)^-
\end{array}
&\quad \mbox{ on } \Tff , \\
&\begin{array}{rl}
\jmp{\tvev}_\rFS=& (\ven \cdot (\vev^+ - \tvev^-))\ven \\
\jmp{\tP}_\rFS=&\ven \cdot (\tsC \tsE)^+ \cdot \ven - (\tP)^-
\end{array}
&\quad \mbox{ on } \Tfs
\end{align}
\end{subequations}
in the fluid regions, using interior boundary continuity conditions. A
similar formulation for Maxwell's equations, using the central flux,
can be found in 
\cite[Chapter 10, Page 434]{Hesthaven2007}.

\subsection{Energy function of central flux}

We consider a time-dependent energy function comprising both the solid
and fluid regions, $\Energy_h = \Energy_{\mathrm S,h} +
\Energy_{\mathrm F,h}$, with
\begin{equation}\label{eq:energy}
\begin{split}
   \Energy_{\mathrm S,h}
         &= \Half\SumS\int_\DeS
                   (\tsE_h : (\tsC \tsE_h) + \rho \, \vev_h
                        \cdot \vev_h) \dd\Omega ,
\\
   \Energy_{\mathrm F,h}
         &= \Half\SumF\int_\DeF
   \left( \tP_h^2 + \trho \, \tvev_h \cdot
                      \tvev_h \right) \dd\Omega.
\end{split}
\end{equation}
The functions in (\ref{eq:energy}) define a norm both in the solid and
in the fluid regions.
Taking the time derivative
and noting that $\tsC$ is symmetric, we have
\begin{align}
   \DDt{\Energy_{\mathrm S,h}}
         &= \SumS\int_\DeS
   \left( \ddt{\tsE_h} : (\tsC \tsE_h) + \rho \, \ddt{\vev_h}
                 \cdot \vev_h \right) \dd\Omega ,
\label{eq:energyS}
\\
   \DDt{\Energy_{\mathrm F,h}}
          &= \SumF\int_\DeF 
   \left( \ddt{\tE_h} \tP_h + \trho \, \ddt{\tvev_h}
                \cdot \tvev_h \right) \dd\Omega .
\label{eq:energyF}
\end{align}
Starting from (\ref{eq:DG1}) and (\ref{eq:DG2}) 
and carrying out the summation over all
the elements yields 
\begin{equation}
\label{eq:central energy}
   \DDt{\Energy_h} =  \SumS\int_\DeS \vef_h \cdot \vev_h \dd\Omega
            - \SumF\int_\DeF \tf_h \, \tE_h \dd\Omega .
\end{equation}
This property is obtained as follows:

In (\ref{eq:DG1}) and (\ref{eq:DG2}) we let $\tsH_h=\tsE_h, \vew_h=\vev_h, 
\tH_h=\tE_h, \tvew_h=\tvev_h$, and obtain elementwise 
\begin{equation}\label{eq:solid VolE to SurfE}
\begin{split}
   & \int_{\DeS} \tfrac12 (\nabla \vev_h + \nabla \vev_h^T) :
                 (\tsC \tsE_h) \dd\Omega
   + \int_{\DeS} (\nabla \cdot (\tsC \tsE_h)) \cdot \vev_h \dd\Omega
\\
 = & \int_{\Tss\cup\Tsf} \vev_h^\mm \cdot 
            (\ven \cdot (\tsC \tsE_h)^\mm) \dd\Sigma,
\end{split}
\end{equation}
and similarily
\begin{equation}\label{eq:fluid VolE to SurfE}
\begin{split}
   & \int_{\DeF} (\nabla \cdot \tvev_h) \, \tP_h \dd\Omega
   + \int_{\DeF} \nabla (\tP_h) \cdot \tvev_h \dd\Omega
\\
 = & \int_{\Tff\cup\Tfs} \ven\cdot\tvev_h^\mm \,
             (\tP_h)^\mm \dd\Sigma.
\end{split}
\end{equation}
From (\ref{eq:DG1}), (\ref{eq:jump_solid}), (\ref{eq:energyS}) and
(\ref{eq:solid VolE to SurfE}),
\begin{align}
  &
  \DDt{\Energy_{\mathrm S,h}}
  =
  \SumS\int_\DeS \vef_h \cdot \vev_h \dd\Omega 
  \nonumber\\
  &+
  \Sumsf\tfrac12\int_{\Tsf}\left(
  (\jmp{\vev_h}_\rSF+\vev_h^\mm) \cdot (\ven \cdot (\tsC \tsE_h)^\mm) +
  \ven\cdot(\jmp{\tsC\tsE_h}_\rSF+(\tsC\tsE_h)^\mm)\cdot \vev_h^\mm
  \right)\dd\Sigma 
  \tag{$\Theta_1$}\label{Theta 1} \\
  &+
  \Sumss\tfrac12\int_{\Tss}\left(
  (\jmp{\vev_h}_\rSS+\vev_h^\mm) \cdot (\ven \cdot (\tsC \tsE_h)^\mm) + 
  \ven\cdot(\jmp{\tsC\tsE_h}_\rSS+(\tsC\tsE_h)^\mm)\cdot \vev_h^\mm
  \right)\dd\Sigma.
  \tag{$\Theta_2$}\label{Theta 2} 
\end{align}
In the above,
\begin{equation}
	\Theta_2=
  \Sumss\tfrac12\int_{\Tss}\left(
  \vev_h^\pp \cdot(\ven \cdot (\tsC \tsE_h)^\mm) +
  \ven\cdot(\tsC\tsE_h)^\pp \cdot \vev_h^\mm
  \right)\dd\Sigma =0.
\end{equation}
The surface integration terms cancel out when summed from both sides
of the solid-solid interfaces because of the continuity condition 
(\ref{eq:BC1}) and the opposite outer normal directions. 
We are left with the contributions from solid-fluid inner faces, 
$\Theta_1$,
\begin{equation}
	\DDt{\Energy_{\mathrm S,h}}=
	\Sumsf\tfrac12\int_{\Tsf}\left(
	\tvev_h^\pp \cdot(\ven \cdot (\tsC \tsE_h)^\mm)+
	(\tP)^\pp \ven\cdot \vev_h^\mm
	\right)\dd\Sigma 
	+ \SumS\int_\DeS \vef_h \cdot \vev_h \dd\Omega.
	\label{eq:summation1}
\end{equation}
A similar result in the fluid region obtained from (\ref{eq:DG2}), 
(\ref{eq:jump_fluid}), (\ref{eq:energyF}) and (\ref{eq:fluid VolE to SurfE})
yields
\begin{equation} 
 	\DDt{\Energy_{\mathrm F,h}}=
	\Sumfs\tfrac12\int_{\Tfs}\left( 
	(\ven \cdot \vev_h^\pp)(\tP)^\mm +
	\ven \cdot (\tsC \tsE_h)^\pp \cdot\tvev_h^\mm  
	\right)\dd\Sigma
	-\SumF\int_\DeF \tf_h \, \tE_h \dd\Omega,
    \label{eq:summation2}
\end{equation}
and the surface integration terms on the solid-fluid and fluid-solid 
interfaces in (\ref{eq:summation1}) and (\ref{eq:summation2}) cancel out 
due to (\ref{eq:BC2}).
Therefore (\ref{eq:central energy}) is obtained. We note that the surface 
integration along solid-fluid interfaces 
$\int_{\Tsf} {\tfrac12} \ven\cdot(\jmp{\tsC\tsE_h}_\rSF)
  \cdot \vew_h^\mm \dd\Sigma$ and 
$ \int_{\Tfs} {\tfrac12} (\ven \cdot \jmp{\tvev_h}_\rFS) \,
  (\tQ_h)^\mm \dd\Sigma$
are essential to guarantee energy conservation.

\subsection{Nodal basis functions}
\label{sec:nodal}

\begin{figure}
\centering
\begin{tabular}{ccc}
\includegraphics[width=1.5in]{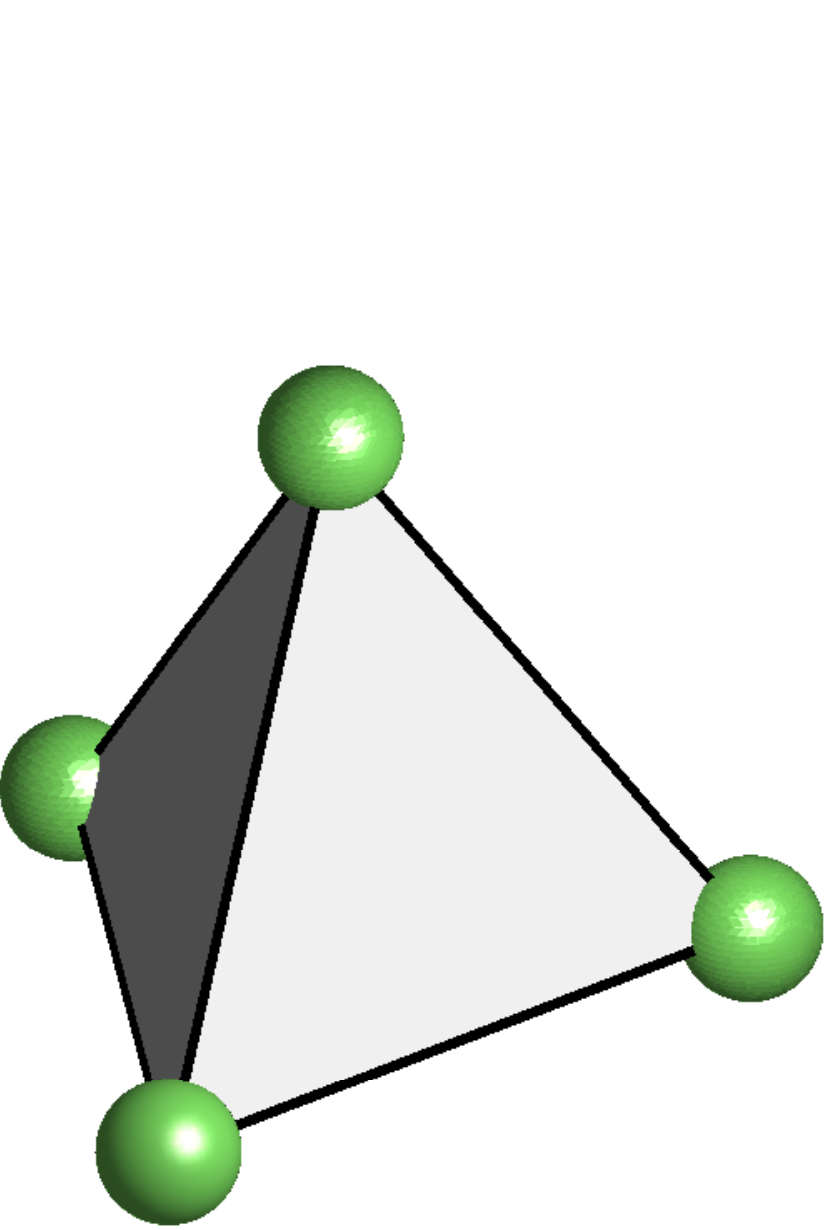} &
\includegraphics[width=1.5in]{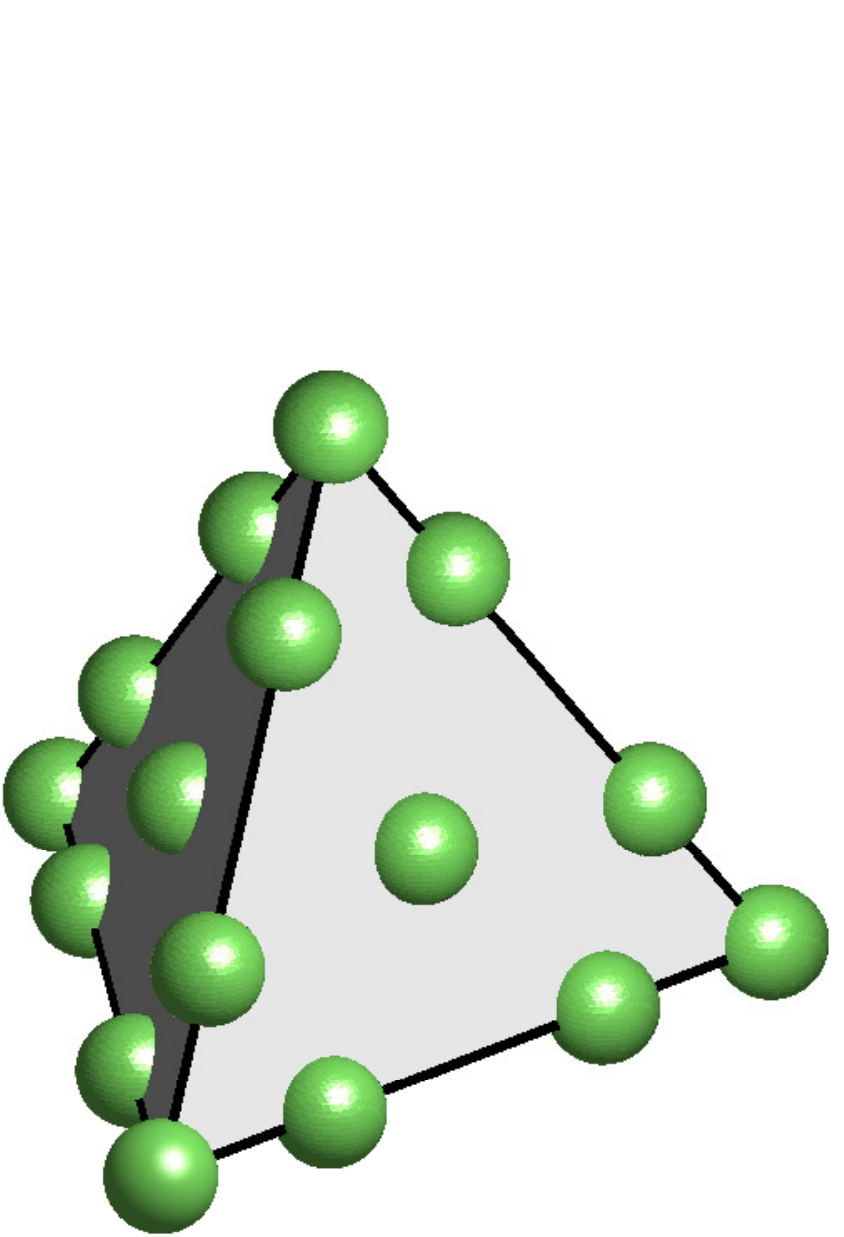} &
\includegraphics[width=1.5in]{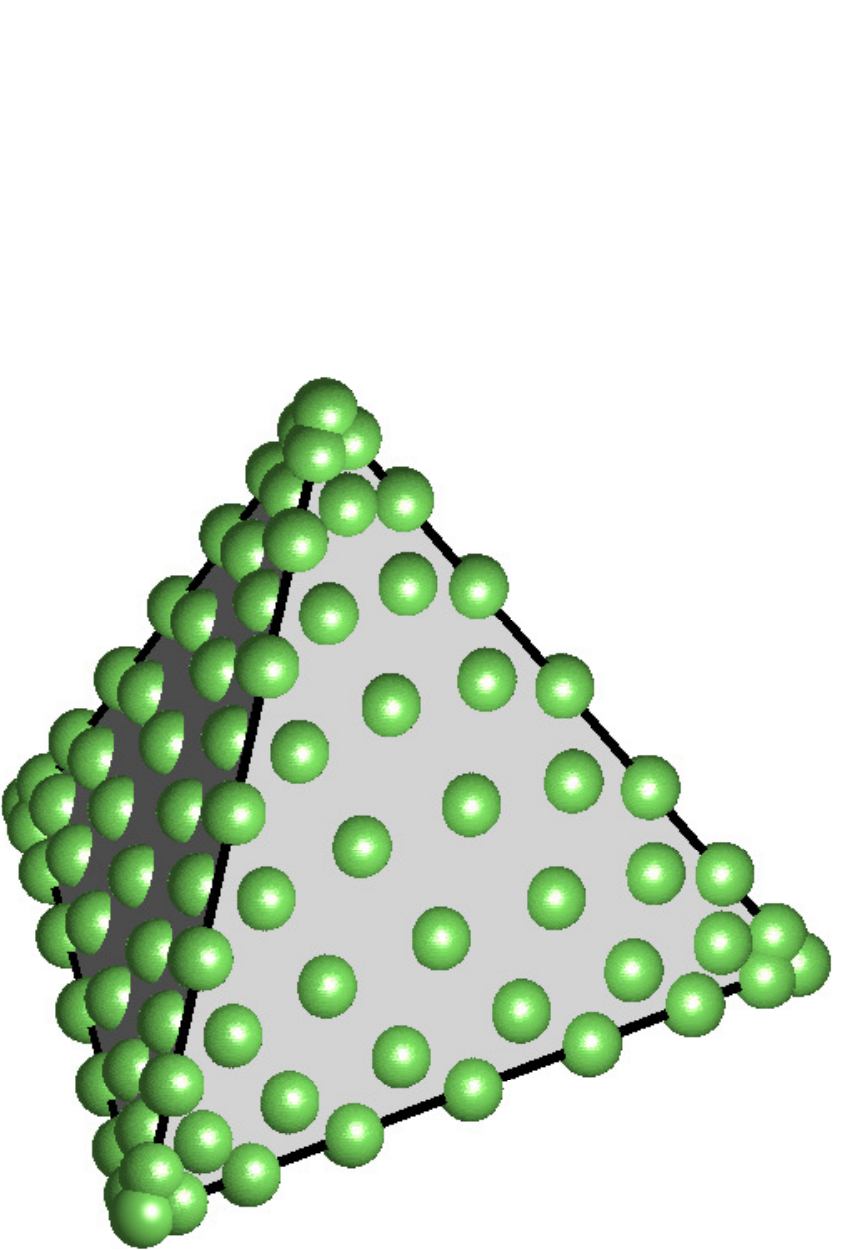} \\
$ N_p $ = 1 & $ N_p $ = 3 & $ N_p $ = 8
\end{tabular}
\caption{Warp \& blend tetrahedral nodal point distribution for $ N_p
  $ = 1, 3, 8. For clarity only facial nodes are illustrated.}
\label{fig:nodal np}
\end{figure}

The discretized solution follows an expansion, componentwise, 
into $\Ndof = \Ndof(N_p)$ nodal trial basis functions of order $N_p$,
as is in
\cite{Hesthaven2007},
\begin{equation}\label{eq:expand-E}
\begin{split}
(\tsE_h)_{ij}(\mathbi{x},t)=&\bigoplus_{\De}\sum_{n=1}^{\Ndof} 
(\tsE_{h,n}^{\De})_{ij}(t)\phi_n(\mathbi{x}) ,\\
\mbox{with } (\tsE_{h,n}^{\De})_{ij}(t)=&(\tsE_h)_{ij}
(\mathbi{x}_n,t) , n=1,2,\cdots,\Ndof,
\end{split}
\end{equation}
and similarly for the other fields, $\vev_h, \tvev_h, \tE_h$.
The superscript, ${\centerdot}^{D^e}$, 
indicates a local expansion within element
$D^e$. In the above, ${\{ \phi_n(\mathbi{x}) \}}_{n = 1}^{\Ndof}$
is a set of three-dimensional Lagrange polynomials
associated with the nodal points, $ {\{ \mathbi{x}_n \}}_{n = 1}^{\Ndof} $
(see Figure~\ref{fig:nodal np}), with each polynomial defined as
\[
	\phi_k(\mathbi{x})=\prod_{j=1,j\neq k}^{\Ndof}
	\frac{\mathbi{x}-\mathbi{x}_j}{\mathbi{x}_k-\mathbi{x}_j}.
\]
We use the warp \& blend method~[\cite{Warburton2006}] to determine the
coordinates of nodal points in the tetrahedron by numerically minimizing 
the Lebesgue constant of interpolation.  
For an order $N_p$ interpolation there are
$\Ndof = \frac16(N_p + 1)(N_p + 2)(N_p + 3)$ nodal points.

The medium coefficients are expanded in a likewise manner 
\begin{equation}\label{eq:expand-c}
\begin{split}
   (\tsC_h )_{ij}(\mathbi{x}) = &\bigoplus_{\De}\sum_{n=1}^{N_p}
   (\tsC_{h,n}^{\DeS})_{ij} \phi_n(\mathbi{x}) ,\\
\mbox{with }(\tsC_{h,n}^{\DeS})_{ij} =& (\tsC_h)_{ij}
(\mathbi{x}_n) , n=1,2,\cdots,\Ndof,
\end{split}
\end{equation}
and similarly for $\rho, \trho, \tlambda$. 
When refining a mesh, we expect an increase in number of elements $\De$
with decreased size.

\subsection{The system of equations in matrix form}
\label{sec:matrix form}

To simplify the notation in the further development of a numerical
scheme, we introduce a joint matrix form of the system of equations.
We map the components of $\tsE, \vev$ and $\tE, \tvev$ to $9
\times 1$ and $4 \times 1$ matrices, respectively,
\begin{equation}\label{eq:unknown vec1}
   \veq=(E_{11},E_{22},E_{33},E_{23},E_{13},E_{12},v_1,v_2,v_3)^T
\quad\text{and}\quad
   \tveq=(\tE,\widetilde{v}_1,\widetilde{v}_2,\widetilde{v}_3)^T ,
\end{equation}
and, correspondingly, the components of body forces $ \vef $ 
and $ \tf $ to the matrix
\[ 
   \veg = (0,0,0,0,0,0,f_1,f_2,f_3)^T 
\quad\text{and}\quad
   \tveg = \left(-\frac{\tf}{\tlambda},0,0,0\right)^T
\quad.
\]
Equations (\ref{eq:elasticPDE}) and (\ref{eq:acousticPDE}) attain the form
\begin{equation}
   \tsQ \, \ddt{\veq} - \nabla \cdot (\tsA \veq) = \veg 
   \quad\mbox{and}\quad
   \ttsQ \, \ddt{\tveq} - \nabla \cdot (\ttsA \tveq) 
            =\tveg,
\end{equation}
where 
\[ \tsQ=\left(
\begin{array}{ccc}I_{6\times 6} & \vline & 0 \\ \hline 0 & \vline & 
\rho I_{3\times 3}\end{array}\right) \quad\text{and}\quad 
\ttsQ=\left(\begin{array}{ccc} {1} & \vline & 0 \\ \hline 0 & 
\vline & \trho I_{3\times 3}\end{array}\right) 
\] 
and
\[ \tsA=(A_1,A_2,A_3) \quad\text{and}\quad \ttsA=(\widetilde{A}_1,
\widetilde{A}_2,\widetilde{A}_3)  ,
\]
that is,
{\ColorRed
\[
	\begin{split}
   (\nabla \cdot (\tsA \veq))_l
   = \partial_{x_k} ( (\tsA_{k})_{lm} \veq_m)
   \quad\text{and}\quad
   (\nabla \cdot (\ttsA \tveq))_l
   = \partial_{x_k} ( (\ttsA_{k})_{lm} \tveq_m)
   ,\\
   k=1,2,3,\quad l,m=1,\cdots,9 \mbox{ or } 1,\cdots,4
	\end{split}
\]
}
with
\[
A_1=\left(\hspace*{-2mm}
\begin{array}{ccc}
  {\displaystyle \mathbf{0}} & \hspace*{-4mm} \vline \hspace*{-4mm} &
  \begin{array}{ccc}
	  1 & 0 & 0\\[-2mm]
	  0 & 0 & 0\\[-2mm]
	  0 & 0 & 0\\[-2mm]
	  0 & 0 & 0\\[-2mm]
	  0 & 0 & \frac{1}{2}\\[-2mm]
	  0 & \frac{1}{2} & 0
  \end{array}\\
  \hline
  \begin{array}{cccccc}
	  C_{11} & C_{12} & C_{13} & 2C_{14} & 2C_{15} & 2C_{16} \\[-2mm]
	  C_{16} & C_{26} & C_{36} & 2C_{46} & 2C_{56} & 2C_{66} \\[-2mm]
	  C_{15} & C_{25} & C_{35} & 2C_{45} & 2C_{55} & 2C_{56} 
  \end{array} & \hspace*{-4mm} \vline \hspace*{-4mm} &
  {\displaystyle\mathbf{0}}
\end{array}
\hspace*{-2mm}\right)
\quad \text{and}\quad 
\widetilde{A}_1=\left(\hspace*{-2mm}
\begin{array}{ccc}
  {\displaystyle \mathbf{0}} & \hspace*{-2mm} \vline \hspace*{-4mm} &
  \begin{array}{ccc}
  1 & 0 & 0\\
  0 & 0 & 0\\
  0 & 0 & 0
  \end{array}\\
  \hline
  {\tlambda} 
  & \hspace*{-2mm} \vline \hspace*{-4mm} &
  {\displaystyle\mathbf{0}}
\end{array}
\hspace*{-2mm}\right) ,
\]
\[
A_2=\left(\hspace*{-2mm}
\begin{array}{ccc}
  {\displaystyle \mathbf{0}} & \hspace*{-4mm} \vline \hspace*{-4mm} &
  \begin{array}{ccc}
	  0 & 0 & 0\\[-2mm]
	  0 & 1 & 0\\[-2mm]
	  0 & 0 & 0\\[-2mm]
	  0 & 0 & \frac{1}{2}\\[-2mm]
	  0 & 0 & 0\\[-2mm]
	  \frac{1}{2} & 0 & 0
  \end{array}\\
  \hline
  \begin{array}{cccccc}
	  C_{16} & C_{26} & C_{36} & 2C_{46} & 2C_{56} & 2C_{66} \\[-2mm]
	  C_{12} & C_{22} & C_{23} & 2C_{24} & 2C_{25} & 2C_{26} \\[-2mm]
	  C_{14} & C_{24} & C_{34} & 2C_{44} & 2C_{45} & 2C_{46} 
  \end{array} & \hspace*{-4mm} \vline \hspace*{-4mm} &
  {\displaystyle\mathbf{0}}
\end{array}
\hspace*{-2mm}\right)
\quad \text{and}\quad 
\widetilde{A}_2=\left(\hspace*{-2mm}
\begin{array}{ccc}
  {\displaystyle \mathbf{0}} & \hspace*{-2mm} \vline \hspace*{-4mm} &
  \begin{array}{ccc}
  0 & 0 & 0\\
  0 & 1 & 0\\
  0 & 0 & 0
  \end{array}\\
  \hline
  {\tlambda} 
  & \hspace*{-2mm} \vline \hspace*{-4mm} &
  {\displaystyle\mathbf{0}}
\end{array}
\hspace*{-2mm}\right) ,
\]
\[
A_3=\left(\hspace*{-2mm}
\begin{array}{ccc}
  {\displaystyle \mathbf{0}} & \hspace*{-4mm} \vline \hspace*{-4mm} &
  \begin{array}{ccc}
	  0 & 0 & 0\\[-2mm]
	  0 & 0 & 0\\[-2mm]
	  0 & 0 & 1\\[-2mm]
	  0 & \frac{1}{2} & 0\\[-2mm]
	  \frac{1}{2} & 0 & 0\\[-2mm]
	  0 & 0 & 0
  \end{array}\\
  \hline
  \begin{array}{cccccc}
	  C_{15} & C_{25} & C_{35} & 2C_{45} & 2C_{55} & 2C_{56} \\[-2mm]
	  C_{14} & C_{24} & C_{34} & 2C_{44} & 2C_{45} & 2C_{46} \\[-2mm]
	  C_{13} & C_{23} & C_{33} & 2C_{34} & 2C_{35} & 2C_{36} 
  \end{array}&
  \hspace*{-4mm} \vline \hspace*{-4mm} &
  {\displaystyle\mathbf{0}}
\end{array}
\hspace*{-2mm}\right)
\quad \text{and}\quad 
\widetilde{A}_3=\left(\hspace*{-2mm}
\begin{array}{ccc}
  {\displaystyle \mathbf{0}} & \hspace*{-2mm} \vline \hspace*{-4mm} &
  \begin{array}{ccc}
  0 & 0 & 0\\
  0 & 0 & 0\\
  0 & 0 & 1
  \end{array}\\
  \hline
  {\tlambda} 
  & \hspace*{-2mm} \vline \hspace*{-4mm} &
  {\displaystyle\mathbf{0}}
\end{array}
\hspace*{-2mm}\right) .
\]
We define the coefficient matrices $ \tsA_n $ in the normal
directions $ \ven=(\,n_1,n_2,n_3) $ as $ \tsA_n=n_1A_1+n_2A_2+n_3A_3
$, thus $ \tsA_n\veq \equiv \ven\cdot(\tsA\veq) $; similarly, $
\ttsA_n=n_1\tilde A_1+n_2\tilde A_2+n_3\tilde A_3 $.  We can also give
them in the matrix form,
\[
\tsA_n =\left(\begin{array}{cc}0&T_{12}\\T_{21}\cdot \tsC&0\end{array}\right)
\quad\text{and}\quad
\ttsA_n = \left(\begin{array}{cc} 0 & \ven^T \\ 
{\tlambda}\ven & 0 \end{array}\right)
,
\]
with
\[
T_{12}=\left(\begin{array}{cccccc}
n_1 & 0 & 0 & 0 & \tfrac12n_3 & \tfrac12n_2 \\
0 & n_2 & 0 & \tfrac12n_3 & 0 & \tfrac12n_1 \\
0 & 0 & n_3 & \tfrac12n_2 & \tfrac12n_1 & 0 
\end{array}\right)^T,
\quad
T_{21}=\left(\begin{array}{cccccc}
n_1 & 0 & 0 & 0 & n_3 & n_2 \\
0 & n_2 & 0 & n_3 & 0 & n_1 \\
0 & 0 & n_3 & n_2 & n_1 & 0 
\end{array}\right).
\]
We introduce 
\[
   \tsLa = \left(
   \begin{array}{ccc} \tsC & \vline & 0 \\ \hline 0 & \vline & 
   I_{3\times 3}\end{array} \right)
\quad\text{and}\quad 
   \ttsLa = \left(
   \begin{array}{ccc} \tlambda & \vline & 0 \\ \hline 0 & \vline & 
   I_{3\times 3}\end{array} \right)
   .
\]
In the solid regions, we write $ \vep = {
  \left(
  H_{11},H_{22},H_{33},H_{23},H_{13},H_{12},w_1,w_2,w_3
  \right)}^T $, and in the
fluid regions, we write $ \tvep = 
  (\tH,\tilde w_1,\tilde w_2,\tilde w_3)^T 
  $. 
The inner product $\innprod{\veq,\vep}_\Omega$ indicates the dot 
product of vectors $\veq$ and $\vep$ followed by integration over 
the domain $\Omega$.
Equation (\ref{eq:DG1}) is then rewritten, regarding the supports of basis
functions $\vep_h$ localized to an element $\De_{\mathrm{S},\mathrm{F}}$,
as
\begin{equation}
\label{eq:semi_disc_DG_S} 
\begin{split}
\innprodDeS{\tsQ_h\ddt{\veq_h},{\tsLa_h}\vep_h}
-&\innprodDeS{\nabla \cdot (\tsA_h\veq_h),{\tsLa_h} \vep_h} 
-\tfrac12\innprod{
    \jmp{\tsA_{n,h}\veq_h}_\rSS,
    (\tsLa_h\vep_h)^\mm}_\Tss\\
-&\tfrac12\innprod{
    \jmp{\tsA_{n,h}\veq_h}_\rSF,
    (\tsLa_h\vep_h)^\mm}_\Tsf
= \innprodDeS{\veg,{\tsLa_h}\vep_h} ,
\end{split}
\end{equation}
\begin{equation}
\label{eq:semi_disc_DG_F} 
\begin{split}
\innprodDeF{\ttsQ_h\ddt{\tveq_h},{\ttsLa_h}\tvep_h}
-&\innprodDeF{\nabla \cdot (\ttsA_h\tveq_h),{\ttsLa_h} \tvep_h} 
-\tfrac12\innprod{
    \jmp{\ttsA_{n,h}\tveq_h}_\rFF,
    (\ttsLa_h\tvep_h)^\mm}_\Tff\\
-&\tfrac12\innprod{
    \jmp{\ttsA_{n,h}\tveq_h}_\rFS,
    (\ttsLa_h\tvep_h)^\mm}_\Tfs
= \innprodDeF{\tveg,{\ttsLa_h}\tvep_h}.
\end{split}
\end{equation}
In the above we identify the central flux as
\begin{equation}\label{eq:central flux}
\Fluxc_{\mathrm S*}=\tfrac12\innprod{\jmp{\tsA_n \veq}_{\mathrm S*}, 
(\tsLa \vep)^-}_{\Sigma_{\mathrm S*}^\mathrm{e}},
\quad
\tFluxc_{\mathrm F*}=\tfrac12\innprod{\jmp{\ttsA_n \tveq}_{\mathrm F*}, 
(\ttsLa \tvep)^-}_{\Sigma_{\mathrm F*}^\mathrm{e}},
\quad
*\in\{\mathrm{S,F}\},
\end{equation}
in which we redefine
\begin{equation}\label{eq:fluid-solid projection}
\begin{split}
  &  \jmp{\tsA_n\veq}_\rSS=(\tsA_n\veq)^\pp-(\tsA_n\veq)^\mm,\quad
    \jmp{\tsA_n\veq}_\rSF=O^T(\ttsA_n\tveq)^\pp-(\tsA_n\veq)^\mm,\\
  &  \jmp{\ttsA_n\tveq}_\rFF=(\ttsA_n\tveq)^\pp-(\ttsA_n\tveq)^\mm,\quad
  \jmp{\ttsA_n\tveq}_\rFS=O^{\phantom{*}}(\tsA_n\veq)^\pp-(\ttsA_n\tveq)^\mm,
\end{split}
\end{equation}
with the map $O: \mathbb{R}^9 \to \mathbb{R}^4$ given by
\[
   O \veq = \left(\begin{array}{c}
            \ven \cdot \tsE \cdot \ven \\
            (\ven \cdot \vev)\ven
            \end{array}\right),
\quad \mbox{ and its adjoint } \quad
   O^T \tveq = \left(\begin{array}{c}
            ( \ven\ven )\tE \\
            (\ven \cdot \tvev)\ven
            \end{array}\right),
\]
which can also be explicitly given in the matrix form
\[
	O=\left(\hspace*{-2mm}
	\begin{array}{ccc}
	\begin{array}{cccccc}
	n_1n_1 & n_2n_2 & n_3n_3 & n_2n_3 & n_1n_3 & n_1n_2 
	\end{array} & \hspace*{-4mm}
	\vline \hspace*{-4mm} & {\displaystyle \mathbf{0}} \\
	\hline
	{\displaystyle \mathbf{0}} & \hspace*{-4mm} \vline \hspace*{-4mm} &
	\begin{array}{ccc}
	n_1n_1 & n_1n_2 & n_1n_3 \\[-2mm]
	n_1n_2 & n_2n_2 & n_2n_3 \\[-2mm]
	n_1n_3 & n_2n_3 & n_3n_3 
	\end{array}
	\end{array}
	\hspace*{-2mm}\right).
\]

\section{
The Boundary Condition penalized numerical flux and stability}
\label{sec:BCpenalty}

Here, we construct our penalized numerical flux. The flux is designed
such that the penalized discrete counterpart of the
weak form (\ref{eq:semi_disc_DG_S}) and (\ref{eq:semi_disc_DG_F}) 
satisfies 
the condition of non-increasing energy 
and guarantees a proper error estimate.  
We replace the central fluxes, $ \Fluxc $ and $\tFluxc$, in 
(\ref{eq:central flux}), by penalized fluxes, $ \Fluxp $ and $\tFluxp$,
by adding penalty terms, that is:
\begin{equation}\label{eq:penalty flux}
\begin{split}
\Fluxp_{\mathrm S*}=&
\tfrac12\innprod{\jmp{\tsA_n \veq}_{\mathrm S*}, 
(\tsLa \vep)^-}_{\Sigma_{\mathrm S*}^\mathrm{e}}
+\alpha\innprod{\tsA_n^{T,\mm}\jmp{\tsA_n \veq}_{\mathrm S*}, 
\vep^-}_{\Sigma_{\mathrm S*}^\mathrm{e}}
\\
&=\tfrac12\innprod{\jmp{\tsA_n \veq}_{\mathrm S*}, 
(\tsLa \vep)^-}_{\Sigma_{\mathrm S*}^\mathrm{e}}
+\alpha\innprod{\jmp{\tsA_n \veq}_{\mathrm S*}, 
(\tsA_n\vep)^-}_{\Sigma_{\mathrm S*}^\mathrm{e}},
\\
\tFluxp_{\mathrm F*}=&
\tfrac12\innprod{\jmp{\ttsA_n \tveq}_{\mathrm F*}, 
(\ttsLa \tvep)^-}_{\Sigma_{\mathrm F*}^\mathrm{e}}
+\alpha\innprod{\ttsA_n^{T,\mm}\jmp{\ttsA_n \tveq}_{\mathrm F*}, 
\tvep^-}_{\Sigma_{\mathrm F*}^\mathrm{e}}
\\
&=\tfrac12\innprod{\jmp{\ttsA_n \tveq}_{\mathrm F*}, 
(\ttsLa \tvep)^-}_{\Sigma_{\mathrm F*}^\mathrm{e}}
+\alpha\innprod{\jmp{\ttsA_n \tveq}_{\mathrm F*}, 
(\ttsA_n\tvep)^-}_{\Sigma_{\mathrm F*}^\mathrm{e}}, 
\quad *\in\{\mathrm S,\mathrm F\}
\end{split}
\end{equation}
with $\alpha $ some positive constant scalar.
With this modification, (\ref{eq:semi_disc_DG_S}) and 
(\ref{eq:semi_disc_DG_F}) becomes
\begin{align}
&
\begin{aligned}
\innprodDeS{\tsQ_h\ddt{\veq_h},{\tsLa_h}\vep_h}
-&\innprodDeS{\nabla \cdot (\tsA_h\veq_h),{\tsLa_h} \vep_h} 
-\tfrac12\innprod{
    \jmp{\tsA_{n,h}\veq_h}_{\mathrm S*},
    (\tsLa_h\vep_h)^\mm}_{\Sigma_{\mathrm S*}^\mathrm{e}}
\\
-&\alpha\innprod{
    \jmp{\tsA_{n,h}\veq_h}_{\mathrm S*},
    (\tsA_{n,h}\vep_h)^\mm}_{\Sigma_{\mathrm S*}^\mathrm{e}}
= \innprodDeS{\veg,{\tsLa_h}\vep_h} ,
\end{aligned}
\label{eq:penalty_DG_S} 
\\
&
\begin{aligned}
\innprodDeF{\ttsQ_h\ddt{\tveq_h},{\ttsLa_h}\tvep_h}
-&\innprodDeF{\nabla \cdot (\ttsA_h\tveq_h),{\ttsLa_h} \tvep_h} 
-\tfrac12\innprod{
    \jmp{\ttsA_{n,h}\tveq_h}_{\mathrm F*},
    (\ttsLa_h\tvep_h)^\mm}_{\Sigma_{\mathrm F*}^\mathrm{e}}
\\
-&\alpha\innprod{
    \jmp{\ttsA_{n,h}\tveq_h}_{\mathrm F*},
    (\ttsA_{n,h}\tvep_h)^\mm}_{\Sigma_{\mathrm F*}^\mathrm{e}}
= \innprodDeF{\tveg,{\ttsLa_h}\tvep_h},
\quad *\in\{\mathrm S,\mathrm F\}.
\end{aligned}
\label{eq:penalty_DG_F} 
\end{align}
In Appendix \ref{sec:stability} we provide a guideline how to choose an 
$\alpha$ based on an error analysis. We set $\alpha=1/2$, in which case 
the energy function with the penalty terms coincides with the one
using an upwind flux \cite[]{Warburton2013}.
For the convergence analysis, we follow \cite[Section 5.1]{Warburton2013}
while obtaining an error estimate. 

{\ColorRed
Following the matrix form in Subsection \ref{sec:matrix form}, we immediately
rewrite the definition of energy functions (\ref{eq:energy}) in solid
and fluid region as 
\begin{equation}
	\begin{split}
	\Energy_{\mathrm S,h}=&
	\Half\SumS \innprod{\tsQ_h\veq_h,\tsLa_h\veq_h}_\DeS 
	= \Half\SumS \norm{\veq}_{L^2(\DeS;\tsQ_h,\tsLa_h)}\\
	\Energy_{\mathrm F,h}=&
	\Half\SumF \innprod{\ttsQ_h\tveq_h,\ttsLa_h\tveq_h}_\DeF
	= \Half\SumF \norm{\tveq}_{L^2(\DeF;\ttsQ_h,\ttsLa_h)}.
	\end{split}
	\label{eq:Energy matrixform}
\end{equation}
Here $\Norm{\centerdot}_{L^2(\DeS; \tsQ,\tsLa)}$ and
$\Norm{\centerdot}_{L^2(\DeF; \ttsQ,\ttsLa)}$ are the energy norms in solid
and fluid regions, and we simplify the notification 
without causing ambiguity by 
$\normVS{\centerdot}$ and $\normVF{\centerdot}$, respectively. 
We also define the energy norms in solid-solid, fluid-fluid and 
solid-fluid interfaces similarly as $\normSS{\centerdot}$, 
$\normFF{\centerdot}$ and $\normSF{\centerdot}$, $\normFS{\centerdot}$.
}
Upon taking the penalty terms into consideration, 
equation (\ref{eq:central energy}) is replaced by
{\ColorRed
\begin{equation}\label{eq:DG-penalty-energy}
\begin{split}
&\DDt{\Energy_h} 
+\frac\alpha 2
\Bigl (
\Sumss \Norm{
    \jmp{\tsA_{n,h}\veq_h}_\rSS
  }_{L^2(\Tss)}^2
+ \Sumff \Norm{
    \jmp{\ttsA_{n,h}\tveq_h}_\rFF
  }_{L^2(\Tff)}^2 
  \\
  &\hspace{1cm}
+ 2\Sumsf \Norm{
    \jmp{\tsA_{n,h}\veq_h}_\rSF
    }_{L^2(\Tsf)}^2 \Bigr )
=  \SumS\int_\DeS \veg_h \cdot \tsLa_h\veq_h \dd\Omega
            + \SumF\int_\DeF \tveg_h \cdot \ttsLa_h \tveq_h \dd\Omega .
\end{split}
\end{equation}
}
To obtain this result, in (\ref{eq:penalty_DG_S}) -- (\ref{eq:penalty_DG_F}),
we let $\vep=\veq, \tvep=\tveq$. 
Taking the summation over all penalty terms on solid-solid interfaces 
yields
{\ColorRed
\begin{equation}\label{eq:penaltyE1}
\begin{split}
\Sumss\innprod{
    \jmp{\tsA_{n,h}\veq_h}_\rSS,
    (\tsA_{n,h}\veq_h)^\mm}_\Tss
&=\hspace{6mm}\Sumss\innprod{
    (\tsA_{n,h}\veq_h)^\pp-(\tsA_{n,h}\veq_h)^\mm,
    (\tsA_{n,h}\veq_h)^\mm}_\Tss
\\&
=-\Half\Sumss\Norm{
    \jmp{\tsA_{n,h}\veq_h}_\rSS
  }_{L^2(\Tss)}^2 
\end{split}
\end{equation}
}
Taking the summation over all penalty terms on fluid-fluid interfaces yields
{\ColorRed
\begin{equation}\label{eq:penaltyE2}
\Sumff\innprod{
    \jmp{\ttsA_{n,h}\tveq_h}_\rFF,
    (\ttsA_{n,h}\tveq_h)^\mm}_\Tff 
=-\Half\Sumff\Norm{
    \jmp{\ttsA_{n,h}\tveq_h}_\rFF
 }_{L^2(\Tff)}^2.
\end{equation}
}
We rewrite the penalty terms on fluid-solid interface from the solid side as
\begin{equation}
	\begin{split}
	&\hspace{-1cm}\innprod{
	\jmp{\tsA_{n,h}\veq_h}_\rSF,
	(\tsA_{n,h}\veq_h)^\mm}_\Tsf \\
	=&\innprod{
	O^T(\ttsA_{n,h}\tveq_h)^\pp,(\tsA_{n,h}\veq_h)^\mm}_\Tsf
	-\innprod{(\tsA_{n,h}\veq_h)^\mm,
	(\tsA_{n,h}\veq_h)^\mm}_\Tsf ,
	\end{split}
	\label{eq:sf energy}
\end{equation}
and from the fluid side as
\begin{equation}
	\begin{split}
	&\hspace{-1cm}\innprod{
	\jmp{\ttsA_{n,h}\tveq_h}_\rFS,
	(\ttsA_{n,h}\tveq_h)^\mm}_\Tfs \\
	=&\innprod{
	O(\tsA_{n,h}\veq_h)^\pp,(\ttsA_{n,h}\tveq_h)^\mm}_\Tfs
	-\innprod{(\ttsA_{n,h}\tveq_h)^\mm,
	(\ttsA_{n,h}\tveq_h)^\mm}_\Tfs \\
	=&\innprod{
	(\tsA_{n,h}\veq_h)^\pp,O^T(\ttsA_{n,h}\tveq_h)^\mm}_\Tfs
	-\innprod{O^T(\ttsA_{n,h}\tveq_h)^\mm,
	O^T(\ttsA_{n,h}\tveq_h)^\mm}_\Tfs,
	\end{split}
	\label{eq:fs energy}
\end{equation}
in which the property $O O^T=I_{4\times 4}$ is used. 
Changing from the fluid to the solid sides yields
\begin{equation}
	\begin{split}
	&\hspace{-1cm}\innprod{
	\jmp{\ttsA_{n,h}\tveq_h}_\rFS,
	(\ttsA_{n,h}\tveq_h)^\mm}_\Tfs \\
	=&\innprod{
	(\tsA_{n,h}\veq_h)^\mm,O^T(\ttsA_{n,h}\tveq_h)^\pp}_\Tsf
	-\innprod{O^T(\ttsA_{n,h}\tveq_h)^\pp,
	O^T(\ttsA_{n,h}\tveq_h)^\pp}_\Tsf.
	\end{split}
	\label{eq:SF energy}
\end{equation}
Summation over all fluid-solid interfaces with
(\ref{eq:sf energy}) and (\ref{eq:SF energy}),
{\ColorRed
\begin{equation}\label{eq:penaltyE3}
\begin{split}
&\Sumsf\innprod{
    \jmp{\tsA_{n,h}\veq_h}_\rSF,
    (\tsA_{n,h}\veq_h)^\mm}_\Tsf 
+\Sumfs\innprod{
    \jmp{\ttsA_{n,h}\tveq_h}_\rFS,
    (\ttsA_{n,h}\tveq_h)^\mm}_\Tfs
\\
&\hspace{1cm}=-\Sumsf\Norm{
    O^T(\ttsA_{n,h}\tveq_h)^\pp-(\tsA_{n,h}\veq_h)^\mm
    }_{L^2(\Tsf)}^2
\\
&\hspace{1cm}
    =-\Sumsf\Norm{
    \jmp{\tsA_{n,h}\veq_h}_\rSF
    }_{L^2(\Tsf)}^2
.
\end{split}
\end{equation}
}
Thus we obtain (\ref{eq:DG-penalty-energy}).

Our approach is reminiscent of earlier work, in which an upwind flux
is defined by the Riemann solutions which are obtained by
diagonalizing $\tsA_n$, that is, $\tsA_n = R D R^T$, on the faces of
each element [\cite{Wilcox2010}], and $D$ is the diagonal matrix of
eigenvalues of $\tsA_n$. 
The upwind flux takes the form,
\begin{equation}\label{eq:upwind flux}
\begin{split}
&\Flux^{\mathrm U}_{\mathrm S*}=
\innprod{\jmp{\tsA_n \veq}_{\mathrm S*}, 
(\tsLa \vep)^-}_{\Sigma_{\mathrm S*}^\mathrm{e}}
+\innprod{\jmp{(R |D| R^T) 
    \veq}_{\mathrm S*}, 
(\tsLa \vep)^-}_{\Sigma_{\mathrm S*}^\mathrm{e}},
\\
&\tFlux^{\mathrm U}_{\mathrm F*}=
\innprod{\jmp{\ttsA_n \tveq}_{\mathrm F*}, 
(\ttsLa \tvep)^-}_{\Sigma_{\mathrm F*}^\mathrm{e}}
+\innprod{\jmp{(\widetilde R |\widetilde D| \widetilde R^T) 
    \tveq}_{\mathrm F*}, 
(\ttsLa \tvep)^-}_{\Sigma_{\mathrm F*}^\mathrm{e}}, 
\quad *\in\{\mathrm{S},\mathrm{F}\},
\end{split}
\end{equation}
where $|\centerdot|$ stands for the operaton of taking the absolute
value of each entry of the diagonal matrix, that is, $|D|_{ij}=|D_{ij}|$.
Our approach avoids this diagonalization, allowing general 
heterogeneous media with anisotropy.

\section{Time Discretization}

In this section, we discuss a time discretization that is
computationally efficient for complex domains. 
Often, the computational meshes used to model the subsurface
must contain regions where the characteristic lengths of the elements
drop far below that of a wavelength because the subsurface contains
very complex geometries and discontinuities. As a result, the time
steps must be equally reduced to produce a stable solution. We follow
two different time discretization schemes: (1) for non-complex
domains, it is advantageous to use a traditional Runge--Kutta (RK)
method and (2) for complex domains, a semi implicit--explicit (IMEX)
method is used. The IMEX method enables the solver to perform implicit
time integration in areas of oversampling, while keeping the
computational efficiency of RK in regions of proper sampling.

\subsection{Explicit Runge--Kutta}

We use an explicit time integration method when the variation in
element size is small. There are a variety of time-stepping methods
available, however, we employ the five stage low-storage explicit
Runge--Kutta (LSERK) method from \cite{Cockburn2001}. 
LSERK is an explicit method the time-step of which is dictated by the
Courant--Friedrichs--Lewy (CFL) condition. 
{\ColorRed
Efforts to define, quantitatively, a stable CFL condition depending on
polynomial order $N_p$, can be found in \cite{Cockburn2001}.
}
The LSERK method is preferred over other methods because it saves
memory at the cost of computation time. 

\subsection{Explicit--Implicit Runge--Kutta}

When the domain in question contains complex geometries within large
domains, such as rough surfaces,
the resulting mesh will contain regions of oversampling relative to
the relevant wavelengths. This hinders the use of an 
{\ColorRed
implicit time-stepping method
}
because its accuracy depends on the size of the time step, which in
turn is dependent on the region of highest spatial sampling. 
A natural approach is
the IMEX method, (e.g. \cite{Ascher1997,Kanevsky2007,Persson2011}),
which allows the regions of oversampling to be integrated in time with
an L-stable third-order and 3-stage Diagonally Implicit Runge--Kutta
(DIRK) method, while using a fast and simple 4-stage third-order ERK
method in the regions of more reasonable sampling (8--10 nodes per
wavelength).

{\ColorRed
The system can be solved without requiring an
interpolation at the boundary of the implicit--explicit regions. 
The intermediate abscissaes of each time step
for implicit Runge--Kutta stages and 
for explicit ones
are selected to equal one another
so as to synchronize the explicit and 
implicit schemes, and the so-called Butcher matrix 
is calculated correspondingly.
The implicit stages 
are solved using a multifrontal factorization.
}

\section{Computational experiments}

Here, we illustrate our DG method by verifying its convergence rate and
carrying out computational experiments.
We use the fourth-order LSERK algorithm for time integration.
For visualization of wavefields or model parameters, 
we write the value in the Visualization Toolkit (VTK)
unstructured mesh format and visualize the result using 
\textit{Paraview}~\cite[]{Henderson2007}.

\subsection{Convergence tests at (interior) boundaries}
\label{sec:convtest}

We carry out computational tests using wave propagation and 
scattering problems in 3-dimensional cubic subdomains.
We first test the propagation of a plane wave
in a homogeneous isotropic elastic medium,
in which periodic boundary conditions are applied.
We also test the free-surface boundary condition with 
a homogeneous isotropic elastic solid,
in which both Rayleigh and Love waves are generated. 
We focus on the Rayleigh wave, 
the particle motion of which is
in the plane perpendicular to the free surface.
A Stoneley wave, generated at a solid-elastic 
interface~\cite[]{achenbach1973} in an unbounded domain
composed of two half spaces with different material properties,
is also simulated and 
compared with the closed-form solution in
\cite[Section 5.2]{Kaufman2005}.
For the test of our DG method at an acousto-elastic interface,
we generate a Scholte wave.
We refer to \cite{Wilcox2010} 
for the closed-form solution. 
The external boundary conditions, beside those mentioned above, 
are imposed 
{\ColorRed 
by using the traction of the exact solution as boundary ``forces''
~\cite[]{Wilcox2010}.
}

\begin{figure}
\centering
\begin{tabular}{cc}
	\includegraphics
	[trim = 4.5mm 0mm 10mm 10mm, clip=true,width=0.5\textwidth]
	{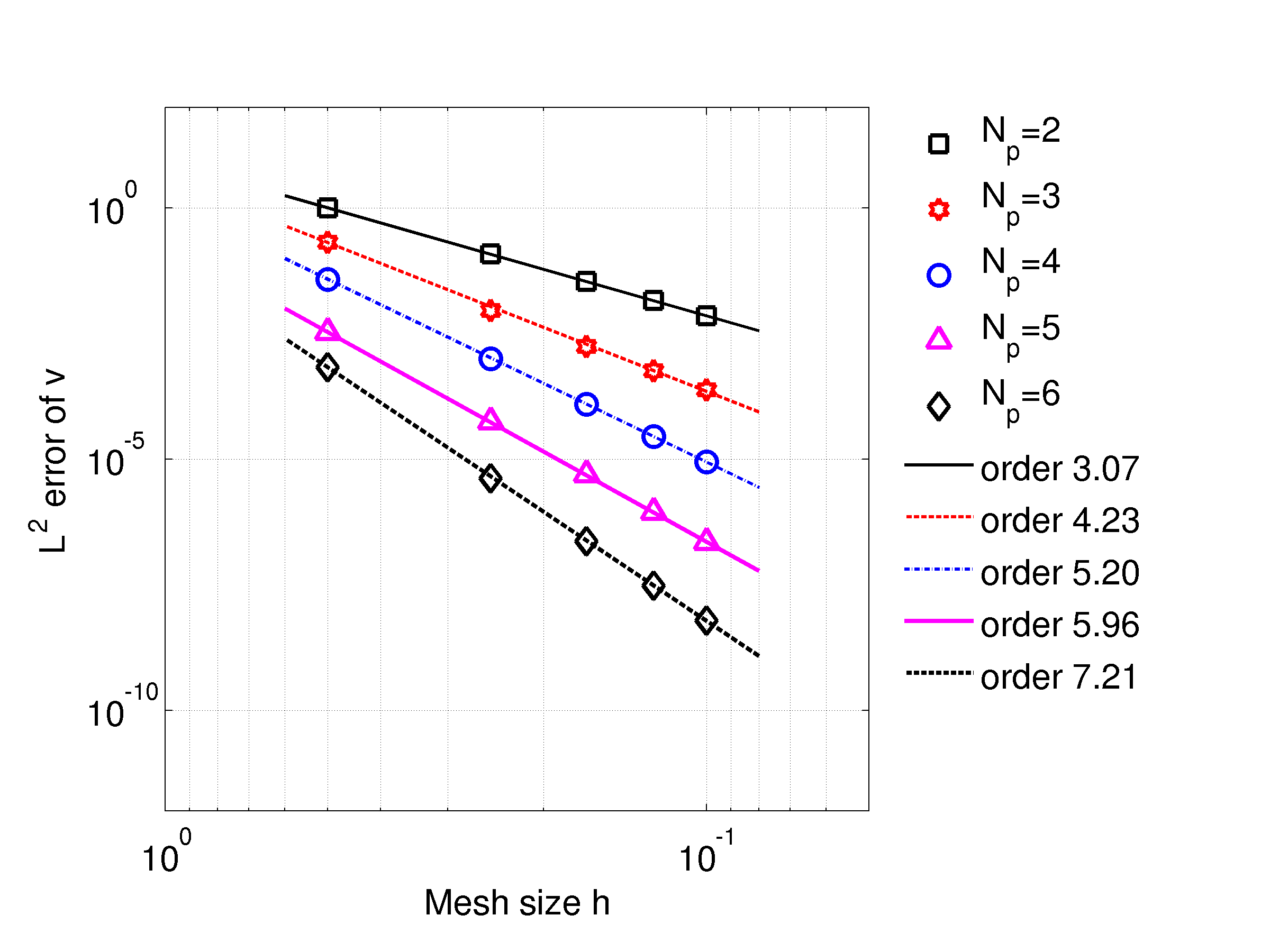}&
	\includegraphics
	[trim = 4.5mm 0mm 10mm 10mm, clip=true,width=0.5\textwidth]
	{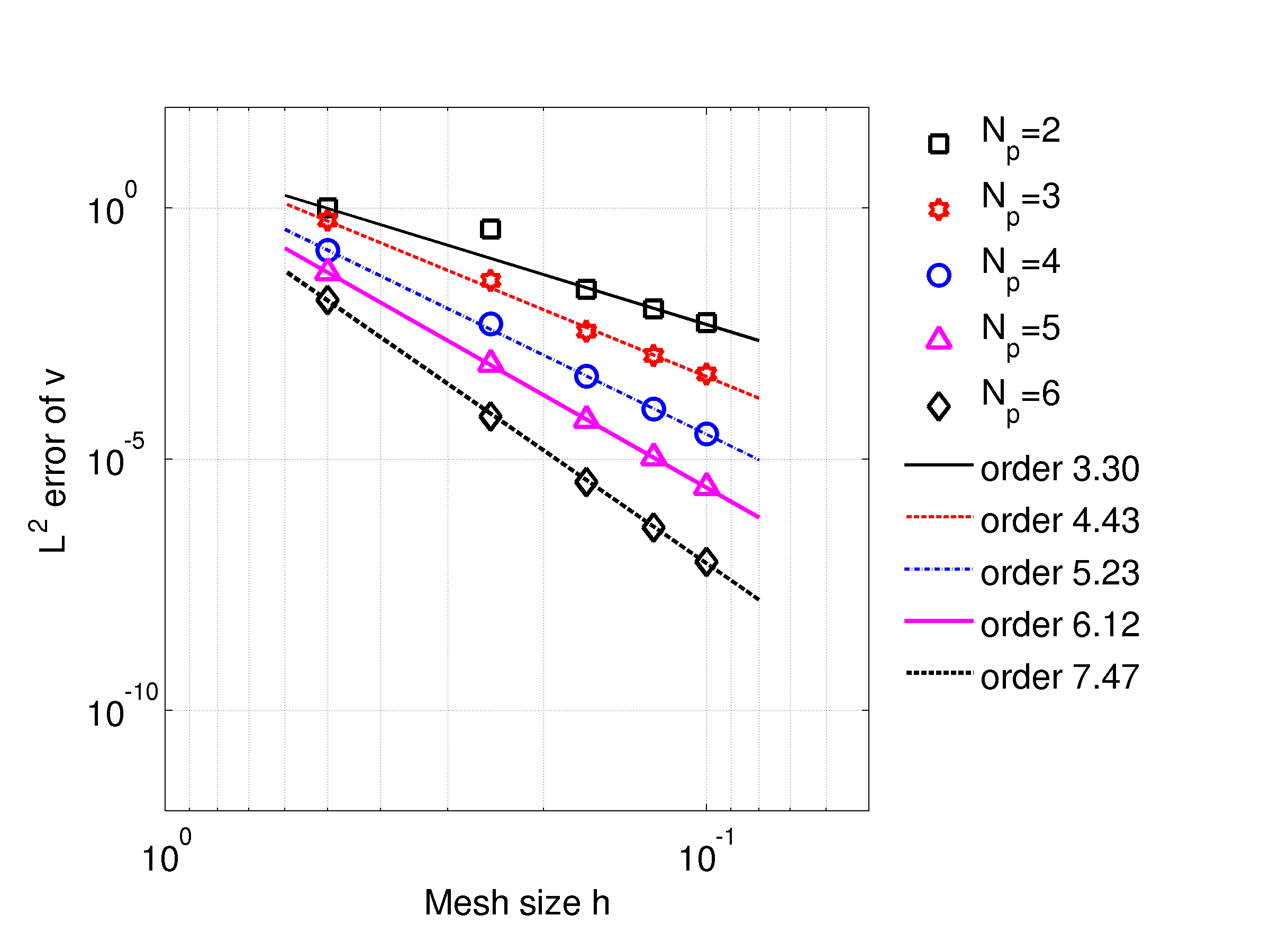}\\
	(A) & (B) \\
	\includegraphics
	[trim = 4.5mm 0mm 10mm 10mm, clip=true,width=0.5\textwidth]
	{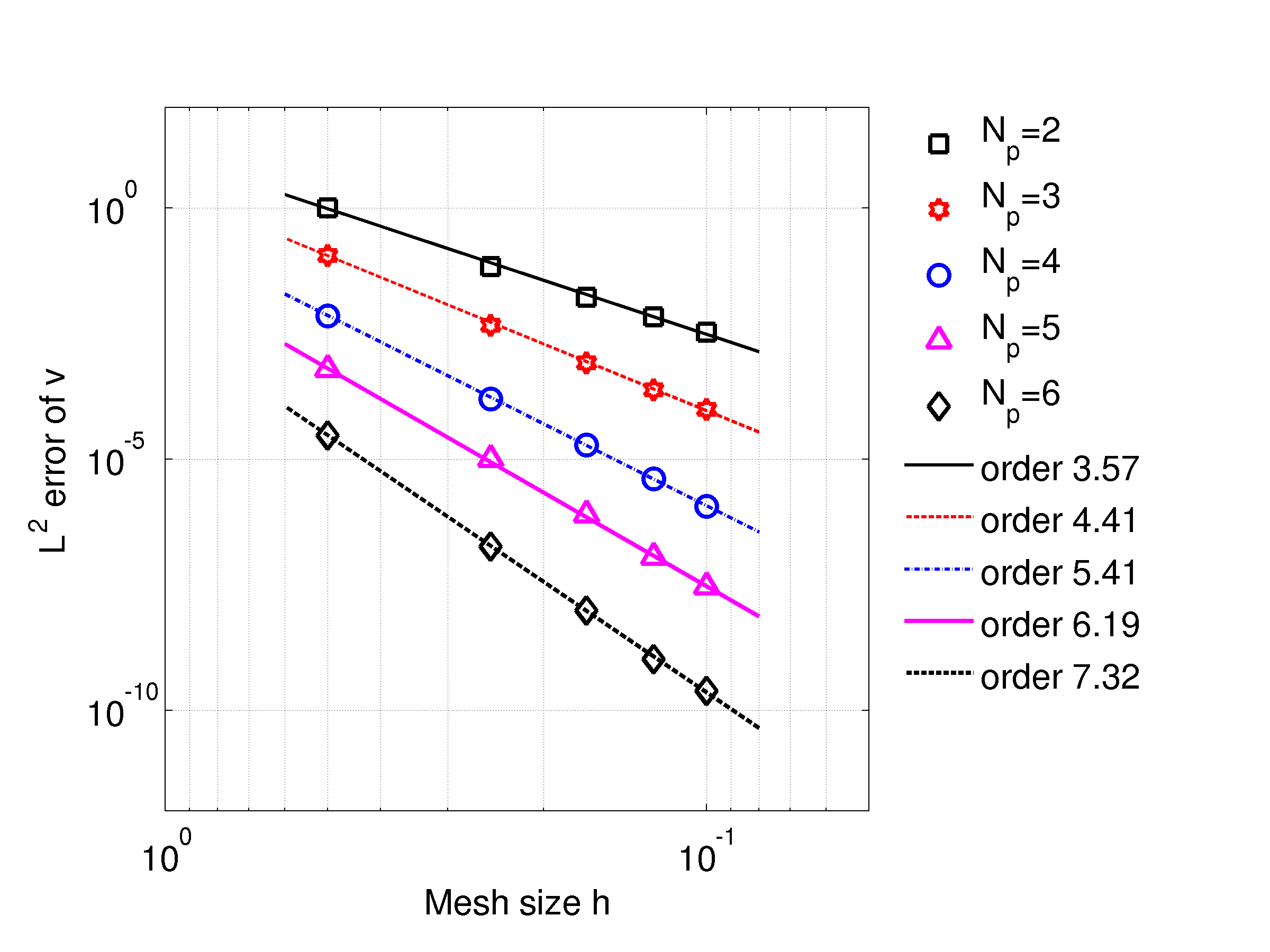}&
	\includegraphics
	[trim = 4.5mm 0mm 10mm 10mm, clip=true,width=0.5\textwidth]
	{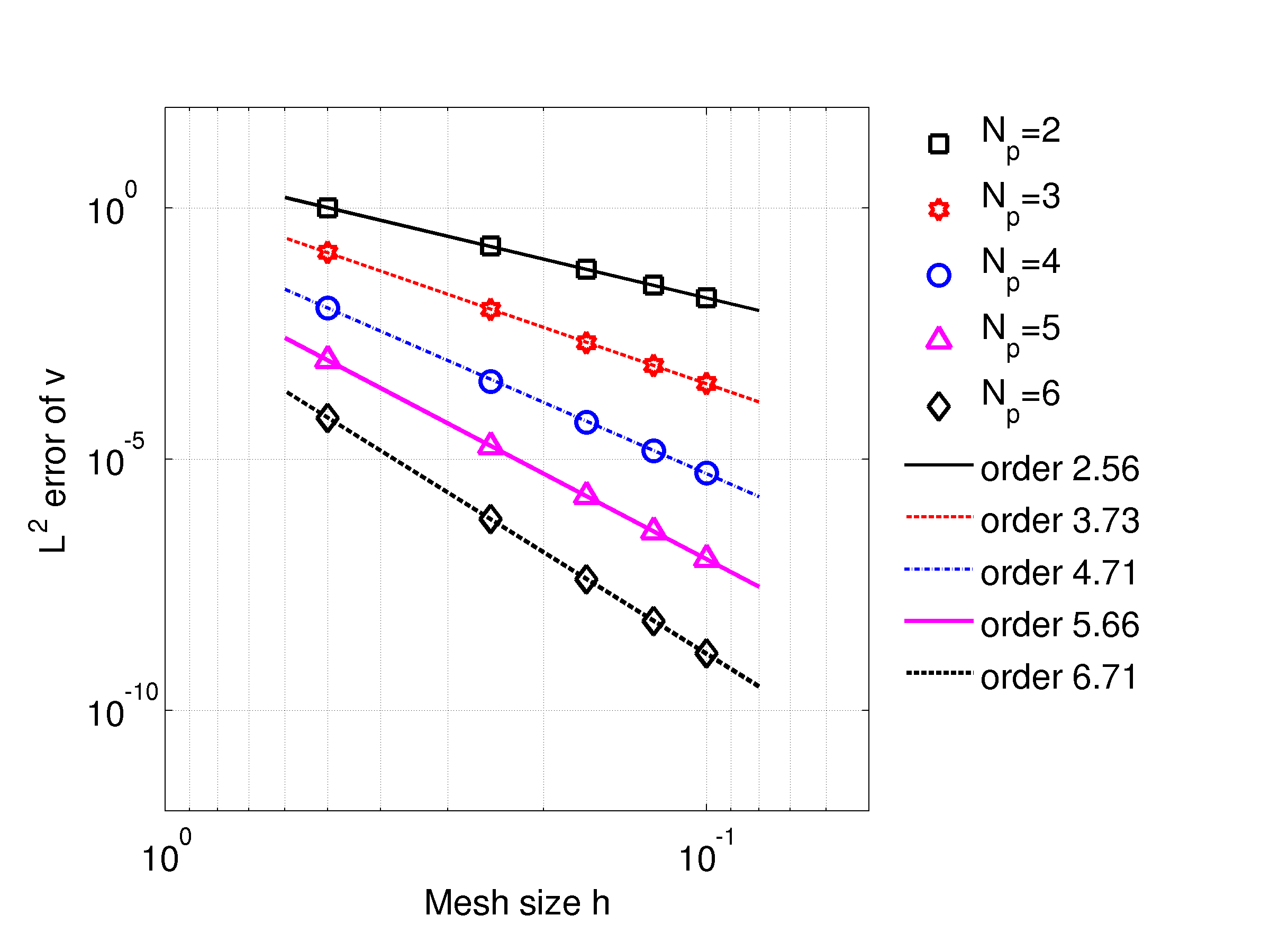}\\
	(C) & (D)
\end{tabular}
\caption{$L^2$ error of partical velocity $\vev$ as a function of 
	mesh size $h$,
  for the simulation of 
  (A) a plane wave,
  (B) a Rayleigh wave,
  (C) a Stoneley wave, and
  (D) a Scholte wave,
  for different orders $N_p = 2,3,\cdots,6 $ . 
  }
\label{fig:convtest}
\end{figure}

The computational domains are discretized as regular tetrahedral meshes.
A sufficiently small constant, $K_\mathrm{CFL}=0.05$, 
was selected during the tests for time stepping,
and a large simulation time (10 s) is choosen for the error computation.
The domain geometry and boundary conditions for each test are given in 
Table \ref{tab:convtest domain}. 
The relevant material parameters, that is, the Lam\'e  
parameters $\lambda$ and $\mu$, and density $\rho$, are given in 
Table \ref{tab:conv test material}. 
We calculate the $L^2$ errors for the particle velocity of the 
numerical solutions, 
which are discretized by $N_p$ order polynomials. 
The magnitudes of the numerical errors at time $t =$ 10 s are shown in 
Figure~\ref{fig:convtest}, as a function of mesh size $h$ for different
values of $N_p$, and least-squares fits to lines, 
with the estimated convergence order for each line shown in the legend.
We observe that the $L^2$ error of our numerical scheme
achieves a convergence rate higher than $N_p+\frac12$.
We also show a comparison of accuracies and convergence rates
tested with the wave types described in this section for the
upwind flux, the central flux and our penalty flux in Appendix
\ref{App:flux_cmp}.

\begin{table}
	\centering
	\begin{tabular}{l@{\hskip 1cm}ll}
		\hline\\[-1mm]
		wave type & domain range (in km) & 
		\hspace{5mm}
		boundary conditions \\[3mm]
		\hline\\[-1mm]
		plane wave & $[-1,1]\times[-1,1]\times[-1,1]$ & 
		\hspace{5mm}
		periodic boundaries \\[5mm]
		Rayleigh wave & $[-1,1] \times [-1,1] \times [0,2]$ &
		\begin{tabular}{@{}l@{}}
		free surface boundary at $x_3=0$, \\
		exact boundary \ColorRed ``force'' at $x_3=2$,\\
		periodic boundaries otherwise
		\end{tabular}\\[7mm]
		Stoneley wave & $[-2,2] \times [-2,2] \times [-2,2]$ &
		\begin{tabular}{@{}l@{}}
		exact boundary \ColorRed ``force'' at $x_3=\pm 2$,\\
		periodic boundaries otherwise
		\end{tabular}\\[5mm]
		Scholte wave & $[-2,2] \times [-2,2] \times [-2,2]$ &
		\begin{tabular}{@{}l@{}}
		exact boundary \ColorRed ``force'' at $x_3=\pm 2$,\\
		periodic boundaries otherwise
		\end{tabular}\\[5mm]
		\hline
	\end{tabular}
	\caption{Geometry and boundary conditions 
		for the four wave types in the convergence tests.
		}
	\label{tab:convtest domain}
\end{table}

\begin{table}
	\centering
	\begin{tabular}{l@{\hskip 1cm}l}
		\hline\\[-1mm]
		wave type & 
		\hspace{5mm}
		material properties 
		\\[3mm]
		\hline\\[-1mm]
		plane wave & 
		\hspace{5mm}
		$\lambda=2.00$ GPa, $\mu=1.00$ GPa,
		$\rho=1.00$ g/cm\superscript{3} 
		\\[4mm]
		Rayleigh wave & 
		\hspace{5mm}
		$\lambda=2.00$ GPa, $\mu=1.00$ GPa,
		$\rho=1.00$ g/cm\superscript{3} 
		\\[4mm]
		Stoneley wave & 
		\begin{tabular}{@{}lc}
		$\lambda=1.20$ Gpa, $\mu=1.20$ GPa,
		$\rho=1.20$ g/cm\superscript{3}, &
		for $x_3>0$ \\
		$\lambda=3.00$ Gpa, $\mu=1.20$ GPa,
		$\rho=4.00$ g/cm\superscript{3}, &
		for $x_3<0$
		\end{tabular}
		\\[5mm]
		Scholte wave & 
		\begin{tabular}{@{}lc}
		$\lambda=1.20$ Gpa, $\mu=1.30$ GPa,
		$\rho=1.10$ g/cm\superscript{3}, &
		for $x_3>0$ \\
		$\lambda=1.11$ Gpa, $\mu=0.00$ GPa,
		$\rho=1.32$ g/cm\superscript{3}, &
		for $x_3<0$
		\end{tabular}
		\\[4mm]
		\hline
	\end{tabular}
	\caption{Material parameters
		for the four wave types in the convergence tests.
		}
	\label{tab:conv test material}
\end{table}

\subsection{Homogeneous orthorhombic solid: Caustics}

\begin{figure}
\centering
\begin{tabular}{ccc}
\includegraphics[trim = 2.45in 2in 2.15in 1in, clip=true,height=1.6in]
{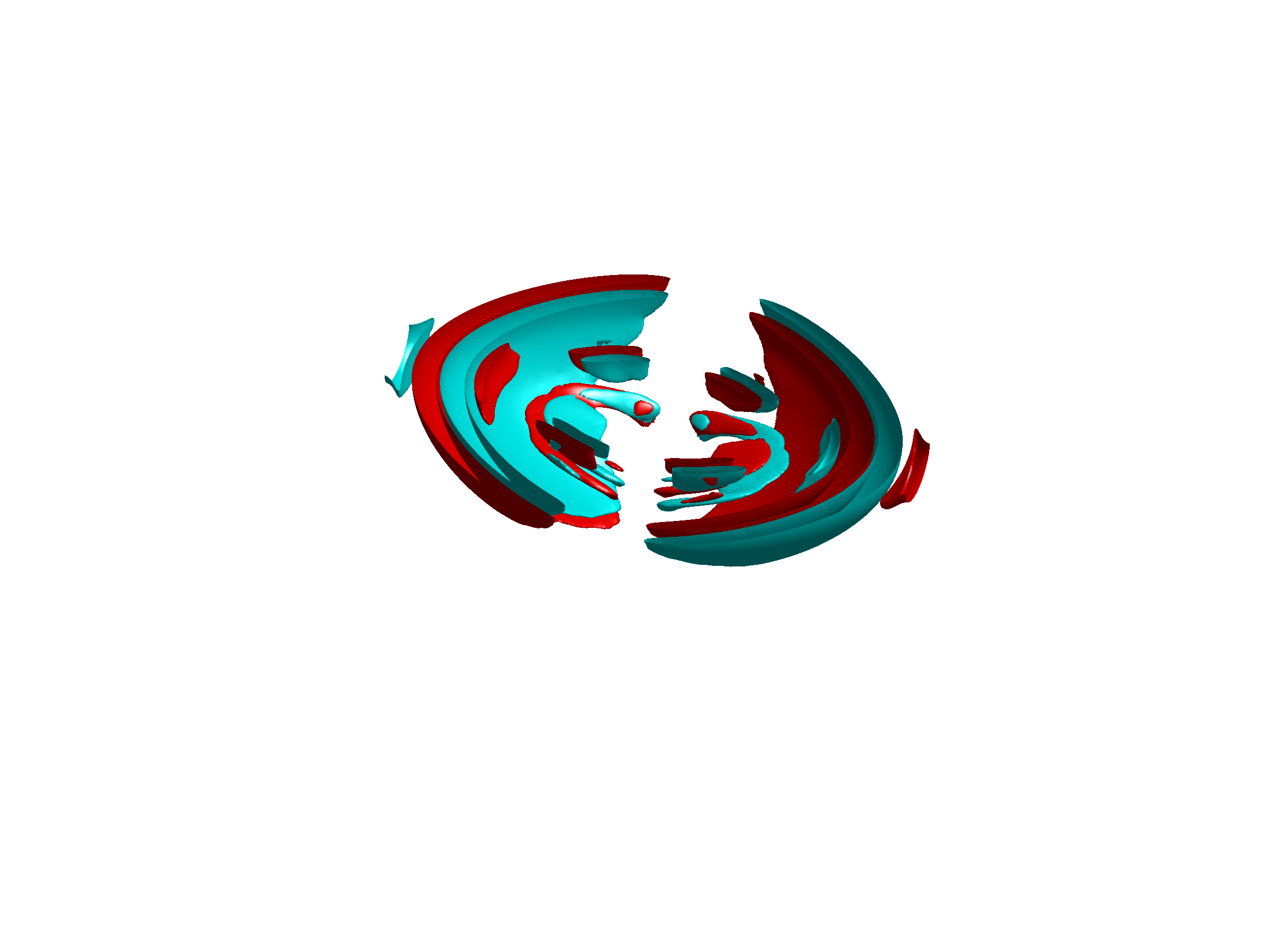} &
\includegraphics[trim = 2.45in 2in 2.15in 1in, clip=true,height=1.6in]
{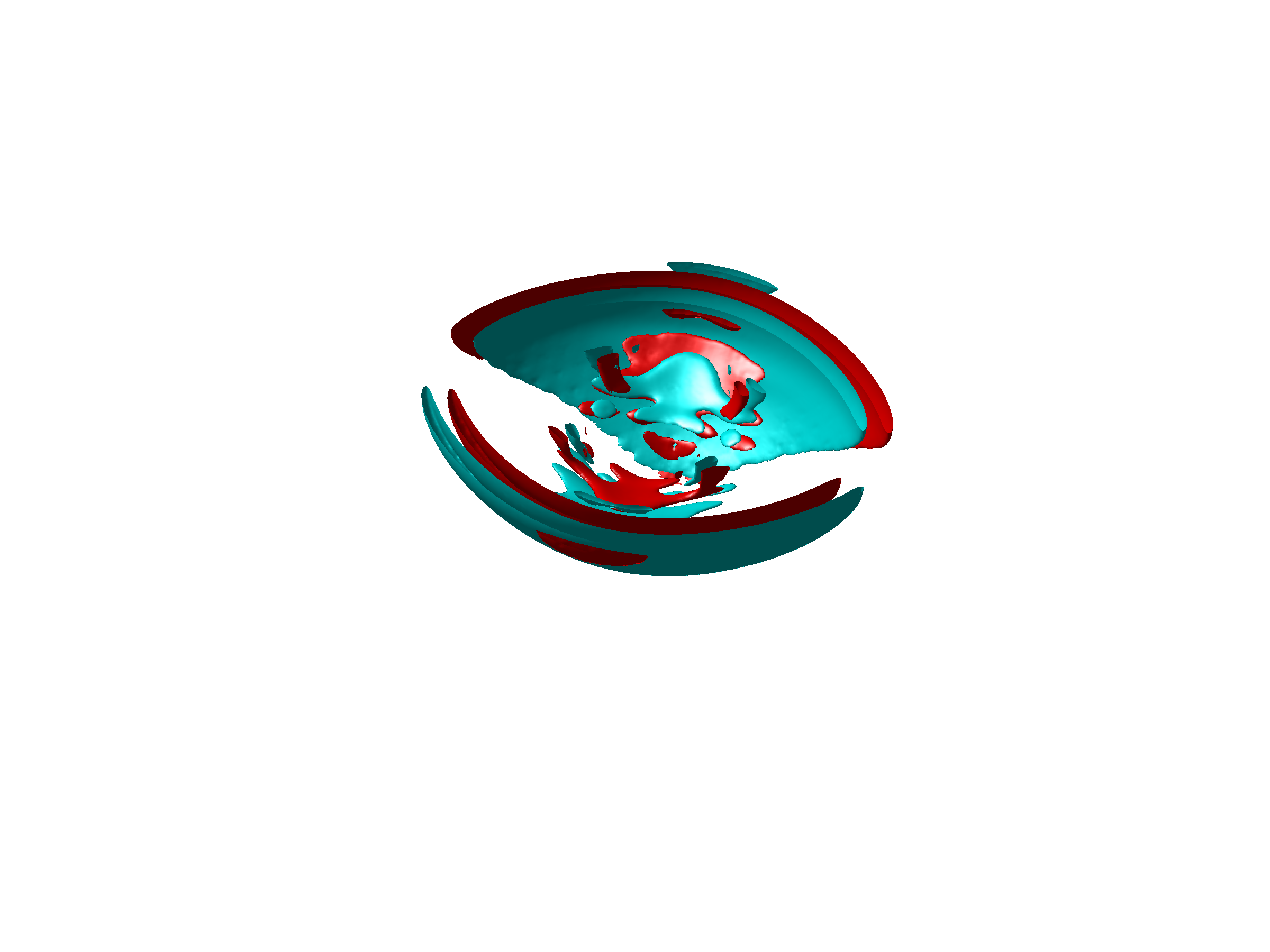} &
\includegraphics[trim = 0.5in 0in 0in 0.5in, clip=true,height=1.2in]
{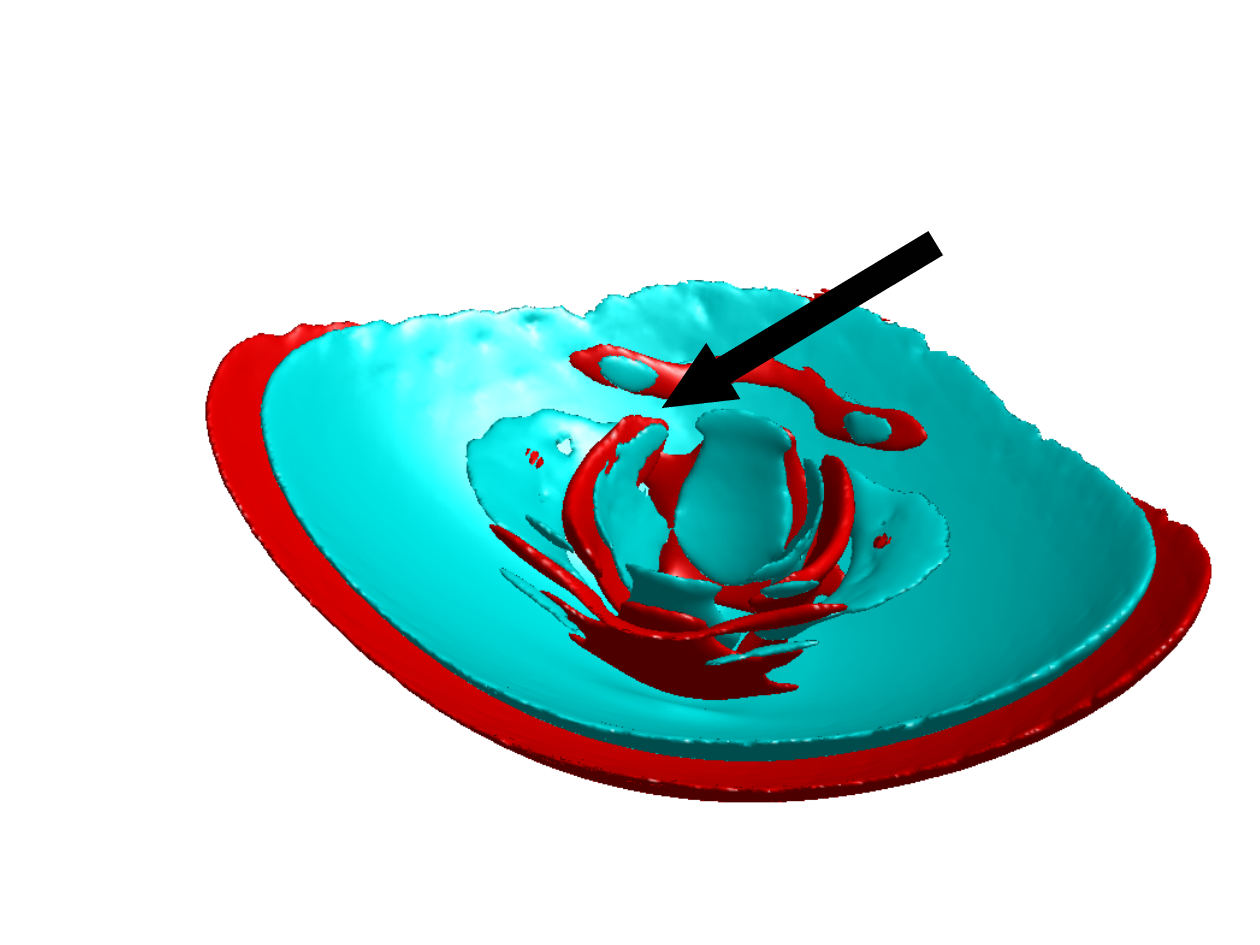} \\
(A) & (B) & (C)
\end{tabular}
\caption{Snapshots of the contours for the {\ColorRed particle velocity}
	(A) $v_1$, (B) $v_2$, and (C) $v_3$ at $t=0.45 s$. 
The black arrow in (C) indicates
the shear wave front forming caustics.}
\label{Fig_orthorhombic}
\end{figure}

Here, we simulate a band-limited fundamental solution in an anisotropic
elastic medium, forming caustics. The medium is orthorhombic and
homogeneous. 
Several minerals in Earth's mantle
have orthorhombic symmetry; this symmetry also appears in regions of
sedimentary basins where fracture sets are commonly found in sandstone
beds, shales, and granites.
The material properties are selected as follows,\\[5mm]
\begin{tabular}{clccccccccccl}
\hline
$\rho$ & &\vline & $C_{11}$& $C_{22}$& $C_{33}$& $C_{44}$& $C_{55}$& $C_{66}$
& $C_{23}$& $C_{13}$& $C_{12}$ & \\
\hline
1.0 & \hspace{-3mm}(g/cm\superscript{3}) &\vline & 30.40 & 19.20 & 16.00 
& 4.67 & 10.86 & 12.82 & 4.80 & 4.00 & 6.24 & (GPa) \\ 
\hline 
\end{tabular}
\\[5mm]
which produce a medium
whose P phase velocities are 5.51 km/s, 4.38 km/s, and 4.00 km/s and
S phase velocities are 2.16 km/s, 3.26 km/s, and 3.58 km/s in the
principal directions (perpendicular to the symmetry planes).
The computational domain is a $5 \times 5
\times 5$ (in km) cube. We place an explosive Gaussian source 
at the center of the cube, using a Ricker wavelet with a
center frequency of 5Hz. 
Images of isosurfaces of the different components of the particle velocity
are shown in Figure~\ref{Fig_orthorhombic}. 
We note the presence of caustics in one of the shear polarizations. 

\subsection{Flat isotropic fluid-solid interface: 
            Propagation of Scholte wave}

\begin{figure}
\centering
\includegraphics[trim = 1.2in 1.6in 0.3in 1.1in, clip=true,
      width=0.8\textwidth]{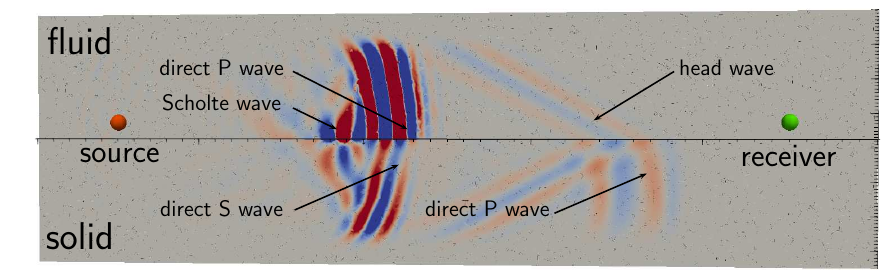}\\[-8mm]
\raggedright \hspace{8mm}(a)\\[10mm]
\centering
\includegraphics[trim = 1.2in 1.6in 0.3in 1.1in, clip=true,
      width=0.8\textwidth]{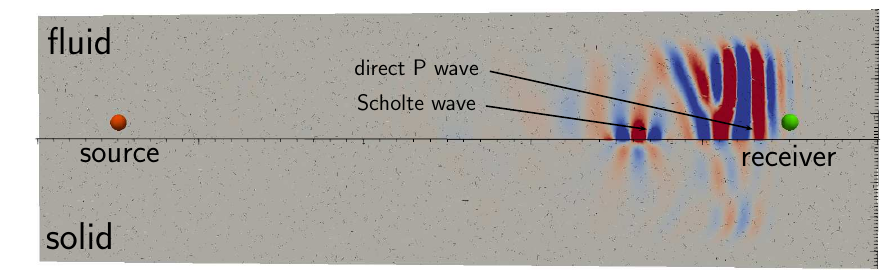}\\[-8mm]
\raggedright \hspace{8mm}(b)\\[10mm]
\caption{ 
	Fluid-solid configuration visualized in the
	$x_1$--$x_3$ plane at $x_2=15.0$, with source and receiver located
	in the fluid. 
	A snapshot at $ t=12s $ is shown in (a), and a snapshot at $ t=26s $ 
	is shown in (b).}
  \label{fig:Scholte snap}
\end{figure}

\begin{figure}
\centering
\includegraphics[width=0.6\textwidth]{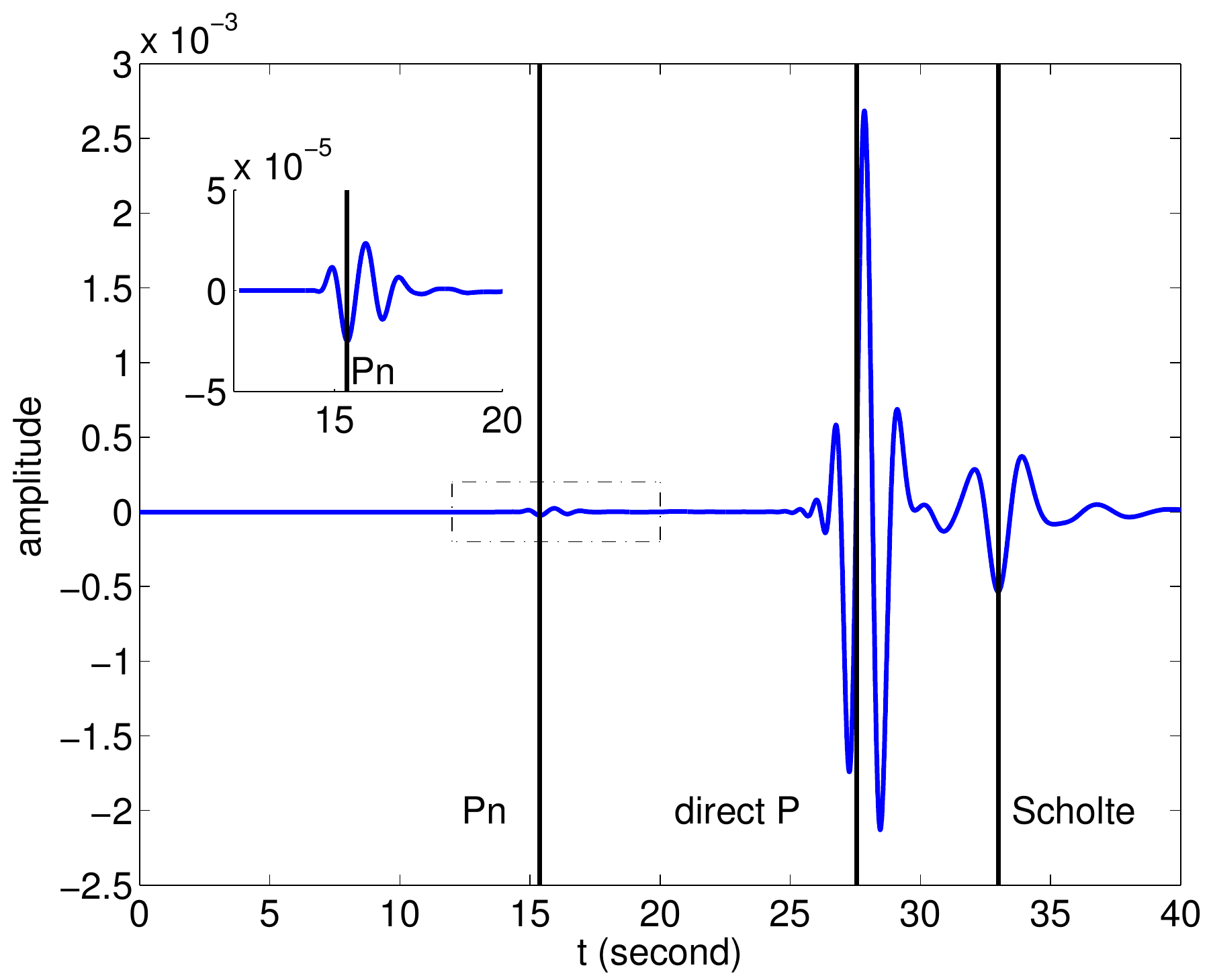}
\caption{Seismic trace from a hydrophone located at $ (40.0,\, 15.0,\,
  6.0) $km in the fluid side. Arrival times of head wave Pn, direct P waves 
  and Scholte waves are indicated by vertical lines. }
  \label{fig:Scholte Phase}
\end{figure}

We present a model with dimensions $[0,50]\times[0,30]\times[0,15]$km
with a flat fluid-solid interface located at $ x_3=7.5 $km. The fluid
side is homogeneous isotropic with an acoustic wave speed $ 1.5 $km/s
and density $ 1.0 $g/cm\superscript{3}. The solid side is homogeneous
isotropic with a P-wave speed $ 3.0 $km/s and S-wave speed $ 1.5
$km/s, and density $ 2.5 $g/cm\superscript{3}. The Scholte wave speed
is computed numerically as $1.2455 $km/s~[e.g., 
\cite{Kaufman2005}].
We place an explosive source in the fluid at location $(5.0,\, 15.0,\,
6.5)$km, using a Ricker wavelet as the source-time series with a central 
frequency of $ 2.0 $Hz. A
receiver is located at $ (45.0,\, 15.0,\, 6.5) $km and records the
synthetic phases for 40 seconds. We apply convolutional perfect matching 
layers (CPMLs)
~[e.g., \cite{Komatitsch2007}]
for all external boundaries of the model, 
highlighting the effects of a fluid-solid internal boundary.

Two snapshots are shown in Figure \ref{fig:Scholte snap}, 
one for the solution at $ t=12 $s 
and the other for the solution 
at $ t=26 $s, in which we observe 
the occurence of a Scholte wave which is well seperated from the body wave
phases at long times. 
The amplitude of the Scholte wave decays exponentially with the 
distance from fluid-solid interface [\cite{Kaufman2005}]. 
Figure \ref{fig:Scholte Phase} shows
the seismogram as well as the arrival times of the head wave Pn, 
the direct P wave and
Scholte wave. The modelled phase arrivals agree well with the travel times 
marked by perpendicular lines.

\subsection{Seismic waves in a geological structure: SEAM model}

\begin{figure}
\centering
\includegraphics[trim = 0.0in 0.0in 2.4in 0.0in, clip=true, 
	width=0.8\textwidth]{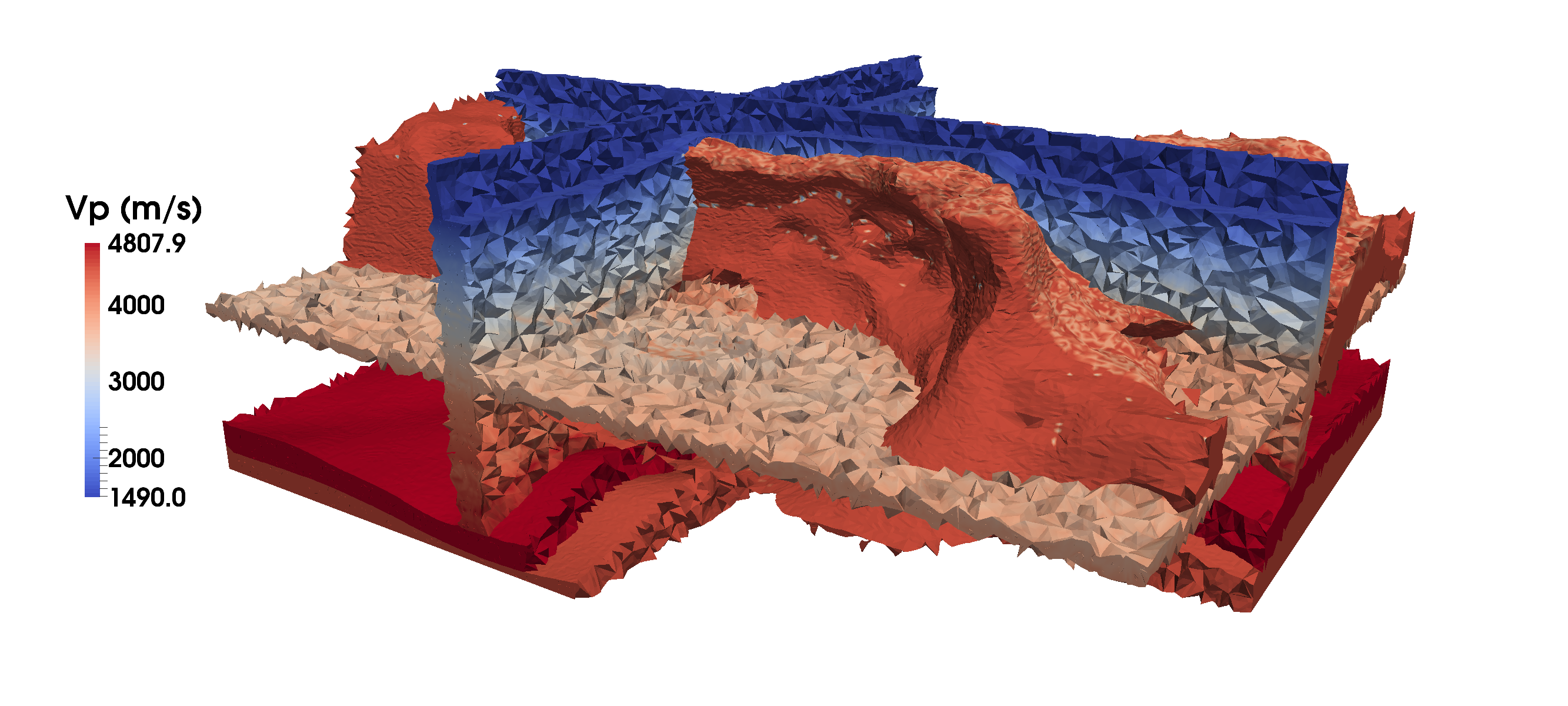}
\caption{\ColorRed A tetrahedral meshing for the 3D SEAM\@ generated by 
	segmentation and mesh deformation techniques.
	The color map shows the P wavespeed $v_p$ interpolation.
  }
\label{fig:seam_mesh}
\end{figure}

In this application, the DG method's ability to model the propagation
and scattering of seismic waves in a field-scale domain with complex
geological structures is demonstrated. The 3D SEAM (SEG Advanced
Modeling) Phase I acoustic model is used that has heterogeneous structures
and represents the sea-bed of the Gulf of Mexico~[\cite{Fehler2011}]. 
It spans a 35 km
by 40 km region of the earth's surface and has a depth of 15 km,
and is discretized as a regular grid 
with 20m $ \times$ 20m $\times$ 10m
sample interval. The model has several geological features that we
will use to test the robustness of the DG method. It contains a
high-velocity salt body that extends through the center of the model
(Figure~\ref{fig:seam_mesh}). The rapid contrast in velocity makes the
model, in the language of partial differential equations, a stiff
domain. Another geometric feature is the sedimentary layering at
approximately 10 km under the surface. These layers will cause multiple
scattering that will lead to constructive and destructive
interference.

\begin{figure}
\centering
\includegraphics[width=0.8\textwidth]{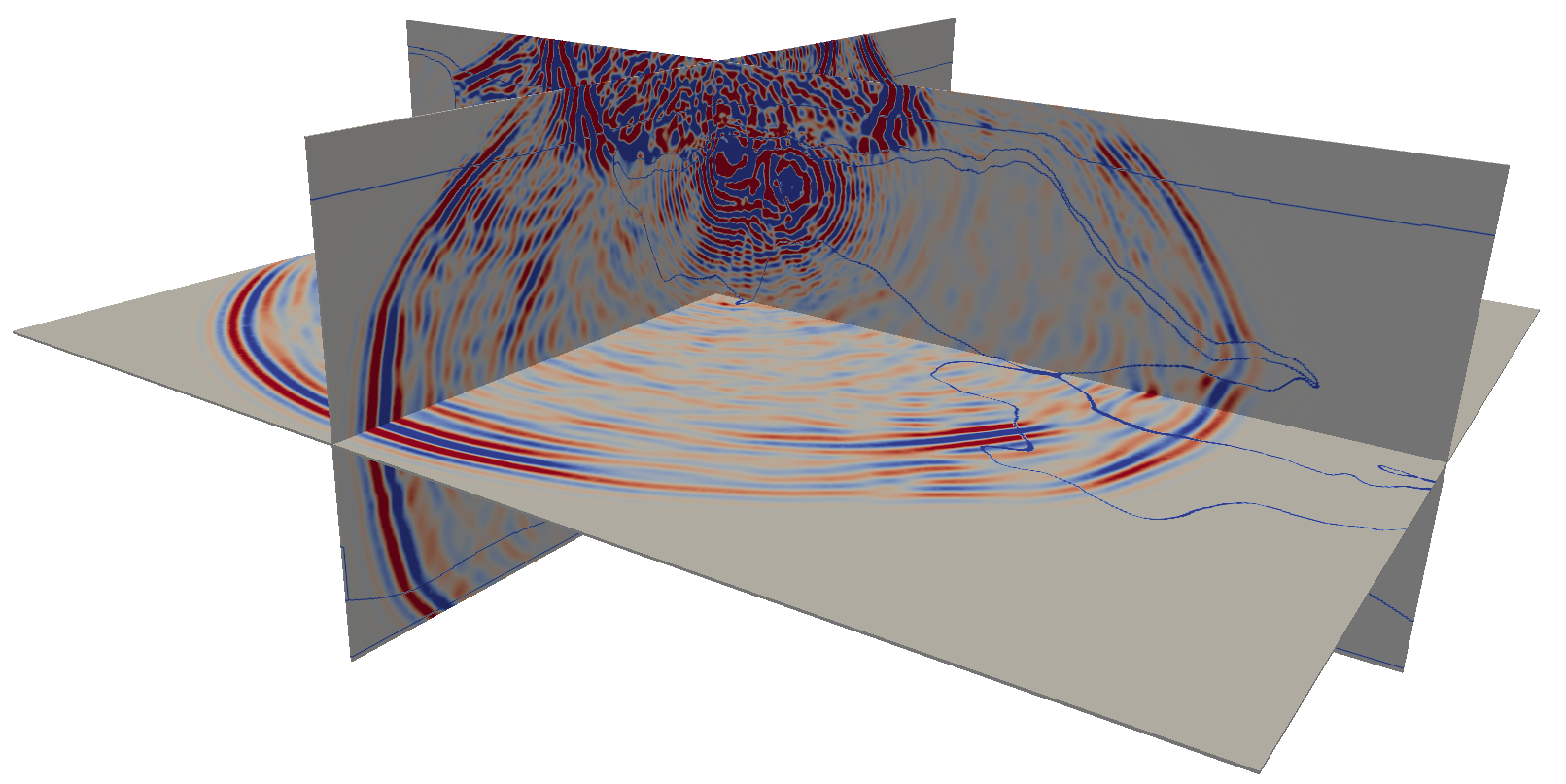}
\caption{Slices of the 3D SEAM acoustic velocity model and snapshot of
	pressure wave field at $t=5.0$s, with the same viewpoint as in 
	Figure~\ref{fig:seam_mesh}.}
\label{fig:seam_data1}
\end{figure}

\begin{figure}
\centering
\includegraphics[trim = 0.2in 1.0in 0.0in 0.5in, clip=true, 
width=1.0\textwidth]{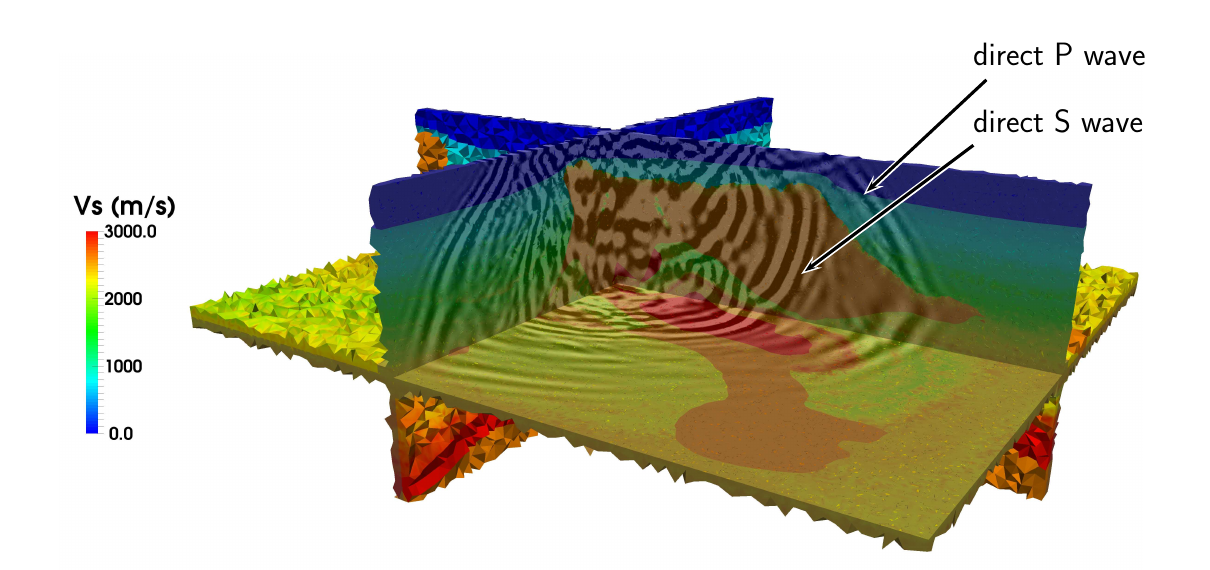}\\[-0.0cm]
\caption{Slices of the isotropic extension of 3D SEAM Phase I shear 
	wavespeed model and snapshot of \ColorRed
	3-component of particle velocity at $t=5.0$s, with the same viewpoint as in
	Figure~\ref{fig:seam_mesh} and \ref{fig:seam_data1}.}
\label{fig:seam_data2}
\end{figure}

A tetrahedral mesh with 863,973 elements of order 3 is generated adaptively
starting from the contours of the wave speed model, including the
rough boundary of the salt body (Figure~\ref{fig:seam_mesh}) 
and selected smooth interfaces associated with the sedimentary layers. 
We generate triangular isosurfaces based on domain partitioning of the 
wavespeed model into four primary subdomains: 
the ocean layer, the salt body, a high-contrast sediment layer and 
the sediment background. We also adaptively add
vertices by tracking the contrasts of wavespeed inside each subregion. 
Using these, a tetrahedral mesh was created using \textit{TetGen}
[\cite{Si2015}]. 
A point source is located at the ocean bottom $(x_1,x_2,x_3)=(17.5 , 15.0 ,
1.45)$km and the source function was a Ricker wavelet with a center
frequency of 10.0 Hz.
{A snapshot of the acoustic pressure wave field
solution is shown in Figure~\ref{fig:seam_data1}.}

We also consider an extension of the SEAM Phase I model to 
isotropic elasiticity as is presented by 
\cite{Oristaglio2012}. We represent, via interpolation,
the S wave speed and density on the unstructured mesh based on
the four distinct subdomains,
and place a point source inside the ocean layer at
$(x_1,x_2,x_3)=(17.5 , 15.0 , 0.10)$km
using a Ricker wavelet with a center frequency of 5.0 Hz.
We apply a pressure-free surface boundary condition
on the ocean surface, and CPMLs elsewhere.
The S wavespeed and $3$-component of the particle velocity are shown 
in Figure~\ref{fig:seam_data2},
in which the shear wave front can be clearly observed after the P arrivals.

\subsection{Scattering from a rough surface: Fractured carbonate}
\label{section rsurf3d}

Here, we model the reflection generated by an explosive point source
from a rough surface embedded in a transversely
isotropic medium. This type of medium closely resembles fractured
samples of carbonate rocks~[\cite{Li2009}]. Carbonates are
abundantly found in nature. They pose many complications when working
with them in the field because the physical properties vary from site
to site and are strongly heterogeneous within the bulk rock.  A
homogeneous transversely isotropic medium can be used to model a
carbonate because a variation in velocity amongst layers is the most
common form of heterogeneity~[\cite{Nurmi1990}].

\begin{figure}
\centering
\begin{tabular}{cc}
	\includegraphics[width=0.4\textwidth]{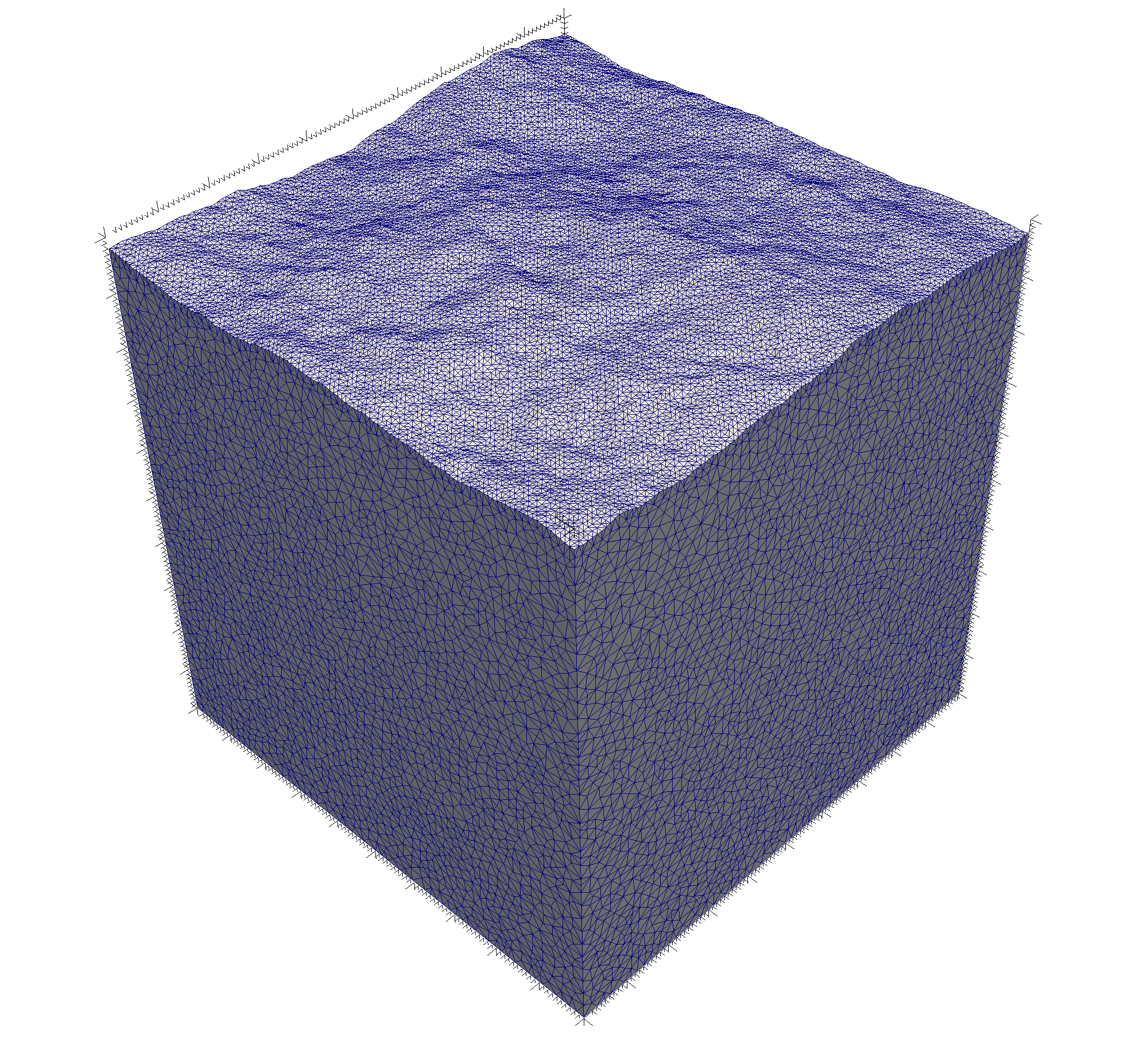}&
	\includegraphics
	[trim = 0in 2.1in 2.1in 0in, clip=true,width=0.4\textwidth]
	{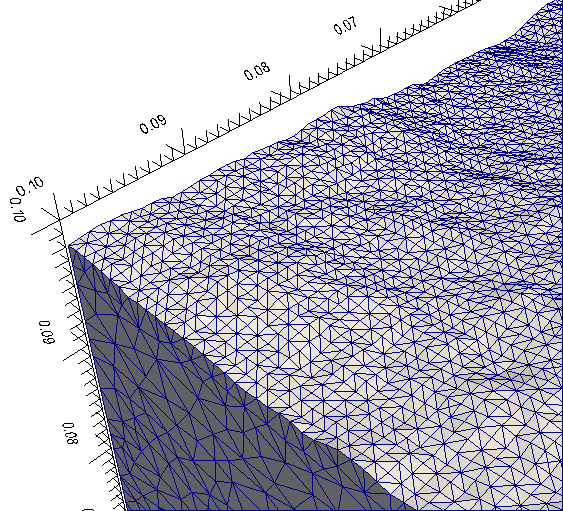}\\
	(A) & (B)
\end{tabular}
\caption{(A) Domain of the digitized rough surface. (B) Zoomed in of the mesh. 
The unit of the axises are in meters.}
\label{fig:3droughsurfdomain}
\end{figure}

\begin{figure}
\centering
\begin{tabular}{cc}
	\includegraphics[width=0.50\textwidth]{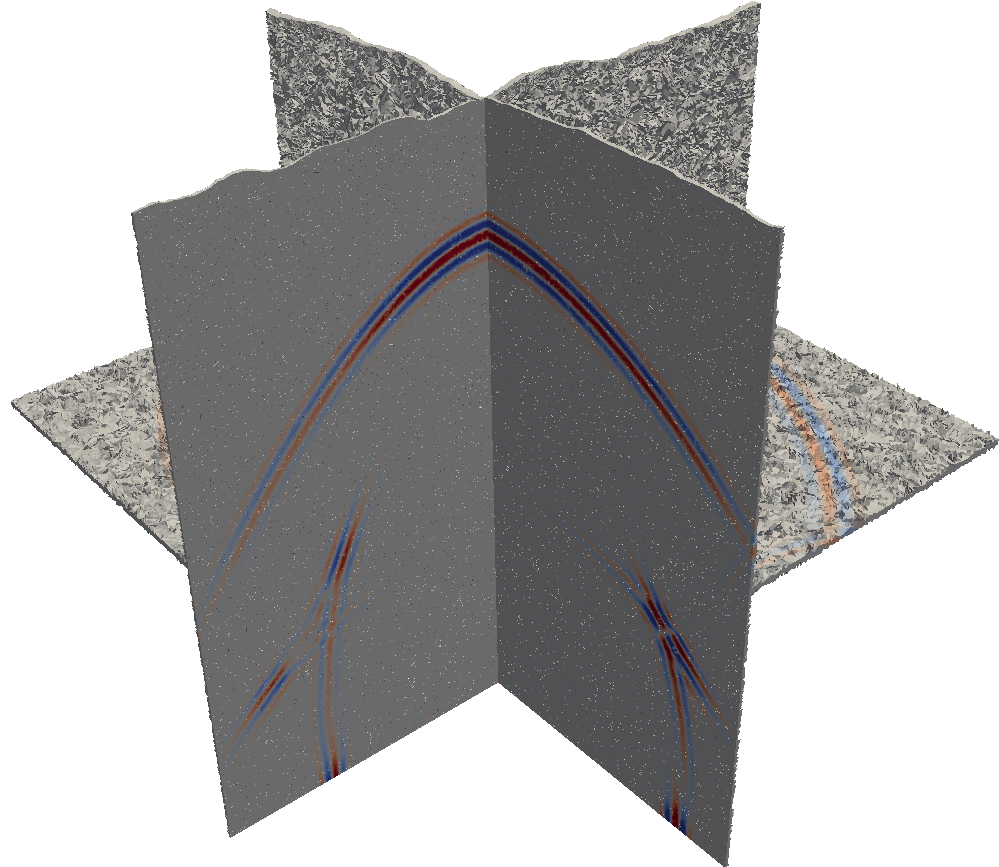}&
	\includegraphics[width=0.50\textwidth]{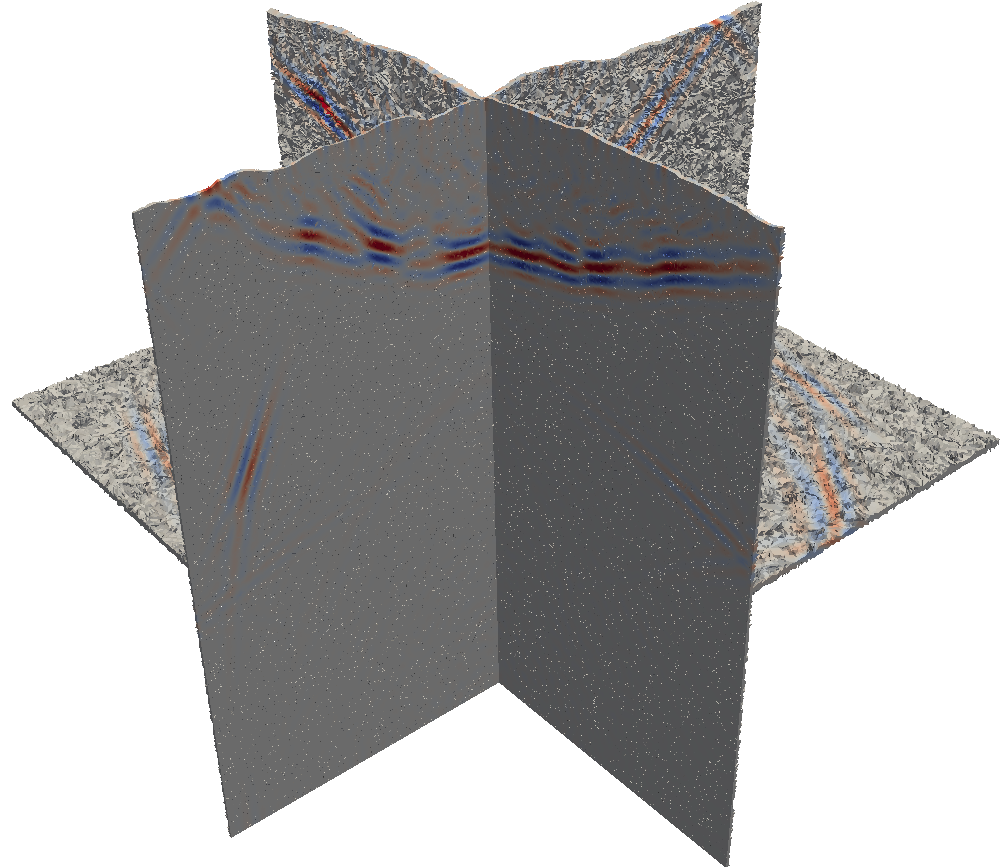}\\
	(A) & (B) \\
	\includegraphics[width=0.50\textwidth]{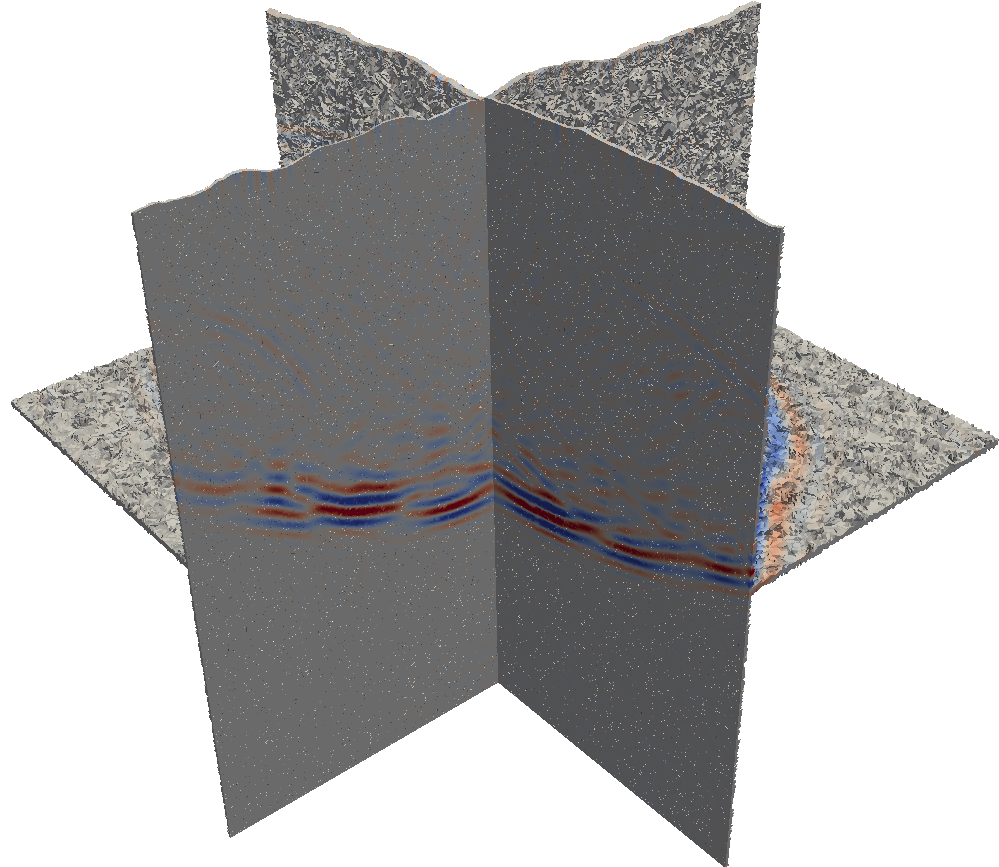}&
	\\ (C) &
\end{tabular}
\caption{Slices of the $V_3$ wave field after (A) 21 $\mu s$, (B) 31 $\mu s$, 
	and (C) 41 $\mu s$ from a 3D rough surface.}
\label{fig:3droughsurfdomain_data}
\end{figure}

Laser profilometry was used to measure the surface roughness of an induced
fracture in Austin Chalk, a carbonate rock sample. 
From these measurements, a profile of the surface was extracted to provide 
a rough boundary in an otherwise cubic domain 
with edge length of 0.1 m. The rough surface was placed
on the top plane of the box, i.e.\
$x_3=0.1$m (Figure~\ref{fig:3droughsurfdomain}). The material properties were
chosen such that the symmetry axis was in the
$(\hat{x}_1,\hat{x}_2,\hat{x}_3)=(0,1,0)$ direction. 
P- and S-phase velocities
along the axis of symmetry are 4000 m/s and 2280 m/s
respectively, and are 4900 m/s and 2000 m/s respectively along the
other two directions.
The following table provides a list of the specific
elastic constants used: \\[5mm]
\begin{tabular}{clccccccccccl}
\hline
$\rho$ & &\vline & $C_{11}$& $C_{22}$& $C_{33}$& $C_{44}$& $C_{55}$& $C_{66}$
& $C_{23}$& $C_{13}$& $C_{12}$ &\\
\hline
1.5 & \hspace{-3mm}(g/cm\superscript{3}) &\vline & 24.00 & 16.00 & 24.00 
& 4.00 & 5.20 & 4.00 & 8.00 & 13.60 & 8.00 & (GPa) \\ 
\hline
\end{tabular}
\\[5mm]
The tetrahedral mesh contains 686,444 elements, with $N_p=4$.
We place an explosive source at $(x_1,x_2,x_3) = (.05,.05,0)$,
using a Ricker wavelet with a central frequency of 1 MHz.
Two snapshots of the wave field were taken of the
$3$-component of the particle velocity (Figure
\ref{fig:3droughsurfdomain_data}) that display the formation of shear-wave
caustics due to anisotropy at $t = 21 \mu s$, 
and the solutions 
of scattering at $t = 31 \mu s$ and $t= 41 \mu s$, respectively.

\subsection{Heterogeneous anisotropic solid-fluid boundary with
            topography}

\begin{figure}
\centering
\includegraphics[trim = 0in 0in 0in 0.3in, clip=true,width=4.5in]
          {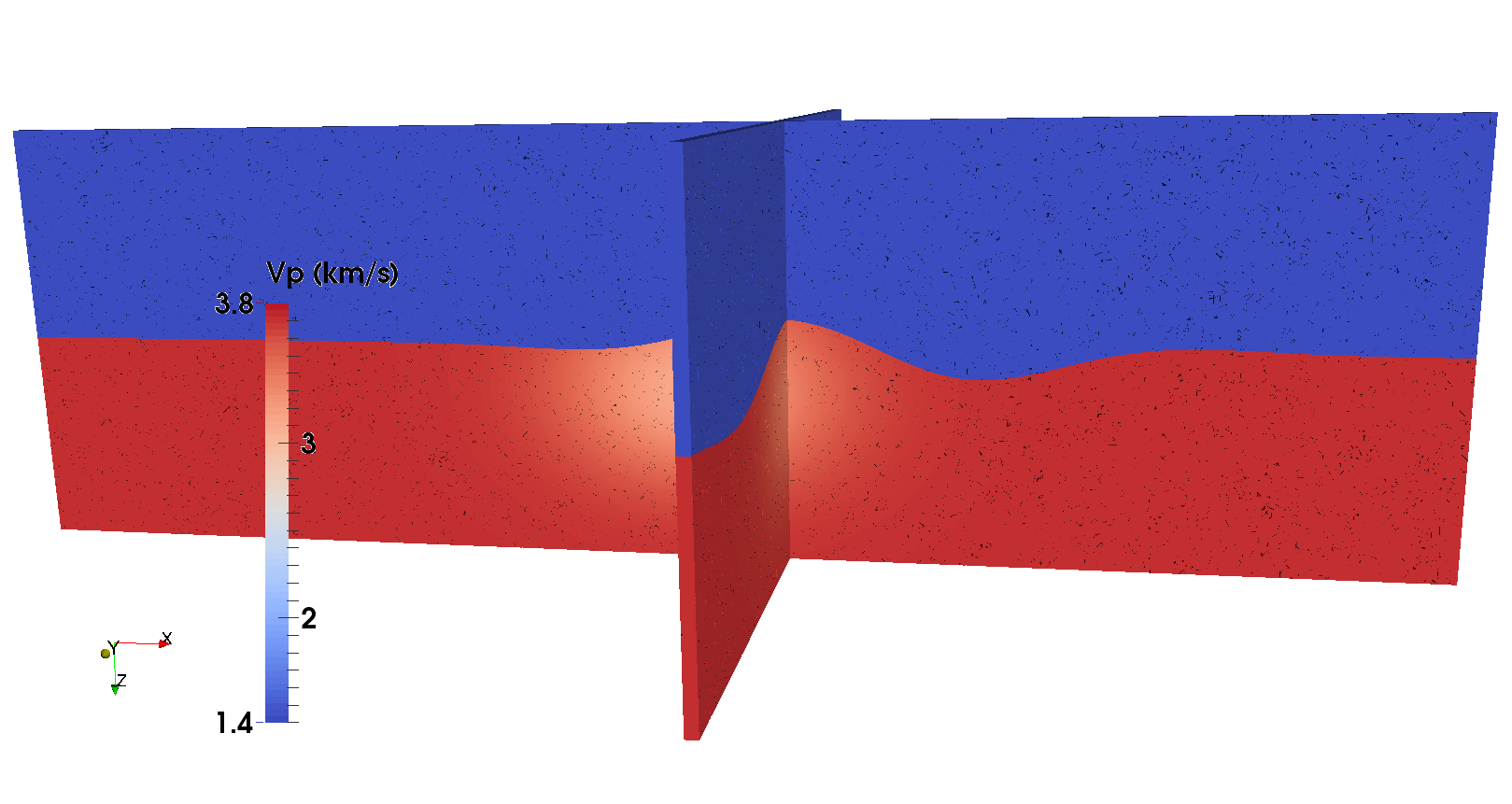} \\[-3cm]
\raggedleft \includegraphics[width=2.0in]{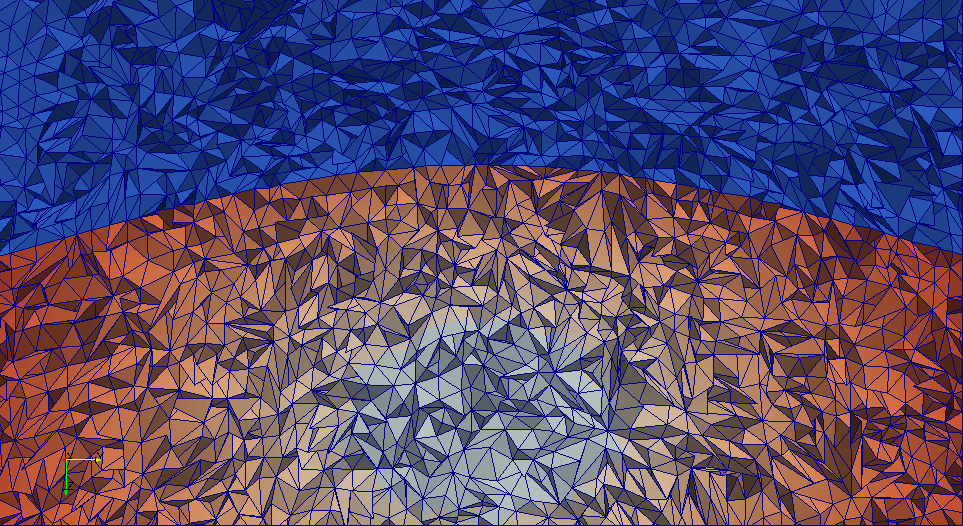}\\[-1cm]
\raggedright (a)\\[1cm]
\centering
\includegraphics[trim = 0in 1in 0.5in 0.2in, clip=true,width=4.5in]
          {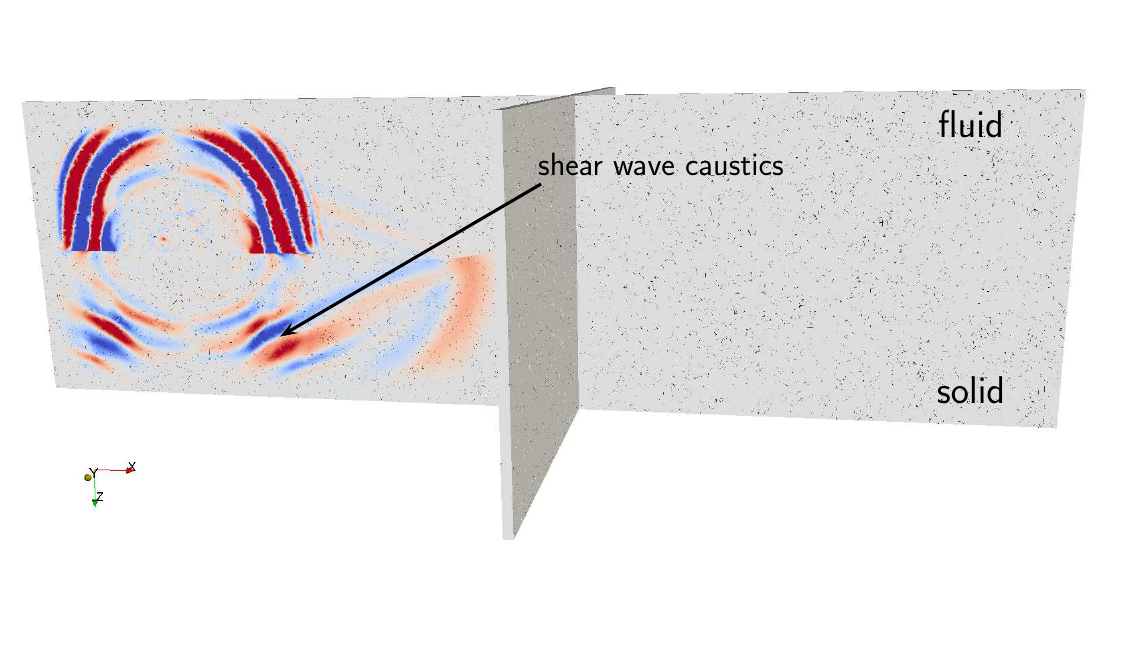} \\[-2cm]
\raggedright (b)\\[1.5cm]
\centering
\includegraphics[trim = 0in 1in 0.5in 0.2in, clip=true,width=4.5in]
          {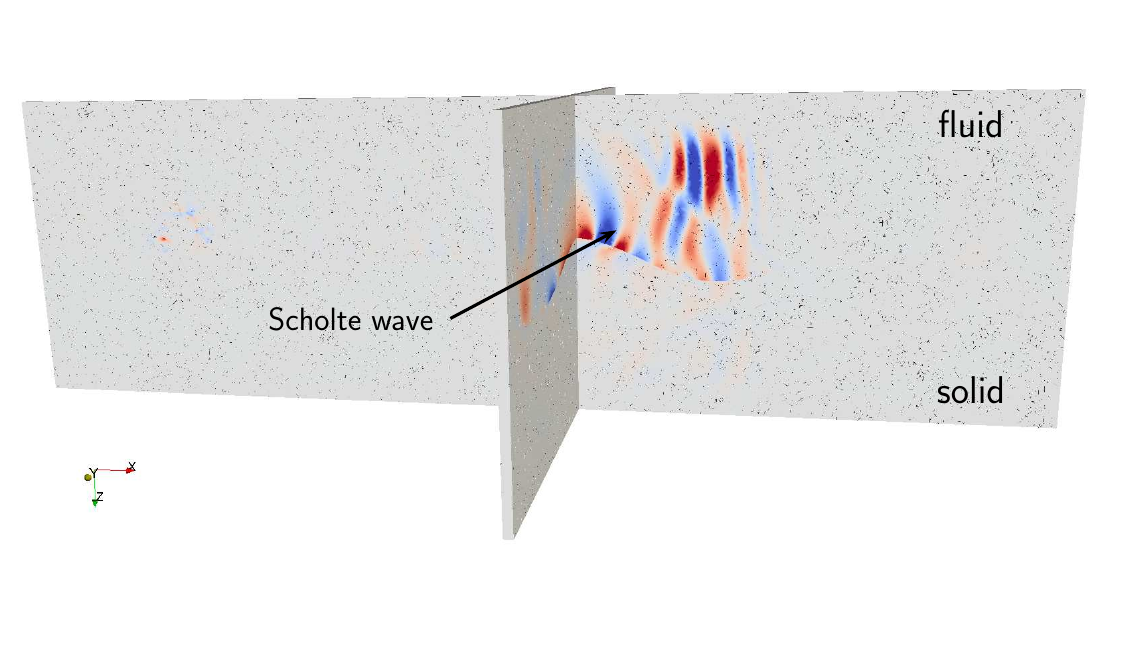} \\[-2cm]
\raggedright (c)\\[1.5cm]
\centering 
\caption{Heterogeneous HTI solid-fluid boundary with topography. 
(a) 3D model setting, with color indicating quasi-P wavespeed; (b) snapshot at t=4.0s; (c)  snapshot at t=18.0s. }
\label{fig:fslayer}
\end{figure}

Here, we use our DG method to simulate the wave propagation and
scattering in a heterogeneous anisotropic solid-fluid
configuration. {The solid-fluid boundary has topography, which is well 
described by adaptively fitting an unstructured mesh (see Figure 
\ref{fig:fslayer}(a)).} The model has
dimensions $[0,50] \times [0,30] \times [0,15]$km. The fluid side is
homogeneous isotropic with an acoustic wave speed $ 1.5 $km/s and
density $ 1.0 $g/cm\superscript{3}.  The solid side consists of a reference
HTI medium component with elastic parameters given by $
C_{11}=33.75,\,C_{22}=22.50,\, C_{33}=13.85,\,
C_{23}=13.85,\,C_{13}=11.44,\,C_{12}=11.44,\,
C_{44}=4.327,\,C_{55}=5.625,\,C_{66}=5.625
$ (GPa), and $ \rho=2.5 $g/cm\superscript{3}. A low-velocity lens is
superimposed with its center located at $ (25,\,15,\,9) $km. We place an 
explosive source in the fluid at location $(8.0,\, 15.0,\,6.5)$km, 
using a Ricker wavelet as source-time series with a central frequency of 
$ 1.0 $Hz. We apply convolutional perfect matching layers (CPMLs) for all 
external boundaries of the model with the thickness of approximately two 
central wavelengths.

The waves are propagated for 40 seconds. Two snapshots in time
of the wave field are shown; the solution at $ t=4 $s (Figure 
\ref{fig:fslayer} (b)) and the solution at $ t=14 $s (Figure 
\ref{fig:fslayer} (c)), with the occurence of a Scholte 
wave and seperation from body waves while propagating. 
We note the fomation of caustics in the solid region, 
caused by the anisotropy and the low-velocity lens.

\section{Discussion}

We develop a DG-method based numerical approach to simulate
acousto-elastic wave phenomena. We demonstrate
its ability to generate accurate solutions in domains with
heterogeneous and complex geometries for long-time simulation. We
briefly discuss the specifics of and differences between our and
earlier developed DG methods for general acousto-elastic wave
problems.

Most of the existing DG
discretizations for solving the acousto-elastic system of equations in
the first-order formulation make use of an upwind numerical flux
derived from the elementwise solution of a Riemann problem. 
In \cite{Dumbser2006}, a Godunov upwind flux is applied upon
diagonalizing the coefficient matrix in the stress-velocity
formulation at element-element interfaces. Specifically, they use a
``one-sided'' upwind numerical flux and, to avoid elementwise
numerical integration and make use of pre-calculated matrices instead,
restrict the coefficients to be 
constant in each element. 
Steger-Warming flux-vector splitting in \cite{Smith2010} 
is another way to obtain an exact Riemann solution for
the linear system
with flexibly parameterized isotropic elastic media, 
allowing variable coefficient within elements.
The velocity-strain formulation introduced by \cite{Wilcox2010} uses
the Rankine-Hugoniot jump condition to obtain an upwind flux for isotropic
solid-fluid interfaces while designing a uniform conservative
formulation for coupled elasto-acoustic systems.

Meanwhile, there are penalty based DG schemes designed to solve numerically 
the second-order system of equations for the displacement. 
The interior penalty Galerkin method is used by 
\cite{Rivi`ere2000} to solve a nonlinear parabolic system, and a symmetric
interior penalty term was employed by 
\cite{Grote2006} 
{to make the stiffness matrix symmetric positive definite}. 
\cite{DeBasabe2008}
studies the dispersion and convergence of these interior penalty
DG-method based schemes for the second-order elliptic
Lam\'e system.
\cite{Warburton2013} defines for a general hyperbolic system a flux that
penalizes the fields based on their continuity.

In our DG-method based scheme, we introduce a penalized numerical flux
the form of which is motivated by the interior boundary continuity
conditions. The fluid-solid boundary conditions are accounted for in
the coupling of elements through the fluxes. 
Our penalty weight
does not depend on the normal direction of the interior faces of the
elements, and moreover, unlike the interior penalty scheme in the
second-order displacement formulation, does not depend on the mesh
size either.

\section{Acknowledgment}
This research was supported in part by the members, BGP, ExxonMobil, PGS, 
Statoil and Total, of the Geo-Mathematical Imaging Group.

\appendix
\section{Convergence analysis}\label{sec:stability}

In this section we consider the $L^2$ error of numerical solutions $\veq_h$
and $\tveq_h$, which satisfy (\ref{eq:penalty_DG_S})--(\ref{eq:penalty_DG_F}) 
for any $\vep_h$ and $\tvep_h \in V_h^{N_p}$. 
We denote by $\pi_h^{N_p}:L^2\mapsto V_h^{N_p}$ the $L^2$ projection onto
the polynomial space of order $N_p$.
We assume that $\vef-\vef_h=0$ and $\tf-\tf_h=0$, and no
error occurs for $L^2$ projection of coefficient matrices, that is,
$\tsA-\tsA_h=0, \tsQ-\tsQ_h=0$ and $ \tsLa-\tsLa_h=0$.
We define $\verr:=\veq-\veq_h$ and $\tverr:=\tveq-\tveq_h$, 
where $\veq$ and $\tveq$ are the exact solutions.
We also denote $\vxi:=\veq_h-\pi_h^{N_p}\veq\, , \,\,
 \tvxi:=\tveq_h-\pi_h^{N_p}\tveq \,$, 
and $\veta:=(1-\pi_h^{N_p})\veq\, , \,\, \tveta:=(1-\pi_h^{N_p})\tveq \,$;
thus $\verr=\veta-\vxi, \, \tverr=\tveta-\tvxi$. 
We define the volume residuals
\begin{equation}\label{eq:vol_res}
\resv(\veq_h):=\tsLa^T\left(\tsQ\ddt{\veq_h}
	-\nabla\cdot(\tsA\veq_h)\right),
\quad
\tresv(\tveq_h):=\ttsLa^T\left(\ttsQ\ddt{\tveq_h}
	-\nabla\cdot(\ttsA\tveq_h)\right),
\end{equation}
and surface residuals
\begin{equation}\label{eq:surf_res}
\begin{split}
&
\ressS(\veq_h):=\tfrac12(\tsLa^\mm)^T\jmp{\tsA_n\veq_h}_\rSS
        +\alpha(\tsA_n^\mm)^T\jmp{\tsA_n\veq_h}_\rSS,
\\&
\tressS(\tveq_h):=\tfrac12(\ttsLa^\mm)^T\jmp{\ttsA_n\tveq_h}_\rFS
          +\alpha(\ttsA_n^\mm)^T\jmp{\ttsA_n\tveq_h}_\rFS,
\\&
\ressF(\veq_h):=\tfrac12(\tsLa^\mm)^T\jmp{\tsA_n\veq_h}_\rSF
	+\alpha(\tsA_n^\mm)^T\jmp{\tsA_n\veq_h}_\rSF,
\\&
\tressF(\tveq_h):=\tfrac12(\ttsLa^\mm)^T\jmp{\ttsA_n\tveq_h}_\rFF
          +\alpha(\ttsA_n^\mm)^T\jmp{\ttsA_n\tveq_h}_\rFF.
\end{split}
\end{equation}
Using (\ref{eq:penalty_DG_S})--(\ref{eq:penalty_DG_F}),
it follows that $(\verr,\tverr)$ satisfy
\begin{equation}\label{eq:numerical solution}
\begin{split}
\SumS \int_{\DeS}\resv(\verr)\cdot\vep_h\dd \Omega 
- \Sumss\int_{\Tss}\ressS(\verr)\cdot\vep_h^\mm \dd \Sigma 
- \Sumsf\int_{\Tsf}\ressF(\verr)\cdot\vep_h^\mm \dd \Sigma =&0,
\\
\SumF \int_{\DeF}\tresv(\tverr)\cdot\tvep_h\dd \Omega 
- \Sumfs\int_{\Tfs}\tressS(\tverr)\cdot\tvep_h^\mm \dd \Sigma 
- \Sumff\int_{\Tff}\tressF(\tverr)\cdot\tvep_h^\mm \dd \Sigma =&0,
\end{split}
\end{equation}
upon setting $\tsQ_h=\tsQ$ and $\tsA_h=\tsA\,$.
We take inner products of (\ref{eq:vol_res}) and (\ref{eq:surf_res}) with 
corresponding test functions, and immediately get, 
after summing up all the terms,
\begin{align}
&\begin{aligned}
  &\SumS\int_{\DeS} \tsQ\ddt{\veq_h} \cdot \tsLa\vep_h\dd \Omega
 - \SumS\int_{\DeS} (\nabla\cdot(\tsA\veq_h))\cdot\tsLa\vep_h\dd \Omega 
 \\&\hspace{1cm}
 - \Half\Sumss\int_{\Tss} \JmpRd\jmp{\tsA_n\veq_h}_{\rSS}
        \cdot (\tsLa\vep_h)^\mm \dd \Sigma 
 - \Half\Sumsf\int_{\Tsf} \JmpRd\jmp{\tsA_n\veq_h}_{\rSF}
        \cdot (\tsLa\vep_h)^\mm \dd \Sigma 
 \\&\hspace{1cm}
 - \alpha\Sumss\int_{\Tss} \JmpRd\jmp{\tsA_n\veq_h}_{\rSS}
        \cdot (\tsA_n\vep_h)^\mm \dd \Sigma 
 - \alpha\Sumsf\int_{\Tsf} \JmpRd\jmp{\tsA_n\veq_h}_{\rSF}
        \cdot (\tsA_n\vep_h)^\mm \dd \Sigma \\
 =&\SumS\int_{\DeS}\resv(\veq_h)\cdot\vep_h\dd \Omega 
 - \Sumss\int_{\Tss}\ressS(\veq_h)\cdot\vep_h^\mm \dd \Sigma 
 - \Sumsf\int_{\Tsf}\ressF(\veq_h)\cdot\vep_h^\mm \dd \Sigma, \\
\end{aligned}
	\label{eq:numerical residual elastic}
	\\[5mm]&
\begin{aligned}
  &\SumF\int_{\DeF} \ttsQ\ddt{\tveq_h} \cdot \ttsLa\tvep_h\dd \Omega
 - \SumF\int_{\DeF} (\nabla\cdot(\ttsA\tveq_h))\cdot\ttsLa\tvep_h\dd \Omega 
 \\&\hspace{1cm}
 - \Half\Sumff\int_{\Tff}\JmpRd\jmp{\ttsA_n\tveq_h}_{\rFF}
        \cdot (\ttsLa\tvep_h)^\mm \dd \Sigma 
 - \Half\Sumfs\int_{\Tfs}\JmpRd\jmp{\ttsA_n\tveq_h}_{\rFS}
        \cdot (\ttsLa\tvep_h)^\mm \dd \Sigma 
 \\&\hspace{1cm}
 - \alpha\Sumff\int_{\Tff}\JmpRd\jmp{\ttsA_n\tveq_h}_{\rFF}
        \cdot (\ttsA_n\tvep_h)^\mm \dd \Sigma 
 - \alpha\Sumfs\int_{\Tfs}\JmpRd\jmp{\ttsA_n\tveq_h}_{\rFS}
        \cdot (\ttsA_n\tvep_h)^\mm \dd \Sigma \\
 =&\SumF\int_{\DeF}\tresv(\tveq_h)\cdot\tvep_h\dd \Omega 
 - \Sumff\int_{\Tff}\tressF(\tveq_h)\cdot\tvep_h^\mm \dd \Sigma 
 - \Sumfs\int_{\Tfs}\tressS(\tveq_h)\cdot\tvep_h^\mm \dd \Sigma.\\
	\label{eq:numerical residual acoustic}
\end{aligned}
\end{align}
We let $\veq_h=\vep_h=\vxi$, $\tveq_h=\tvep_h=\tvxi$, when equations
(\ref{eq:numerical residual elastic}) and 
(\ref{eq:numerical residual acoustic}) become
\begin{align}
&
\begin{aligned}
&\hspace{-3mm}\frac12\DDt{}\SumS\Norm{\vxi}_{L^2(\DeS)}^2 
 - \alpha\Sumss\int_{\Tss}\JmpRd\jmp{\tsA_n\vxi}_{\rSS}
        \cdot (\tsA_n\vxi)^\mm \dd \Sigma 
 - \alpha\Sumsf\int_{\Tsf}\JmpRd\jmp{\tsA_n\vxi}_{\rSF}
 \cdot (\tsA_n\vxi)^\mm \dd \Sigma \\[-2mm]
 &\hspace{5mm} - \left(
   \SumS\int_{\DeS} (\nabla\cdot(\tsA\vxi)) \cdot \tsLa\vxi\dd \Omega  
 + \Half\Sumss\int_{\Tss}\JmpRd\jmp{\tsA_n\vxi}_{\rSS}
        \cdot (\tsLa\vxi)^\mm \dd \Sigma 
	\right.\\[-4mm]
	& \hspace{8cm}\left.
 + \Half\Sumsf\int_{\Tsf}\JmpRd\jmp{\tsA_n\vxi}_{\rSF}
        \cdot (\tsLa\vxi)^\mm \dd \Sigma \right)\\
& =\SumS\int_{\DeS}\resv(\vxi)\cdot\vxi \dd \Omega 
 - \Sumss\int_{\Tss}\ressS(\vxi)\cdot\vxi^\mm \dd \Sigma 
 - \Sumsf\int_{\Tsf}\ressF(\vxi)\cdot\vxi^\mm \dd \Sigma ,
\end{aligned}
\label{eq:error elastic}
\\[5mm]&
\begin{aligned}
&\hspace{-3mm}\frac12\DDt{}\SumF\Norm{\tvxi}_{L^2(\DeF)}^2 
 - \alpha\Sumff\int_{\Tff}\JmpRd\jmp{\ttsA_n\tvxi}_{\rFF}
        \cdot (\ttsA_n\tvxi)^\mm \dd \Sigma 
 - \alpha\Sumfs\int_{\Tfs}\JmpRd\jmp{\ttsA_n\tvxi}_{\rFS}
 \cdot (\ttsA_n\tvxi)^\mm \dd \Sigma \\[-2mm]
 &\hspace{5mm} - \left(
 \SumF\int_\DeF ( \nabla\cdot(\ttsA\tvxi) ) \cdot \ttsLa\tvxi\dd \Omega  
 + \Half\Sumff\int_{\Tff}\JmpRd\jmp{\ttsA_n\tvxi}_{\rFF}
        \cdot (\ttsLa\tvxi)^\mm \dd \Sigma 
	\right. \\[-4mm]
	& \hspace{8cm}\left.
 + \Half\Sumfs\int_{\Tfs}\JmpRd\jmp{\ttsA_n\tvxi}_{\rFS}
        \cdot (\ttsLa\tvxi)^\mm \dd \Sigma \right)\\
& =\SumF\int_\DeF\tresv(\tvxi)\cdot\tvxi \dd \Omega 
 - \Sumff\int_{\Tff}\tressF(\tvxi)\cdot\tvxi^\mm \dd \Sigma 
 - \Sumfs\int_{\Tfs}\tressS(\tvxi)\cdot\tvxi^\mm \dd \Sigma .
\end{aligned}
\label{eq:error acoustic}
\end{align}
Adding (\ref{eq:error elastic}) and (\ref{eq:error acoustic}), and using
the energy result in Section \ref{sec:BCpenalty}, 
the terms in between parentheses on 
the left-hand sides of both equations cancel one another, and the penalty
terms turn into quadratic forms, that is,
\begin{equation}\label{eq:energy estimate}
\begin{split}
& \hspace{-3mm}
  \frac12\DDt{}\left(\SumS\normVS{\vxi}^2 
  +\SumF\normVF{\tvxi}^2 \right) \\[-2mm]
  & \hspace{2mm} +\frac\alpha2\left(
   \Sumss\normSS{\jmp{\tsA_n\vxi}_{\rSS}}^2
  +\Sumff\normFF{\jmp{\ttsA_n\tvxi}_{\rFF}}^2
  +2\Sumsf\normSF{\jmp{\tsA_n\vxi}_{\rSF}}^2 \right)\\
  &\hspace{-3mm} 
  =\SumS\int_{\DeS}\resv(\vxi)\cdot\vxi \dd \Omega 
 - \Sumss\int_{\Tss}\ressS(\vxi)\cdot\vxi^\mm \dd \Sigma 
 - \Sumsf\int_{\Tsf}\ressF(\vxi)\cdot\vxi^\mm \dd \Sigma \\[-2mm]
 &  \hspace{2mm} + 
 \SumF\int_{\DeF}\tresv(\tvxi)\cdot\tvxi \dd \Omega 
 - \Sumff\int_{\Tff}\tressF(\tvxi)\cdot\tvxi^\mm \dd \Sigma 
 - \Sumfs\int_{\Tfs}\tressS(\tvxi)\cdot\tvxi^\mm \dd \Sigma .
\end{split}
\end{equation}
Let $\vep_h=\vxi$ in (\ref{eq:numerical solution}), and subtract it from the
right-hand side of (\ref{eq:energy estimate}). We note that 
$\verr=\veta-\vxi, \, \tverr=\tveta-\tvxi$, and obtain
\begin{equation}\label{eq:energy estimate1}
\begin{split}
& \hspace{-3mm}
\frac12\DDt{}\left(\SumS\normVS{\vxi}^2 
+\SumF\normVF{\tvxi}^2 \right) \\[-1mm]
  & \hspace{2mm} +\frac\alpha2\left(
   \Sumss\normSS{\jmp{\tsA_n\vxi}_{\rSS}}^2
  +\Sumff\normFF{\jmp{\ttsA_n\tvxi}_{\rFF}}^2
  +2\Sumsf\normSF{\jmp{\tsA_n\vxi}_{\rSF}}^2 \right) \\
  &\hspace{-5mm} 
  =\SumS\int_{\DeS}\resv(\veta)\cdot\vxi \dd \Omega 
 - \Sumss\int_{\Tss}\ressS(\veta)\cdot\vxi^\mm \dd \Sigma 
 - \Sumsf\int_{\Tsf}\ressF(\veta)\cdot\vxi^\mm \dd \Sigma \\[-2mm]
 & \hspace{2mm} + \SumF\int_{\DeF}\tresv(\tveta)\cdot\tvxi \dd \Omega 
 - \Sumff\int_{\Tff}\tressF(\tveta)\cdot\tvxi^\mm \dd \Sigma 
 - \Sumfs\int_{\Tfs}\tressS(\tveta)\cdot\tvxi^\mm \dd \Sigma .
\end{split}
\end{equation}
Having the energy result (\ref{eq:energy estimate1}),
which corresponds with Equation (5.10) in
\cite{Warburton2013}, we follow the same process
as described in the reference. 

We apply integration by parts:
\begin{equation}
	\begin{split}
		& 
		\int_\DeS (\nabla\cdot(\tsA\veq))\cdot(\tsLa\vep)\dd\Omega
		= 
		\int_\DeS 
		(\nabla\cdot\tsS)\cdot\vew 
		+\tfrac12(\nabla\vev+\nabla\vev^T):(\tsC\tsH)
		\dd\Omega
		\\ &\hspace{1cm} =
		-\int_\DeS 
		\tsS:\tfrac12(\nabla\vew+\nabla\vew^T)
		+\vev\cdot(\nabla\cdot(\tsC\tsH))
		\dd\Omega
		\\[-3mm] &\hspace{2.4cm}
		+\int_{\Tss\cup\Tsf} 
		(\ven\cdot\tsS)^\mm\cdot\vew^\mm
		+\vev^\mm\cdot(\ven\cdot(\tsC\tsH))^\mm
		\dd\Sigma
		\\ &\hspace{1cm} =
		\boxed{
		- \int_\DeS (\nabla\cdot(\tsA\vep))\cdot(\tsLa\veq)\dd\Omega
		}
		+ \int_{\Tss\cup\Tsf}(\tsA_n\veq)^\mm\cdot(\tsLa\vep)^\mm
		\dd\Sigma,
	\end{split}
	\label{eq:matform intbyparts 1}
\end{equation}
and similarly
\begin{equation}
	\hspace{-5mm}
	\int_\DeF (\nabla\cdot(\ttsA\tveq))\cdot(\ttsLa\tvep)\dd\Omega
	= \boxed{
	- \int_\DeF (\nabla\cdot(\ttsA\tvep))\cdot(\ttsLa\veq)\dd\Omega
	}
	+ \int_{\Tfs\cup\Tff}(\ttsA_n\tveq)^\mm\cdot(\ttsLa\tvep)^\mm
		\dd\Sigma.
	\label{eq:matform intbyparts 2}
\end{equation}
We set $\veq=\veta,\,\vep=\vxi$ in (\ref{eq:matform intbyparts 1})
and $\tveq=\tveta,\,\tvep=\tvxi$ in (\ref{eq:matform intbyparts 2}).
The boxed terms in (\ref{eq:matform intbyparts 1}) and 
(\ref{eq:matform intbyparts 2}) vanish as the projection errors $\veta$ 
and $\tveta$ are orthogonal to the spatial derivatives of the polynomial 
solutions $\veq_h$ and $\tveq_h$
by Galerkin approximation,
and then the right-hand side of (\ref{eq:energy estimate1}) becomes
\begin{align}
	&\begin{aligned}
 & \hspace{-5mm}
   \SumS\int_{\DeS}\resv(\veta)\cdot\vxi \dd \Omega 
 - \Sumss\int_{\Tss}\ressS(\veta)\cdot\vxi^\mm \dd \Sigma 
 - \Sumsf\int_{\Tsf}\ressF(\veta)\cdot\vxi^\mm \dd \Sigma 
 \\[-2mm] & \hspace{2mm} 
 + \SumF\int_{\DeF}\tresv(\tveta)\cdot\tvxi \dd \Omega 
 - \Sumff\int_{\Tff}\tressF(\tveta)\cdot\tvxi^\mm \dd \Sigma 
 - \Sumfs\int_{\Tfs}\tressS(\tveta)\cdot\tvxi^\mm \dd \Sigma 
	\end{aligned}
	\nonumber\\
	&\hspace{5mm}\boxed{
   \hspace{-8mm}
 = \hspace{4mm}
 \SumS\int_{\DeS}\tsQ\ddt{\veta}\cdot(\tsLa\vxi) \dd\Omega
 + \SumF\int_{\DeF}\ttsQ\ddt{\tveta}\cdot(\ttsLa\tvxi) \dd\Omega
	}\tag{$\Xi_1$}\label{Xi 1}\\
	&\hspace{5mm}\boxed{
	\begin{aligned}
 & \hspace{0mm}
 - \Sumss\int_{\Tss}\avg{\tsA_n\veta}_\rSS\cdot(\tsLa\vxi)^\mm \dd \Sigma
 - \Sumsf\int_{\Tsf}\avg{\tsA_n\veta}_\rSF\cdot(\tsLa\vxi)^\mm \dd \Sigma
 \\ & \hspace{0mm}
 - \Sumff\int_{\Tff}\avg{\ttsA_n\tveta}_\rFF\cdot(\ttsLa\tvxi)^\mm \dd \Sigma
 - \Sumfs\int_{\Tfs}\avg{\ttsA_n\tveta}_\rFS\cdot(\ttsLa\tvxi)^\mm \dd \Sigma
 \\ & \hspace{0mm}
 - \alpha\Sumss\int_{\Tss}\jmp{\tsA_n\veta}_\rSS\cdot(\tsA_n\vxi)^\mm 
 \dd \Sigma
 - \alpha\Sumsf\int_{\Tsf}\jmp{\tsA_n\veta}_\rSF\cdot(\tsA_n\vxi)^\mm 
 \dd \Sigma
 \\ & \hspace{0mm}
 - \alpha\Sumff\int_{\Tff}\jmp{\ttsA_n\tveta}_\rFF\cdot(\ttsA_n\tvxi)^\mm 
 \dd \Sigma
 - \alpha\Sumfs\int_{\Tfs}\jmp{\ttsA_n\tveta}_\rFS\cdot(\ttsA_n\tvxi)^\mm 
 \dd \Sigma,
	\end{aligned}
	\tag{$\Xi_2$}\label{Xi 2} }
\end{align}
in which we use the following simplified notation for averaging:
\[
	\begin{split}
  &  \avg{ \tsA_n \veq}_\rSS=\tfrac12
  (               ( \tsA_n \veq)^\pp+( \tsA_n \veq)^\mm),\quad
     \avg{ \tsA_n \veq}_\rSF=\tfrac12
  (O^T            (\ttsA_n\tveq)^\pp+( \tsA_n \veq)^\mm),\\
  &  \avg{\ttsA_n\tveq}_\rFF=\tfrac12
  (               (\ttsA_n\tveq)^\pp+(\ttsA_n\tveq)^\mm),\quad
     \avg{\ttsA_n\tveq}_\rFS=\tfrac12
  (O^{\phantom{*}}( \tsA_n \veq)^\pp+(\ttsA_n\tveq)^\mm).
	\end{split}
\]
For the volume integration terms (cf. (\ref{Xi 1})) we obtain the estimate
\begin{equation}
	\begin{split}
 &
	 \SumS \normVS{\vxi} \NormVS{\ddt{\veta}} 
	   + \SumF \normVF{\tvxi}\NormVF{\ddt{\tveta}}
\\
& \hspace{1cm}\leq 
\sqrt{ \SumS \normVS{\vxi}^2+ \SumF \normVF{\tvxi}^2}
\\&\hspace{3cm}
\sqrt{
	 \SumS \NormVS{\ddt{\veta}}^2 
	+\SumF \NormVF{\ddt{\tveta}}^2 
}\quad .
	\end{split}
	\label{eq:err estimate vol}
\end{equation}
For the surface integration terms (cf. (\ref{Xi 2})), 
we use the symmetry in $\tsA$ and $\tsLa$ to find that
\begin{equation}
	(\tsA_n\veq)^\pm \cdot (\tsLa\vep)^\mm = 
	\ven \cdot \tsS^\pm \cdot \vew^\mm +
	\ven \cdot (\tsC\tsH\,)^\mm \cdot \vev^\pm =
	(\tsA_n\vep)^\mm \cdot (\tsLa\veq)^\pm .
	\label{eq:rule 1}
\end{equation}
Thus 
\begin{equation}
	\begin{split}
	&\hspace{-4mm} \Sumss \int_\Tss 
		\avg{\tsA_n\veta}_\rSS \cdot(\tsLa\vxi)^\mm \dd \Sigma
	\\& \hspace{1cm} = 
	\Half\Sumss \int_\Tss 
		(\tsA_n\veta)^\pp
		\cdot(\tsLa\vxi)^\mm \dd \Sigma
		+
	\Half\Sumss \int_\Tss 
		(\tsA_n\veta)^\mm 
		\cdot(\tsLa\vxi)^\mm \dd \Sigma
	\\& \hspace{1cm} = 
	\Half\Sumss \int_\Tss 
		(\tsA_n\vxi)^\mm
		\cdot(\tsLa\veta)^\pp \dd \Sigma
		+
	\Half\Sumss \int_\Tss 
		(\tsA_n\vxi)^\mm 
		\cdot(\tsLa\veta)^\mm \dd \Sigma
	\\& \hspace{1cm} = 
	\Half\Sumss \int_\Tss 
	\hspace{-3mm}-
	(\tsA_n\vxi)^\pp\cdot(\tsLa\veta)^\mm \dd \Sigma
		+
	\Half\Sumss \int_\Tss 
		(\tsA_n\vxi)^\mm 
		\cdot(\tsLa\veta)^\mm \dd \Sigma
	\\& \hspace{1cm} = 
	\Half\Sumss \int_\Tss 
	\hspace{-3mm}-
		\jmp{\tsA_n\vxi}_\rSS \cdot(\tsLa\veta)^\mm \dd \Sigma
	\quad=\,
	-\Half\Sumss \int_\Tss \JmpRd
	\jmp{\tsA_n\vxi}_\rSS \cdot\avg{\tsLa\veta}_\rSS \dd \Sigma,
	\end{split}
	\label{eq:Tss}
\end{equation}
in which the second equality uses (\ref{eq:rule 1}), 
and the third equality is obtained by exchanging the summation order 
of elements between solid-solid interfaces. Similarly, we have
\begin{equation}
	(\ttsA_n\tveq)^\pm \cdot (\ttsLa\tvep)^\mm = 
	(\tlambda\tE\,)^\pm \ven\cdot \tvew^\mm +
	(\tlambda\tH\,)^\mm \ven\cdot \tvev^\pm =
	(\ttsA_n\tvep)^\mm \cdot (\ttsLa\tveq)^\pm ,
	\label{eq:rule 2}
\end{equation}
and
\begin{equation}
		\Sumff \int_\Tff
		\avg{\ttsA_n\tveta}_\rFF \cdot(\ttsLa\tvxi)^\mm \dd \Sigma
	= 
		-\Half\Sumff \int_\Tff \JmpRd
		\jmp{\ttsA_n\tvxi}_\rFF \cdot\avg{\ttsLa\tveta}_\rFF 
		\dd \Sigma.
	\label{eq:Tff}
\end{equation}
{\ColorRed
For fluid-solid interfaces we also have the symmetry
\begin{equation}
	\begin{split}
	O^T(\ttsA_n\tveq)^\pp \cdot (\tsLa\vep)^\mm =\, &
	(\tlambda\tE)^\pp \ven \cdot \vew^\mm +
	(\ven \cdot (\tsC\tsH\,)^\mm \cdot\ven)(\ven\cdot \tvev^\pp) =
	(\tsA_n\vep)^\mm \cdot O^T(\ttsLa\tveq)^\pp, \\[5mm]
	O^{\phantom{*}}\,(\tsA_n\veq)^\pp \cdot (\ttsLa\tvep)^\mm =\, &
	(\ven \cdot \tsS^\pp \cdot\ven) (\tvew^\mm\cdot\ven) +
	(\tlambda\tH)^\mm \ven \cdot \vev^\pp =
	O^T(\ttsA_n\tvep)^\mm \cdot (\tsLa\veq)^\pp ,
	\end{split}
	\label{eq:rule 3}
\end{equation}
and using (\ref{eq:rule 1}), (\ref{eq:rule 2}) and (\ref{eq:rule 3}),
}
\begin{equation}
	\begin{split}
	& \hspace{-3mm}
		\Sumsf \int_\Tsf 
		\avg{\tsA_n\veta}_\rSF \cdot(\tsLa\vxi)^\mm \dd \Sigma \,
	+ 
		\Sumfs \int_\Tfs 
		\avg{\ttsA_n\tveta}_\rFS \cdot(\ttsLa\tvxi)^\mm \dd \Sigma
	\\ & \hspace{1cm} = 
		\Sumsf \int_\Tsf \tfrac12 (
		O^T(\ttsA_n\tveta)^\pp \cdot(\tsLa\vxi)^\mm 
		+(\tsA_n\veta)^\mm \cdot (\tsLa\vxi)^\mm
		) \dd \Sigma \,
	\\[-3mm] & \hspace{3cm} + 
		\Sumfs \int_\Tfs \tfrac12 (
		O^{\phantom{*}}(\tsA_n\veta)^\pp \cdot(\ttsLa\tvxi)^\mm 
		+(\ttsA_n\tveta)^\mm \cdot (\ttsLa\tvxi)^\mm
		) \dd \Sigma \,
	\\ & \hspace{1cm} = 
		\Sumsf \int_\Tsf \tfrac12 (
		(\tsA_n\vxi)^\mm\cdot O^T(\ttsLa\tveta)^\pp
		+(\tsA_n\vxi)^\mm \cdot (\tsLa\veta)^\mm
		) \dd \Sigma \,
	\\[-3mm] & \hspace{3cm} + 
		\Sumfs \int_\Tfs \tfrac12 (
		O^T(\ttsA_n\tvxi)^\mm \cdot(\tsLa\veta)^\pp 
		+(\ttsA_n\tvxi)^\mm \cdot (\ttsLa\tveta)^\mm
		) \dd \Sigma \,
	\\ & \hspace{1cm} = 
		\Sumsf \int_\Tsf \tfrac12 (
		(\tsA_n\vxi)^\mm\cdot O^T(\ttsLa\tveta)^\pp
		+(\tsA_n\vxi)^\mm \cdot (\tsLa\veta)^\mm
	\\[-3mm] & \hspace{3cm} - 
		O^T(\ttsA_n\tvxi)^\pp \cdot(\tsLa\veta)^\mm 
		-O^T(\ttsA_n\tvxi)^\pp \cdot O^T(\ttsLa\tveta)^\pp
		) \dd \Sigma \,
	\\ & \hspace{1cm} = 
		-\Sumsf \int_\Tsf 
		\jmp{\tsA_n\vxi}_\rSF \cdot\avg{\tsLa\veta}_\rSF 
		\dd \Sigma \,.
	\end{split}
	\label{eq:Tsf}
\end{equation}
For the penalty terms in (\ref{Xi 2}), it is straightforward to check that
\begin{align}
	\Sumss\int_{\Tss}\jmp{\tsA_n\veta}_\rSS\cdot
	(\tsA_n\vxi)^\mm\dd\Sigma
	=&-\Half
	\Sumss\int_{\Tss}\jmp{\tsA_n\veta}_\rSS\cdot
	\jmp{\tsA_n\vxi}_\rSS\dd\Sigma,
	\label{eq:alfa Tss}
	\\ 
	\Sumff\int_{\Tff}\jmp{\ttsA_n\tveta}_\rFF\cdot
	(\ttsA_n\tvxi)^\mm\dd\Sigma
	=&-\Half
	\Sumff\int_{\Tff}\jmp{\ttsA_n\tveta}_\rFF\cdot
	\jmp{\ttsA_n\tvxi}_\rFF\dd\Sigma,
	\label{eq:alfa Tff}
\end{align}
and
\begin{equation}
	\begin{split}
	&
	\Sumsf\int_{\Tsf}\jmp{\tsA_n\veta}_\rSF\cdot
	(\tsA_n\vxi)^\mm\dd\Sigma
	+
	\Sumfs\int_{\Tfs}\jmp{\ttsA_n\tveta}_\rFS\cdot
	(\ttsA_n\tvxi)^\mm\dd\Sigma
	\\ & \hspace{1cm}
	=
	\Sumsf\int_{\Tsf} (
	O^T(\ttsA_n\tveta)^\pp\cdot(\tsA_n\vxi)^\mm 
	-
	(\tsA_n\veta)^\mm\cdot(\tsA_n\vxi)^\mm 
	) \dd\Sigma
	\\[-3mm] & \hspace{3.2cm}
	+
	\Sumfs\int_{\Tfs} (
	O^{\phantom{*}}(\tsA_n\veta)^\pp\cdot(\ttsA_n\tvxi)^\mm 
	-
	(\ttsA_n\tveta)^\mm\cdot(\ttsA_n\tvxi)^\mm 
	) \dd\Sigma
	\\ & \hspace{1cm}
	=
	\Sumsf\int_{\Tsf} (
	O^T(\ttsA_n\tveta)^\pp\cdot(\tsA_n\vxi)^\mm 
	-
	(\tsA_n\veta)^\mm\cdot(\tsA_n\vxi)^\mm 
	\\[-3mm] & \hspace{3.2cm}
	+
	(\tsA_n\veta)^\mm\cdot O^T(\ttsA_n\tvxi)^\pp 
	-
	O^T(\ttsA_n\tveta)^\pp\cdot O^T(\ttsA_n\tvxi)^\pp 
	) \dd\Sigma
	\\ & \hspace{1cm}
	=-
	\Sumsf\int_{\Tsf}\jmp{\tsA_n\veta}_\rSF\cdot
	\jmp{\tsA_n\vxi}_\rSF\dd\Sigma.
	\end{split}
	\label{eq:alfa Tsf}
\end{equation}
Using (\ref{eq:Tss}), (\ref{eq:Tff}), (\ref{eq:Tsf}), (\ref{eq:alfa Tss}),
(\ref{eq:alfa Tff}) and (\ref{eq:alfa Tsf}) in (\ref{Xi 2}) yields
the estimate for (\ref{Xi 2})
\begin{equation}
	\begin{split}
	&
	\Half\Sumss \int_\Tss 
	\jmp{\tsA_n\vxi}_\rSS \cdot 
	\Bigl( 
	\avg{\tsLa\veta}_\rSS 
	+ \alpha 
	\jmp{\tsA_n\veta}_\rSS \Bigr)
	\dd \Sigma
	\\[-0mm] &
	\phantom{=} \hspace{5mm} +
	\Half\Sumff \int_\Tff 
	\jmp{\ttsA_n\tvxi}_\rFF \cdot 
	\Bigl( 
	\avg{\ttsLa\tveta}_\rFF 
	+ \alpha 
	\jmp{\ttsA_n\tveta}_\rFF \Bigr)
	\dd \Sigma
	\\[-0mm] &
	\phantom{=} \hspace{5mm} + \phantom{\tfrac12}
	\Sumsf \int_\Tsf 
	\jmp{\tsA_n\vxi}_\rSF \cdot 
	\,\Bigl( 
	\avg{\tsLa\veta}_\rSF 
	+ \alpha 
	\jmp{\tsA_n\veta}_\rSF \,\Bigr)
	\dd \Sigma 
	\\ & 
	\leq \hspace{5mm}\phantom{+}
	\Half\Sumss 
	\left( 
	\normSS{\jmp{\tsA_n\vxi}_\rSS} \normSS{\avg{\tsLa\veta}_\rSS}
	\right.
	\\[-4mm] & \hspace{4cm}
	\left. 
	+ \alpha 
	\normSS{\jmp{\tsA_n\vxi}_\rSS} \normSS{\jmp{\tsA_n\veta}_\rSS}
	\right)
	\\[-0mm] &
	\phantom{=} \hspace{5mm} +
	\Half\Sumff 
	\left( 
	\normFF{\jmp{\ttsA_n\tvxi}_\rFF} \normFF{\avg{\ttsLa\tveta}_\rFF}
	\right.
	\\[-4mm] & \hspace{4cm}
	\left. 
	+ \alpha 
	\normFF{\jmp{\ttsA_n\tvxi}_\rFF} \normFF{\jmp{\ttsA_n\tveta}_\rFF}
	\right)
	\\[-0mm] &
	\phantom{=} \hspace{5mm} + \phantom{\tfrac12}
	\Sumsf 
	\left( 
	\normSF{\jmp{\tsA_n\vxi}_\rSF} \normSF{\avg{\tsLa\veta}_\rSF}
	\right.
	\\[-4mm] & \hspace{4cm}
	\left. 
	+ \alpha 
	\normSF{\jmp{\tsA_n\vxi}_\rSF} \normSF{\jmp{\tsA_n\veta}_\rSF}
	\right)
	\\ & 
	\leq \frac1{2\beta} \Bigl(
  	 \Sumss\normSS{\jmp{\tsA_n\vxi}_{\rSS}}^2
  	+\Sumff\normFF{\jmp{\ttsA_n\tvxi}_{\rFF}}^2
  	+2\Sumsf\normSF{\jmp{\tsA_n\vxi}_{\rSF}}^2 \Bigr)
	\\&\hspace{1cm} 
  	+\frac\beta4 \Bigl(
	\hspace{6mm}\,\phantom{+2}\Sumss 
	 \left(\normSS{\avg{\tsLa\veta}_{\rSS}}^2
	 \,+\alpha^2\normSS{\jmp{\tsA_n\veta}_{\rSS}}^2
	 \,\right)
	 \\[-1mm]&\hspace{2cm} 
	 \hspace{5mm}+\phantom{2}\Sumff
	 \left(\normFF{\avg{\ttsLa\tveta}_{\rFF}}^2
	 +\alpha^2\normFF{\jmp{\ttsA_n\tveta}_{\rFF}}^2
	 \right)
	 \\[-1mm]&\hspace{2cm} 
	 \hspace{5mm}+2\,\Sumsf
	 \left(\normSF{\avg{\tsLa\veta}_{\rSF}}^2
	 +\alpha^2\normSF{\jmp{\tsA_n\veta}_{\rSF}}^2
	 \,\right)
	 \quad	 \Bigr).
	\end{split}
	\label{eq:err estimate surf0}
\end{equation}
The first inequality is obtained by Cauchy--Schwarz,
and the second one is based on Young's inequality with factor $\beta$ 
(or so-called ``Peter--Paul inequality''). 
{\ColorRed
Since 
\[
	\begin{split}
	&
	\Sumss \normSS{\avg{\tsLa\veta}_{\rSS}}^2 
	= \Sumss \left( \tfrac12
	\normSS{(\tsLa\veta)^\pp+(\tsLa\veta)^\mm}
	\right)^2
	\\ &\hspace{1cm} 
	\leq \Sumss \tfrac14
	\left(\normSS{(\tsLa\veta)^\pp}^2+\normSS{(\tsLa\veta)^\mm}^2
	+2\normSS{(\tsLa\veta)^\pp}\normSS{(\tsLa\veta)^\mm}\right)
	\\ &\hspace{1cm} 
	\leq \Sumss \tfrac12\left( 
	\normSS{(\tsLa\veta)^\pp}^2 + \normSS{(\tsLa\veta)^\mm}^2
	\right)
	= \Sumss \normSS{(\tsLa\veta)^\mm}^2,
	\\ &
	\Sumss \normSS{\jmp{\tsA_n\veta}_{\rSS}}^2 
	= 4\Sumss \left( \tfrac12
	\normSS{(\tsA_n\veta)^\pp+(\tsA_n\veta)^\mm}
	\right)^2
	\\ &\hspace{1cm} 
	\leq 4\Sumss \tfrac14
	\left(\normSS{(\tsA_n\veta)^\pp}^2+\normSS{(\tsA_n\veta)^\mm}^2
	+2\normSS{(\tsA_n\veta)^\pp}\normSS{(\tsA_n\veta)^\mm}
	\right)
	\\ &\hspace{1cm} 
	\leq 4\Sumss \tfrac12\left( 
	\normSS{(\tsA_n\veta)^\pp}^2 + \normSS{(\tsA_n\veta)^\mm}^2
	\right)
	= 4\Sumss \normSS{(\tsA_n\veta)^\mm}^2,
	\end{split}
\]
due to Cauchy--Schwarz followed by Young's inequality, 
}
and
\[
	\begin{split}
	&
	\Sumff \normFF{\avg{\ttsLa\tveta}_{\rFF}}^2 
	\leq \hspace{1.5mm}\Sumff \normFF{(\ttsLa\tveta)^\mm}^2,
	\\ &
	\Sumff \normFF{\jmp{\ttsA_n\tveta}_{\rFF}}^2 
	\leq 4\Sumff \normFF{(\ttsA_n\tveta)^\mm}^2,
	\\ &
	\Sumsf \normSF{\avg{\tsLa\veta}_{\rSF}}^2 
	\leq \Half\Sumsf \normSF{(\tsLa\veta)^\mm}^2
	+ \Half\Sumfs \normFS{(\ttsLa\tveta)^\mm}^2,
	\\ &
	\Sumsf \normSF{\jmp{\tsA_n\veta}_{\rSF}}^2 
	\,\leq 2\Sumsf \normSF{(\tsA_n\veta)^\mm}^2
	+ 2\Sumfs \normFS{(\ttsA_n\tveta)^\mm}^2,
	\end{split}
\]
we get the estimate for (\ref{Xi 2}),
\begin{equation}
	\begin{split}
	\Xi_2 &
	\leq \frac1{2\beta} \left(
	\Sumss\normSS{\jmp{\tsA_n\vxi}_{\rSS}}^2
  	+\Sumff\normFF{\jmp{\ttsA_n\tvxi}_{\rFF}}^2
  	+2\Sumsf\normSF{\jmp{\tsA_n\vxi}_{\rSF}}^2 \right)
	\\&\hspace{1cm} 
  	+\frac\beta4 \left(
	\hspace{5mm}\phantom{+2}\SumS 
	    \left(\Norm{(\tsLa\veta)^\mm}_{L^2(\Tss\cup\Tsf)}^2
	 +4\alpha^2\Norm{(\tsA_n\veta)^\mm}_{L^2(\Tss\cup\Tsf)}^2
	 \right)
	 \right.
	 \\[-3mm]&\hspace{2cm} \left.
	 \hspace{5mm}+\phantom{2}\SumF
	    \left(\Norm{(\ttsLa\tveta)^\mm}_{L^2(\Tff\cup\Tfs)}^2
	 +4\alpha^2\Norm{(\ttsA_n\tveta)^\mm}_{L^2(\Tff\cup\Tfs)}^2
	 \right)
	 \quad	 \right).
	\end{split}
	\label{eq:err estimate surf}
\end{equation}
Using (\ref{eq:err estimate vol}) and (\ref{eq:err estimate surf})
in (\ref{eq:energy estimate1}) yields
\begin{equation}
\begin{split}
  & \frac12\DDt{}\left(
  \SumS\normVS{\vxi}^2 +\SumF\normVF{\tvxi}^2 
  \right) 
  +(\frac\alpha2-\frac1{2\beta})
   \left(
	  \Sumss\Norm{\jmp{\tsA_n\vxi}_\rSS}^2_{L^2(\Sss)}
   \right.
   \\[-0mm]& \hspace{3cm} \left.
	  +
	  \Sumff\Norm{\jmp{\ttsA_n\tvxi}_\rFF}^2_{L^2(\Sff)}
	 + 2\Sumsf\Norm{\jmp{\tsA_n\vxi}_\rSF}^2_{L^2(\Ssf)}
  \right) 
\\& 
\leq 
\sqrt{ \SumS \normVS{\vxi}^2+ \SumF \normVF{\tvxi}^2}
\sqrt{
	 \SumS \NormVS{\ddt{\veta}}^2 
	+\SumF \NormVF{\ddt{\tveta}}^2 
}
\\[-0mm]&\hspace{2cm} 
  	+\frac\beta4 \left(
	\hspace{0mm}\phantom{+2}\SumS 
	    \left(\Norm{(\tsLa\veta)^\mm}_{L^2(\Tss\cup\Tsf)}^2
	 +4\alpha^2\Norm{(\tsA_n\veta)^\mm}_{L^2(\Tss\cup\Tsf)}^2
	 \right)
	 \right.
	 \\[-0mm]&\hspace{2cm} \left.
	 \hspace{11mm}+\phantom{2}\SumF
	    \left(\Norm{(\ttsLa\tveta)^\mm}_{L^2(\Tff\cup\Tfs)}^2
	 +4\alpha^2\Norm{(\ttsA_n\tveta)^\mm}_{L^2(\Tff\cup\Tfs)}^2
	 \right)
	 \quad	 \right).
	\end{split}
	\label{eq:err est final}
\end{equation}
Following \cite[Section 5.1]{Warburton2013}, we can finally obtain the 
required error estimate from (\ref{eq:err est final}). 
We take $\alpha = 1/2$; by choosing $\beta$ sufficiently large in
Young's inequality, we control the error by applying a modified
Gronwall's lemma \cite[p.A2007]{Warburton2013}.

\section{Computational benchmark among types of flux}\label{App:flux_cmp}

In this appendix, we compare three types of numerical fluxes: 
the central flux, the upwind flux proposed by 
\cite{Wilcox2010}, and the boundary condition penalized flux
proposed in our DG method. The comparisons are conducted using plane
waves, Rayleigh waves, Stoneley waves and Scholte waves, 
with the parameter settings as in \ref{sec:convtest}.

\begin{figure}
\centering
\begin{tabular}{cc}
	\includegraphics
	[trim = 4.5mm 0mm 10mm 10mm, clip=true,width=0.5\textwidth]
	{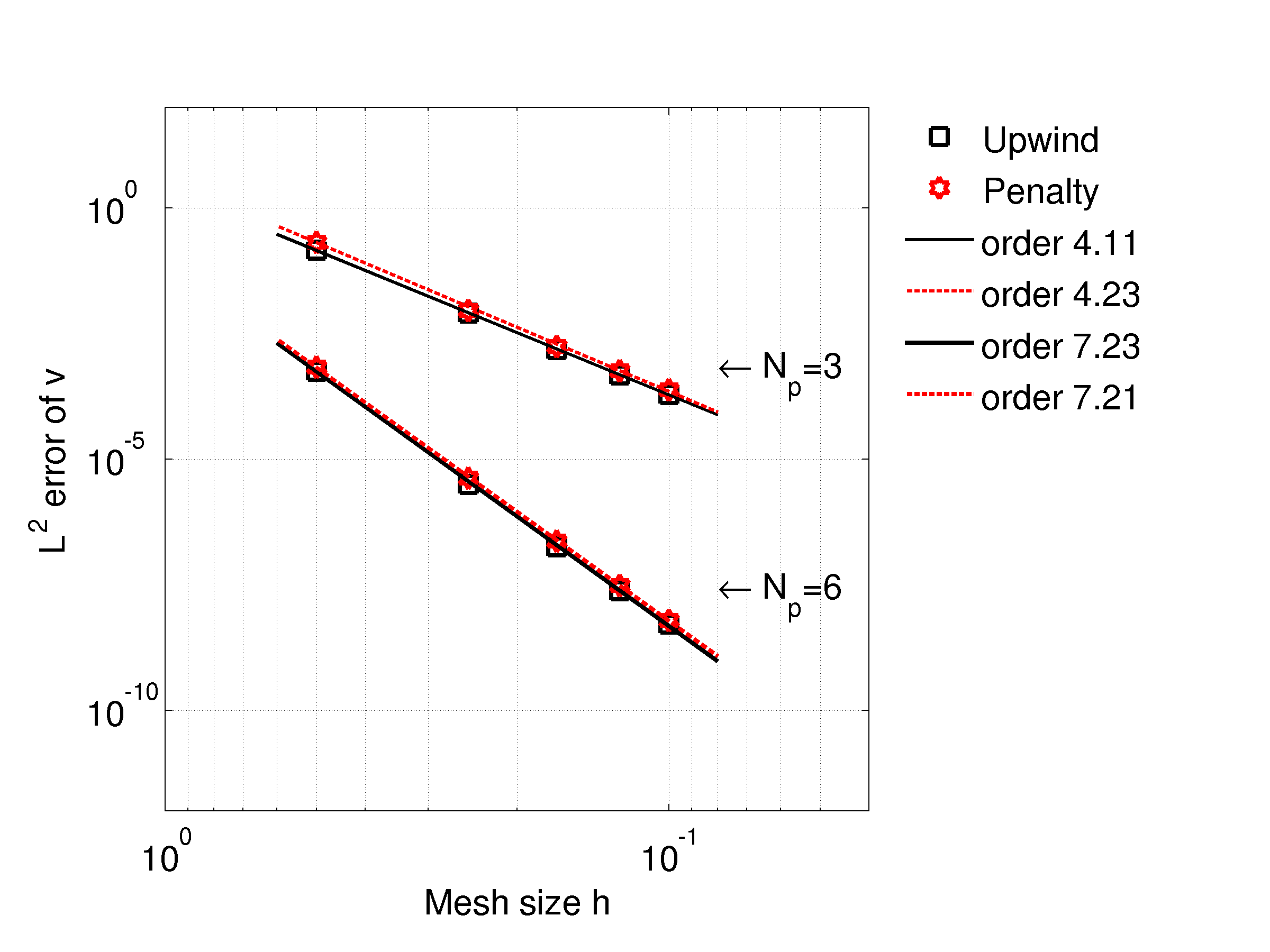}&
	\includegraphics
	[trim = 4.5mm 0mm 10mm 10mm, clip=true,width=0.5\textwidth]
	{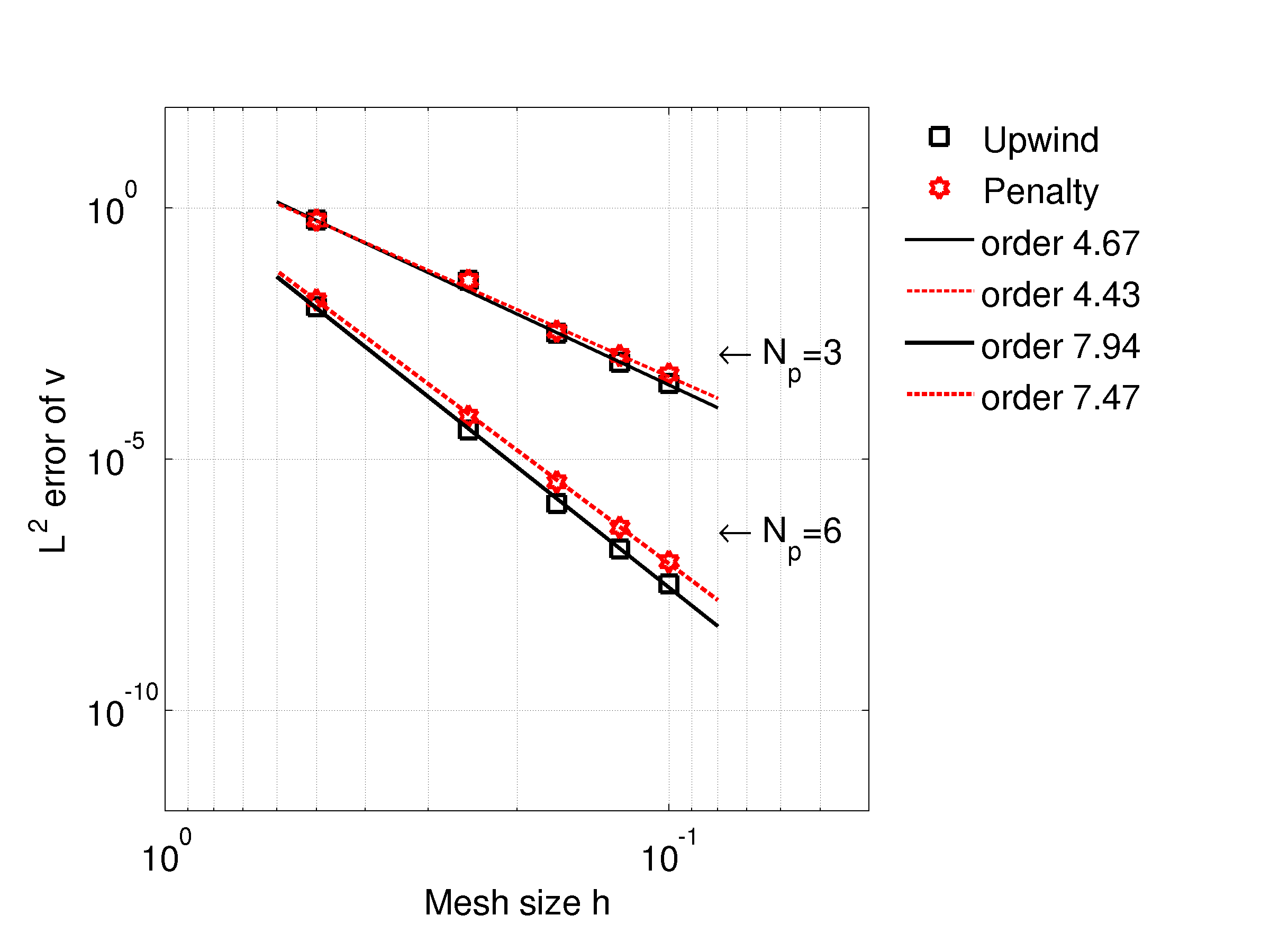}\\
	(A) & (B) \\
	\includegraphics
	[trim = 4.5mm 0mm 10mm 10mm, clip=true,width=0.5\textwidth]
	{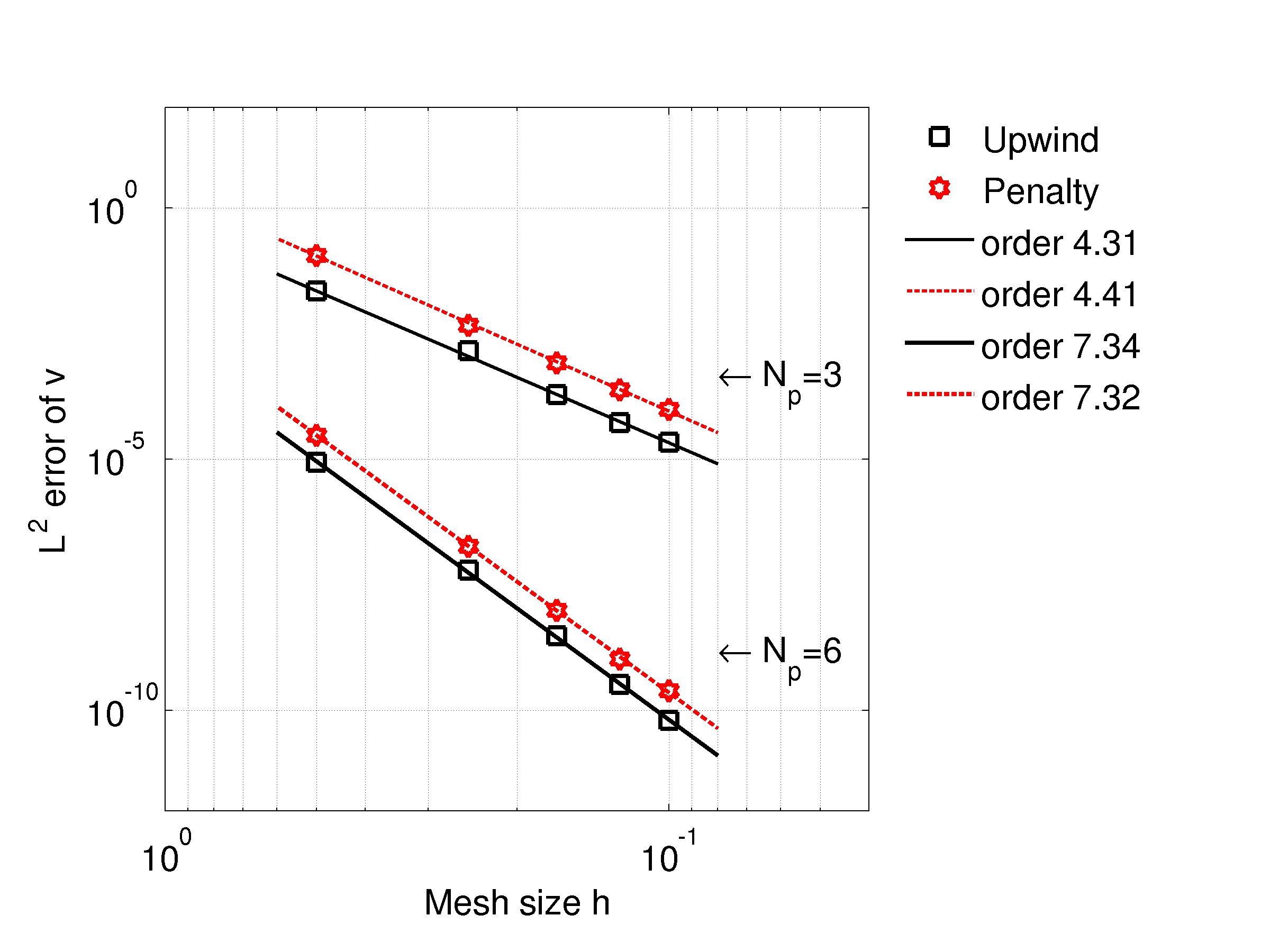}&
	\includegraphics
	[trim = 4.5mm 0mm 10mm 10mm, clip=true,width=0.5\textwidth]
	{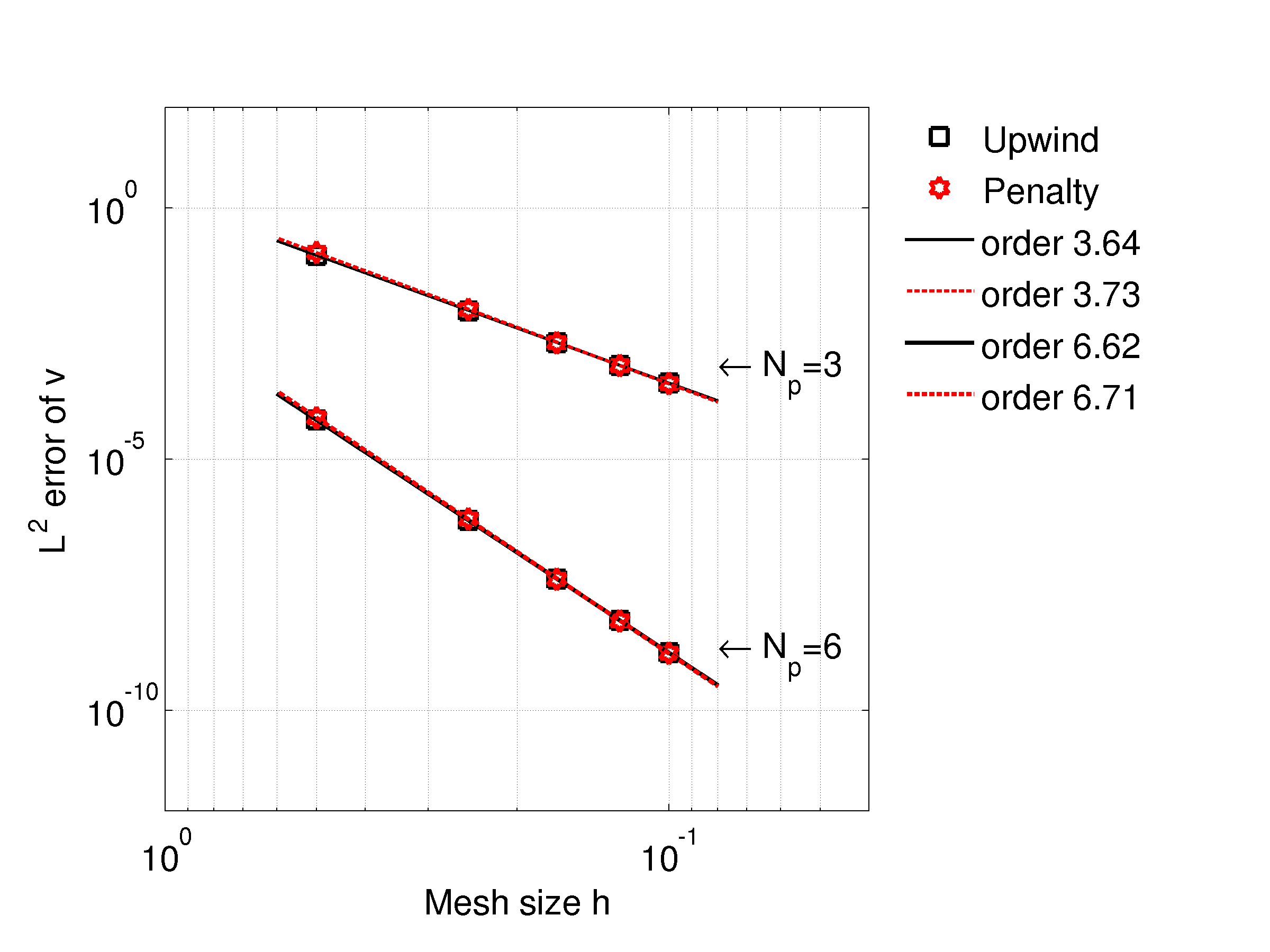}\\
	(C) & (D)
\end{tabular}
\caption{
  comparison of the accuracies and convergence rates between 
  the penalized numerical fluxes and the upwind flux when simulating
  (A) a plane wave,
  (B) a Rayleigh wave,
  (C) a Stoneley wave, and
  (D) a Scholte wave,
  for different orders $N_p = 3$ and $6$. 
  }
\label{fig:conv_cmp}
\end{figure}

\begin{figure}
\centering
\includegraphics[width=0.75\textwidth]{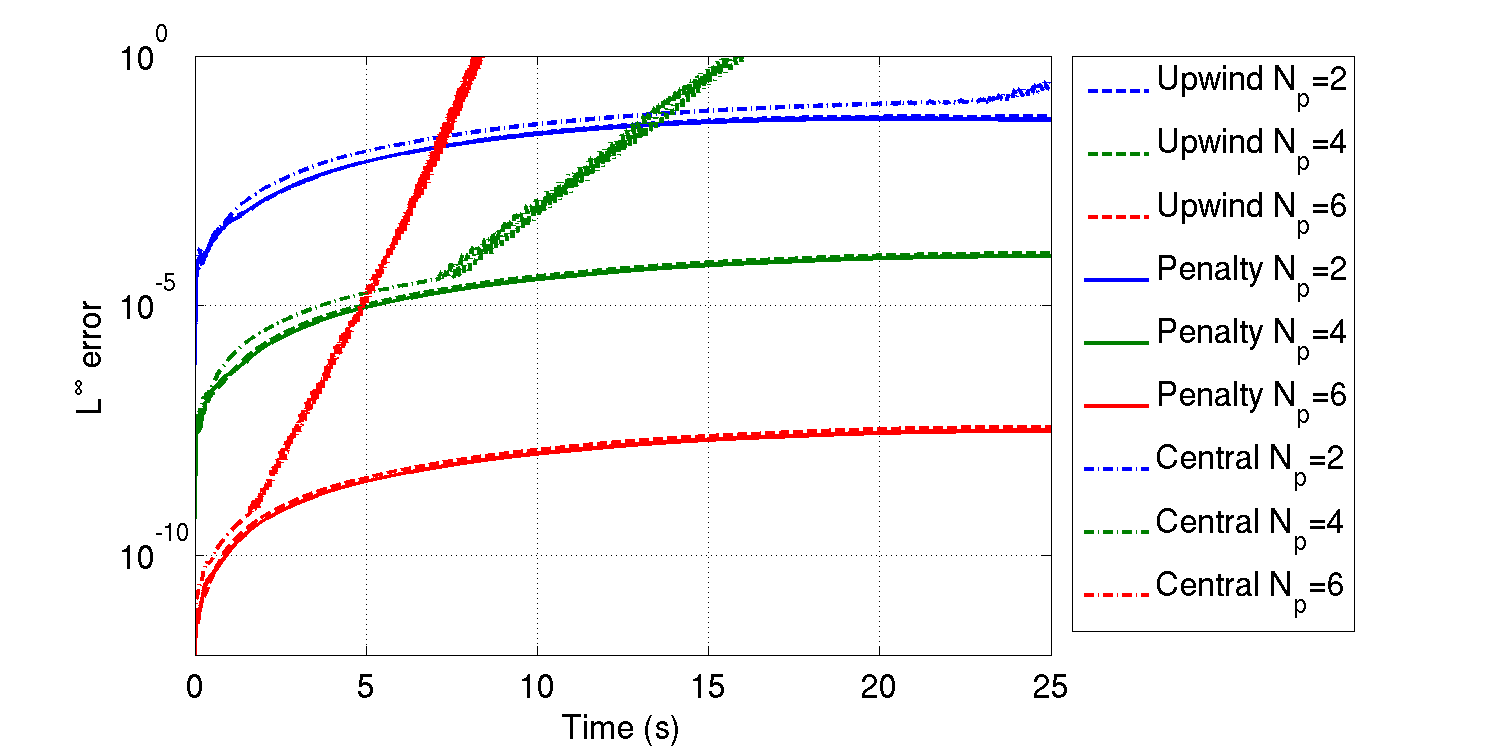}
\caption{Accuracy as a function for propagation time of three types of
  numerical fluxes for the simulation of a Scholte wave. 
  The maximum propagation time is $25$ s, which
  corresponds with a distance of $5.4$ wavelengths; $h = 0.25$km 
  ($0.125$ wavelengths). When the numerical method starts to blow up,
  the error function increases exponentially and the curve becomes jittery.}
\label{fig:Scholte_time_cmp}
\end{figure}

Figure~\ref{fig:conv_cmp} compares the accuracies and convergence rates 
of the penalized numerical fluxes with the upwind flux when simulating
a plane wave (A), a Rayleigh
wave (B), a Stoneley wave (C) and a Scholte wave (D), 
for both the lower-order case ($ N_p=3 $) and 
the relatively higher-order case ($ N_p=6 $).
We observe that the measured errors for the penalized flux 
are sometimes slightly larger than those for the upwind flux,
while the orders of convergence are essentially the same
(and both better than $\mathcal{O}(h^{N_p+\frac12})$). 

Figure~\ref{fig:Scholte_time_cmp} shows the time dependent error measure
for all three types of numerical fluxes. 
We simulate the Scholte wave for 25 seconds, with uniform mesh size
$h=0.125$km and order $N_p=2,4,6$ for each numerical flux.
{The central flux turns out to be unstable with
growing oscillatory numerical errors during time evolution. 
This instability may be caused by the inaccuracy of numerical integration, 
in which case the integration by parts does not necessary hold, 
and the energy-conserving property no longer holds. 
In comparison, the upwind and penalty fluxes
extend the tolerance to inexact numerical integration.
}


\bibliographystyle{gji} 
\bibliography{fluidsolid} 

\end{document}